\documentclass{aa}
\usepackage{graphicx}
\usepackage{natbib}
\newcounter{subfig}
\bibpunct{(}{)}{;}{a}{}{,} 
\usepackage{txfonts}
\usepackage[colorlinks,linkcolor=black,urlcolor=black,citecolor=black]{hyperref}
\begin{document}
\title{A new method for an objective, $\chi^2$-based spectroscopic analysis of early-type stars%
\thanks{Based on observations made with ESO Telescopes at the La Silla Paranal Observatory under program IDs 074.D-0021(A), 088.A-9003(A), and 091.C-0713(A). 
\newline
Based on observations made with the Nordic Optical Telescope, op\-erated by the Nordic Optical Telescope Scientific Association at the Observatorio del Roque de los Muchachos, La Palma, Spain, of the Instituto de Astrofisica de Canarias, proposal 41-027. 
}\fnmsep
\thanks{Figures \ref{fig:spectra_1}--\ref{fig:spectra_9} and \ref{fig:binary_spectra_1}--\ref{fig:binary_spectra_9} are available in electronic form at \newline \url{http://www.aanda.org}.}
}
\subtitle{First results from its application to single and binary B- and late O-type stars}
\author{A.~Irrgang\inst{\ref{remeis}}
        \and
        N.~Przybilla\inst{\ref{innsbruck}}
        \and
        U.~Heber\inst{\ref{remeis}}
        \and
        M.~B\"ock\inst{\ref{bonn}}
        \and
        M.~Hanke\inst{\ref{remeis}}
        \and 
        M.-F.~Nieva\inst{\ref{remeis},\ref{innsbruck}}
        \and 
        K.~Butler\inst{\ref{muenchen}}
       }
\institute{Dr.~Karl~Remeis-Observatory \& ECAP, Astronomical Institute, Friedrich-Alexander University Erlangen-Nuremberg,\\ Sternwartstr.~7, 96049 Bamberg, Germany\\
           \email{andreas.irrgang@fau.de}\label{remeis}
           \and
           Institut f\"ur Astro- und Teilchenphysik, Universit\"at Innsbruck, Technikerstr. 25/8, 6020 Innsbruck, Austria\label{innsbruck}
           \and
           Max-Planck Institut f\"ur Radioastronomie, Auf dem H\"ugel 69, 53121 Bonn, Germany\label{bonn}
           \and
           Universit\"atssternwarte M\"unchen, Scheinerstr.~1, 81679 M\"unchen, Germany\label{muenchen}
           }
\date{Received 2 December 2013 / Accepted 5 March 2014}

\abstract
{A precise quantitative spectral analysis, encompassing atmospheric parameter and chemical elemental abundance determination, is time-consuming due to its iterative nature and the multi-parameter space to be explored, especially when done by the naked eye.}
{A robust automated fitting technique that is as trustworthy as traditional methods would allow for large samples of stars to be analyzed in a consistent manner in reasonable time.}
{We present a semi-automated quantitative spectral analysis technique for early-type stars based on the concept of $\chi^2$ min\-imization. The method's main features are as follows: far less subjectivity than the naked eye, correction for inaccurate continuum normalization, consideration of the whole useful spectral range, and simultaneous sampling of the entire multi-parameter space (effective temperature, surface gravity, microturbulence, macroturbulence, projected rotational velocity, radial velocity, and elemental abundances) to find the global best solution, which is also applicable to composite spectra.}
{The method is fast, robust, and reliable as seen from formal tests and from a comparison with previous analyses.}
{Consistent quantitative spectral analyses of large samples of early-type stars can be performed quickly with very high accuracy.}

\maketitle
%
\section{Introduction}
The chemical evolution of galaxies is dominated by the evolution of early-type stars, since these objects are the progenitors of core-collapse supernovae, and therefore contribute to stellar nucleosynthesis in a pronounced way. In this context, important issues are the effects of rotation, especially for that of rotational mixing, on the evolution of massive stars \citep[e.g.,][]{rotation_heger,rotation_meynet}, as well as spatial and temporal variations of the chemical composition within the Galactic disk \citep[e.g.,][]{chemical_evolution_fuhrmann,chemical_evolution_przybilla}. Quantitative spectroscopic analyses of B- and late O-type stars allow for atmospheric parameters and chemical elemental surface abundances to be inferred with high precision, which directly addresses both of the aforementioned topics \citep[see][]{cas2}. Due to the high frequency of binary systems among early-type stars \citep[see, e.g.,][]{binaries_1,binaries_2}, analysis techniques, which are also able to deal with spectra of double-lined spectroscopic binary systems (SB2) are desirable.

Quantitative spectroscopy is based on the comparison of synthetic and observed spectra. Owing to the multi-dimensionality of the parameter space involved (which include the following for B- and late O-type stars: effective temperature $T_{\mathrm{eff}}$, surface gravity $\log (g\,\mathrm{(cm\,s^{-2})})$, microturbulence $\xi$, macroturbulence $\zeta$, projected rotational velocity $\varv\,\sin(i)$, radial velocity $\varv_{\mathrm{rad}}$, metallicity $Z$, and elemental abundances $\{n(x)\}$), investigations are time-consuming since an iterative approach is required \citep[for details see][]{nieva_iterative}. Starting from initial estimates for the entire set of parameters, individual variables are refined by using spectral indicators that are sensitive to as few parameters as possible to reduce the complexity of the problem. In early-type stars of solar metallicity, for instance, Stark-broadened hydrogen and helium lines are primarily affected by changes in $T_{\mathrm{eff}}$, $\log(g)$, and $n(\mathrm{He})$ while they are comparatively insensitive to all others. Consequently, these features allow for the temperature, surface gravity, and helium abundance to be constrained. The use of multiple ionization equilibria, which requires that spectral lines of different ionization stages of the same element indicate equal abundances, yields further constraints on $T_{\mathrm{eff}}$ and $\log(g)$ but also on $\xi$ and $n(x)$. Matching the strength of spectral lines and their shape allows for $\zeta$ and $\varv\,\sin(i)$ to be derived. Because of the highly non-trivial coupling of different parameters, the adjustment of individual variables involves re-evaluation of most of the available indicators, leading to an iterative procedure. 

Although those iterative steps can be automated to speed up the investigation \citep[see][]{lefever_fitting}, the underlying strategy is still prone to miss the global best solution. On the one hand, not all parameters are varied at the same time but many of them separately so that correlations between them are neglected \citep[see][]{mokiem_fitting}. On the other hand, parameters are constrained from selected spectral indicators or windows instead of exploiting the information encoded in the entire spectrum. A global analysis method, which is a method simultaneously probing all parameters while considering the maximum useful spectral range, is therefore our goal.

Automated fitting techniques are suitable for this purpose. Moreover, automation is far less subjective, since the matching of theory and observation is based on a mathematical measure, such as a $\chi^2$ criterion instead of visual inspection. This is particularly important when one wants to analyze larger samples in a homogeneous manner. The size of the corresponding multi-parameter space, however, requires calculations of numerous synthetic spectra, which is computationally expensive and, therefore, a major obstacle for automated fitting. To minimize the number of calculations involved, synthetic atmospheres may be computed on demand in the course of the fitting process, as realized by \citet{mokiem_fitting}. In this way, spectra are computed only if they are actually used. Nevertheless, even very efficient fitting algorithms can take from several dozens to hundreds of iterations to find the best solution, which implies a non-negligible run-time of the fitting process. This drawback can be overcome by making use of pre-calculated model grids in which interpolation between grid points can be used to evaluate the fitting function. Unfortunately, sufficient sampling of the whole multi-parameter space is typically not possible, given its large dimension. Consequently, grid-based fitting methods are usually restricted to small subspaces by either keeping some parameters fixed when computing tailored grids \citep[see][]{castro_fitting} or limiting the allowed parameter range, thus reducing the advantages of global automated fitting.

However, due to some unique properties of spectra of early-type stars, such as the low density of spectral lines and the continuous opacity that is dominated by hydrogen and helium, many parameters --~in particular, the elemental abundances of the trace elements~-- are independent of each other. Exploiting this fact, it is possible to probe the entire parameter space by computing only a tiny fraction of it. Based on this idea, we have developed a grid-based global fitting method that facilitates quick and precise determinations of the atmospheric parameters of B- and late O-type stars, which takes non-local thermodynamic equilibrium (non-LTE) effects into consideration (Sect.~\ref{sec:setting_up}). Furthermore, it is shown that the accuracy of the analysis is generally not limited by statistics, such as the signal-to-noise ratio (S/N) of the observed spectrum, but rather by systematics, such as the uncertainties in atomic data (Sect.~\ref{sec:tests}). For demonstration purposes, the method is then applied to three well-studied early-type stars in Orion and to three SB2 systems yielding atmospheric and fundamental stellar parameters (Sect.~\ref{sec:analysis}). A discussion of the results obtained (Sect.~\ref{sec:discussion}) is rounded off by a summary (Sect.~\ref{sec:summary}).
\section{Setting up the fitting function}\label{sec:setting_up}
\subsection{Input physics: models and codes}
\begin{table}
\caption{\label{table:model_atoms} Model atoms for non-LTE calculations.}
\begin{tabular}{ll}
\hline\hline
Ion & Model atom \\
\hline
\ion{H}             & \citet{hydrogen_modelatom} \\
\ion{He}{i/ii}      & \citet{helium_modelatom} \\
\ion{C}{ii-iii}     & \citet{carbon_modelatom1,carbon_modelatom2} \\
\ion{N}{ii}         & \citet{nitrogen_modelatom}\tablefootmark{a} \\
\ion{O}{i/ii}       & \citet{oxygen_modelatom1}, \citet{oxygen_modelatom2}\tablefootmark{a} \\
\ion{Ne}{i/ii}      & \citet{neon_modelatom}\tablefootmark{a} \\
\ion{Mg}{ii}        & \citet{magnesium_modelatom} \\
\ion{Al}{ii/iii}    & Przybilla (in prep.) \\
\ion{Si}{ii/iii/iv} & Przybilla \& Butler (in prep.) \\
\ion{S}{ii/iii}     & \citet{sulfur_modelatom}, updated \\
\ion{Ar}{ii}        & Butler (in prep.) \\
\ion{Fe}{ii/iii}    & \citet{iron_modelatom1}, \citet{iron_modelatom2}\tablefootmark{b} \\
\hline
\end{tabular}
\tablefoot{\tablefoottext{a}{Updated, see \citet{cas2} for details.}\tablefoottext{b}{Corrected, see \citet{cas2} for details.}}
\end{table}
Synthetic spectra are calculated by following the hybrid, non-LTE approach discussed by \citet{hybrid1} and \citet{carbon_modelatom1,hybrid2,carbon_modelatom2}. The structure of the atmosphere, such as the stratification of temperature and density, is based on line-blanketed, plane-parallel, homogeneous, and hydrostatic LTE-model atmospheres calculated with {\sc Atlas12} \citep{atlas12}. Deviations from local thermodynamic equilibrium are then accounted for by applying updated versions of {\sc Detail} and {\sc Surface} \citep{detailsurface1,detailsurface2}. The {\sc Detail} code numerically solves the coupled radiative transfer and statistical equilibrium equations to obtain population numbers in non-LTE, which in turn are input to the {\sc Surface} code to compute the final synthetic spectrum that uses more detailed line-broadening data. Comprehensive, state-of-the-art model atoms (see Table~\ref{table:model_atoms}) allow spectral lines to be modeled with high fidelity for chemical elements accessible in the optical spectra of most B- and late O-type stars. Although partly based on LTE concepts, the hybrid method outlined above is consistent with full non-LTE calculations for B-type stars \citep[see][]{hybrid2,hybrid3} but is considerably faster and, most importantly, is able to handle more realistic representations of model atoms. For the first time here, we consistently use the concept of LTE opacity sampling to treat background line opacities (employing the realization of \citealt{atlas12}) in all computational steps, instead of relying on pre-calculated opacity distribution functions.
\subsection{Probing the parameter space}
Stellar atmospheres are described by a variety of physical quantities. The implementation of those parameters in synthetic model computations varies depending on their nature. Within the scope of this paper, it is instructive to distinguish between primary atmospheric parameters, which are quantities whose effects can be modeled solely by solving the coupled radiative and structural equations describing an atmosphere, and secondary ones, which are quantities whose effects can be incorporated after an atmosphere has been built. For OB-type stars near the main sequence and their evolved progeny, BA-type supergiants, primary parameters are $T_{\mathrm{eff}}$, $\log(g)$, $\xi$, $Z$, and $\{n(x)\}$, while $\zeta$, $\varv\,\sin(i)$, and $\varv_{\mathrm{rad}}$ are secondary ones. By definition, the effective parameter space for which model atmospheres have to be calculated is therefore spanned by the primary parameters. Given a grid-based fitting method that  interpolates a regular multi-dimensional mesh of synthetic spectra, the dimension of this parameter space is the product of the individual dimensions of the primary parameters. Due to this multiplicity, each chemical element included in the analysis drastically increases the number of required models. For instance, calculating a grid with ten elements at nine abundances each implies a factor of $9^{10}$, which has to be multiplied with the dimension given by the remaining primary parameters. Since it is infeasible to compute such a large number of models in a reasonable time, a simplification is necessary. To this end, one can exploit the unique properties of (nearby) early-type stars: Due to their young ages, the metallicity of these stars is very well approximated by the cosmic standard \citep{cas2}, in general, so that $Z=0.014$ can be assumed throughout. In massive stars, the latter implies that all metals can be considered as trace elements, which implies that moderate changes of their abundances do not significantly affect the atmospheric structure. This occurs because their respective bound-free opacities are small compared to hydrogen and helium and because line blanketing is relatively insensitive to abundance changes of a specific metal. Altering the abundance of an individual metal has therefore little to no effect on spectral lines of the other metals, except in the case of line blends. However, the density of lines is low in optical spectra of early-type stars, so that intrinsic blends are actually rare (With some more restrictions, this even applies to the ultraviolet region.). Moreover, many of these line blends are caused by macroscopic flux redistributions, which can be, for example, via macroturbulence, stellar rotation, or the finite resolving power of the spectrograph, and are thus not an issue, since they do not affect the microscopic physics governing radiative transfer. Consequently, instead of calculating $9^{10}$ combinations of abundances, it is sufficient to compute models for nine pre-chosen abundance values for each of the ten species. All of these models contain only lines of one specific trace element, while standard abundances are assumed for the other metals to account for their background opacities. Those $10 \times 9$ spectra serve as a basis from which each of the $9^{10}$ combinations can be constructed, such as via multiplication of the normalized base spectra. Note that helium cannot be treated as a trace element, since corresponding abundance changes have serious impacts on the atmospheric structure, for instance, via the mean molecular weight or its contribution to the continuum opacity. Hence, the number of required synthetic models is the product of the dimensions of $T_{\mathrm{eff}}$, $\log(g)$, $\xi$, $n(\mathrm{He})$, and the base dimensions of the trace elements.
\subsection{Numerical implementation of the model}\label{subsection:numerical_implementation}
From the considerations of the previous subsection, the fitting function is constructed as follows: for each trace element of interest, a multi-dimensional regular mesh spanned by effective temperature ($12\,000$ to $34\,000\,\mathrm{K}$, step size: $1000\,\mathrm{K}$), surface gravity (range depending on temperature but typically within $3.0$ to $4.6\,\mathrm{dex}$, $0.2\,\mathrm{dex}$), microturbulence (range depending on temperature but typically within $0$ to $20\,\mathrm{km\,s^{-1}}$, $2\,\mathrm{km\,s^{-1}}$), helium abundance (solar value minus $0.3\,\mathrm{dex}$ to solar value plus $0.5\,\mathrm{dex}$, $0.2\,\mathrm{dex}$), and abundance of the species itself (typically solar value minus $0.8\,\mathrm{dex}$ to solar value plus $0.8\,\mathrm{dex}$, $0.2\,\mathrm{dex}$) has been calculated in advance. This grid can easily be extended whenever needed. Spectra for arbitrary parameters within the mesh are approximated by linear interpolation. At this point, synthetic models contain only spectral lines of hydrogen, helium, and the element under consideration. To combine the normalized spectra of the individual trace elements, each of them is divided by a model with the same specifications but which contains only hydrogen and helium lines. Multiplying the resulting ``corrected spectra'' with each other and with the hydrogen-helium model leads to a model that takes all of the primary parameters into account. This simplification is well justified as long as spectral lines of different metals are not blended due to microscopic flux redistributions, such as microturbulence, natural line width, and thermal- or pressure broadening. Furthermore, even for blends of weak spectral lines, the method is a good approximation as interaction effects are tiny in that case. Secondary atmospheric parameters are incorporated afterwards: macroturbulence and stellar rotation via convolution of the synthetic spectrum with a joint profile function, which is obtained from numerical stellar disk integration \citep{gray} with a linear limb-darkening law \citep[using coefficients by][]{claret_bloemen}, and radial velocity by shifting the entire wavelength scale, according to the Doppler formula.
\subsection{Extension to composite spectra}
Multiplicity is an important issue for massive stars, as the majority of them probably resides in systems of two (or more) components \citep[see][]{binaries_1, binaries_2}. While visual binaries or systems with a much fainter secondary can be treated as single stars in the spectroscopic analysis, this is clearly not the case for double-lined spectroscopic binary systems where features of both stars are visible in the spectrum. Stars classified as chemically peculiar objects based on low-resolution and/or low-S/N spectra might actually be unrecognized SB2 systems where the depths of the absorption lines have been altered due to continuum emission of a companion. 

The fitting method presented here is also capable of dealing with composite spectra in a simplified manner. To do so, models for the normalized single-star spectra are created separately for the primary ($f_{\mathrm{p}}$) and secondary ($f_{\mathrm{s}}$) component following the procedure given in Sect.~\ref{subsection:numerical_implementation}. To obtain the model for their normalized composite spectrum ($f_{\mathrm{comp}}$), these individual contributions have to be summed up while taking their weights into account. The latter are given by the components' continuum fluxes ($f_{\mathrm{p,cont}}$, $f_{\mathrm{s,cont}}$) and by the ratio of their projected, effective\footnote{Spectra received from different surface elements on the star are assumed to be identical in Eq.~(\ref{eq:composite}). Consequently, effects like limb darkening are neglected, and $A_{\mathrm{eff,p}}$ and $A_{\mathrm{eff,s}}$ are more effective than absolute surface areas. However, if those effects are qualitatively similar in the primary and secondary component, they probably cancel each other out when taking the ratio of surface areas yielding $A_{\mathrm{eff,s}}/A_{\mathrm{eff,p}} \approx A_{\mathrm{s}}/A_{\mathrm{p}}$.} surface areas, which is parametrized by one additional free parameter $A_{\mathrm{eff,s}}/A_{\mathrm{eff,p}}$:
\begin{equation}
	f_{\mathrm{comp}} = \frac{ f_{\mathrm{cont,p}} \, f_{\mathrm{p}} + A_{\mathrm{eff,s}}/A_{\mathrm{eff,p}} \, f_{\mathrm{cont,s}} \, f_{\mathrm{s}}}{f_{\mathrm{cont,p}} + A_{\mathrm{eff,s}}/A_{\mathrm{eff,p}} \, f_{\mathrm{cont,s}}}\, . \;
   \label{eq:composite}
\end{equation}
Note that this approach conflates all binary interactions, such as mutual distortions and their respective effects on the emitted spectra into the parameter $A_{\mathrm{eff,s}}/A_{\mathrm{eff,p}}$. Consequently, the geometric interpretation of $A_{\mathrm{eff,s}}/A_{\mathrm{eff,p}}$ as the ratio of the effective surface areas of secondary to primary component is only valid as long as those interactions are negligible, such as for well detached systems. Otherwise, $A_{\mathrm{eff,s}}/A_{\mathrm{eff,p}}$ is a combination of the ratio of effective surface areas and a fudge factor that accounts for all missing interaction effects. Of course, the ability of $A_{\mathrm{eff,s}}/A_{\mathrm{eff,p}}$ to compensate these shortcomings is limited and certainly fails for very close or contact systems where a much more sophisticated method is necessary \citep[see, for instance,][]{binary_spectra_1,binary_spectra_2}. Nevertheless, the approach outlined here offers a very fast and efficient way to derive atmospheric parameters and elemental abundances of SB2 systems from the analysis of a single epoch composite spectrum. 
\subsection{Comparison with observation}
To compare models with observations, the synthetic spectra are finally convolved with the instrumental profile, which is a Gaussian curve whose full width at half maximum $\Delta \lambda = \lambda / R$ is a function of the spectral resolving power $R$.

The goodness of fit of a model is then derived from its absolute $\chi^2$: 
\begin{equation}
   \chi^2 = \sum\limits_{i}\chi_{i}^2 = \sum\limits_{i}\left(\frac{f_i-f_{{\mathrm{model,}}i}}{\delta_i}\right)^2\, . \;
   \label{eq:chisqr}
\end{equation}
Here, $f_i$, $f_{{\mathrm{model,}}i}$ and $\delta_i$ are the observed flux, model flux and uncertainty in the observed flux at data point $i$, respectively. The sum is taken over all pixels $i$ in the spectrum, which either excludes only those lines that have well-known shortcomings, are missing in our models, or are of non-photospheric origin, such as telluric lines and diffuse interstellar bands. The lower the value of $\chi^2$, the better the quality of the fit\footnote{A reduced $\chi^2$ lower than 1 indicates overestimated uncertainties.}. Other ways of measuring the fitness exist but may have different properties. One reason to choose a standard $i$-based over a line-based $\chi^2$ criterion \citep[as, for instance, done by][]{simon_diaz_fitting,castro_fitting} is that it gives lines with more data points a larger weight than those with less information. Moreover, the corresponding $\chi^2$ distribution is well studied, which allows statistical uncertainties of the input parameters to be deduced from the $\chi^2$ statistics (see Sect.~\ref{subsection:uncertainties}).

The analysis is carried out completely within the Interactive Spectral Interpretation System \citep[ISIS,][]{isis}, which simplifies the investigation tremendously. Apart from many useful functions and tools, various minimization algorithms, such as the simplex or gradient methods are available. As shown in the next section, the $\chi^2$ landscape of our problem is generally very well behaved so that the absolute minimum is found after a relatively small number of steps, which allows for quick and efficient analyses.
\section{Formal tests and discussion of uncertainties}\label{sec:tests}
\subsection{Noise estimation}\label{subsection:noise_estimation}
Calculating $\chi^2$ requires proper knowledge of the measurement uncertainties $\delta_i$, which either are systematic in nature and, for instance, caused by an incorrect continuum normalization, or, more importantly, have statistical fluctuations, such as noise $n_i$. The latter can be estimated from an observed spectrum, which does not need to be flux calibrated, in an easy, fast, and robust way when assuming that the noise $n_i$ of data point $i$ obeys a Gaussian probability distribution $p(n_i)$ with a mean value of zero and a priori unknown standard deviation $\sigma_i$:
\begin{equation}
p(n_i) = \frac{1}{\sqrt{2\pi}\sigma_i}\exp\left(-\frac{n_i^2}{2\sigma_i^2}\right)\, . \;
\label{eq:noise_distribution}
\end{equation}
For small regions (several data points), the measured flux $f_i$ as a function of the wavelength $\lambda$ can be approximately written as the sum of a linear function $a + b \lambda_i$, which represents the first two terms in a Taylor expansion of the pure signal and a noise component $n_i$:
\begin{equation}
f_i = a + b \lambda_i + n_i\, . \;
\end{equation}
To estimate the noise level $\sigma_i$, consider the quantity $\Delta_i$ defined as
\begin{eqnarray}
\Delta_i & \equiv & f_i - (w_{i-2}f_{i-2} + w_{i+2}f_{i+2})\nonumber \\
         &    =   & n_i - w_{i-2}n_{i-2} - w_{i+2}n_{i+2} + a + b \lambda_i \nonumber \\
         &        & - w_{i-2}(a + b \lambda_{i-2}) - w_{i+2}(a + b \lambda_{i+2})\label{eq:defining_iota}\, . \;
\end{eqnarray}
Here, $w_{i-2}$ and $w_{i+2}$ are weight factors and chosen such that only the noise terms in Eq.~(\ref{eq:defining_iota}) remain:\footnote{Note: $\lambda_{i+1} = \lambda_i + \Delta \lambda_i^{\mathrm{pixel}}/2 + \Delta \lambda_{i+1}^{\mathrm{pixel}}/2$ with $\Delta \lambda_i^{\mathrm{pixel}} = \Delta \lambda_i/2$ because of Nyquist's sampling theorem. For long-slit spectrographs, $\lambda / \Delta \lambda = R_{\mathrm{long-slit}} \propto \lambda$ so that $\Delta \lambda_i = \mathrm{constant}$ which implies $w_{i-2}=1-w_{i+2}=1/2$ to arrive at Eq.~(\ref{eq:defining_iota2}) from Eq.~(\ref{eq:defining_iota}). For Echelle spectrographs, $\lambda / \Delta \lambda = R_{\mathrm{Echelle}} = \mathrm{const.}$, which yields $\lambda_{i+1} = k \lambda_i$ with $k=(4 R_{\mathrm{Echelle}}+1)/(4 R_{\mathrm{Echelle}}-1)$ which leads to $w_{i-2}=1-w_{i+2}=(k^2-1)/(k^2-k^{-2}) \approx 0.5$.}
\begin{equation}
\Delta_i = n_i - w_{i-2}n_{i-2} - w_{i+2}n_{i+2}\, . \;
\label{eq:defining_iota2}
\end{equation}
The reason for comparing $f_i$ to the (weighted) average of data points $i-2$ and $i+2$ instead of $i-1$ and $i+1$ is that adjacent pixels are likely correlated. For example, this is due to detector cross-talk or actions taken during data reduction like the wavelength calibration. Assuming that there is no correlation with the next neighbor but one implies that Eq.~(\ref{eq:noise_distribution}) is valid for points $i$, $i-2$, and $i+2$, and the probability distribution $p(\Delta_i)$ reads:
\begin{eqnarray}
p(\Delta_i) & = & \int\limits_{-\infty}^{\infty} \int\limits_{-\infty}^{\infty} \int\limits_{-\infty}^{\infty} p(n_i)p(n_{i-2})p(n_{i+2})\\          && \delta\left(\Delta_i - (n_i - w_{i-2}n_{i-2} - w_{i+2}n_{i+2}) \right)\mathrm{d} n_i \mathrm{d} n_{i-2}\mathrm{d} n_{i+2}\, . \; \nonumber
\end{eqnarray}
Here, $\delta$ is the dirac delta function. For adjacent data points, it is well justified to assume $\sigma_i = \sigma_{i-2} = \sigma_{i+2} \equiv \sigma$, so that $p(\Delta_i)$ simplifies to
\begin{equation}
p(\Delta_i) =  \frac{1}{\sqrt{2\pi} \tilde{\sigma}}\exp\left(-\frac{\Delta_i^2}{2\tilde{\sigma}^2}\right), \quad \tilde{\sigma} = \sigma\sqrt{w^2_{i-2}+w^2_{i+2}+1}\, . \;
\label{eq:iota_distribution}
\end{equation}
Consequently, the distribution of $\Delta_i =  f_i - (w_{i-2}f_{i-2} + w_{i+2}f_{i+2})$ is a Gaussian with a standard deviation $\tilde{\sigma}$ defined by Eq.~(\ref{eq:iota_distribution}). Extending the assumption of a constant noise level $\sigma_i = \sigma$ to a statistically significant number of data points allows $\sigma$, which is the statistical component of the uncertainty $\delta_i$ in Eq.~(\ref{eq:chisqr}), to be derived from the measurable distribution of $\Delta_i$. If the reduced $\chi^2$ at the best fit is larger than $1$, it might also be necessary to consider a systematic component of $\delta_i$ (see footnote~\ref{footnote:chisqr}).
\subsection{Performance and reliability of the method}\label{subsection:performance}
Before fitting real spectra, several formal tests were carried out to examine the properties of the automated method. With this aim, mock spectra were constructed from synthetic ones by adding Gaussian-distributed noise that corresponded to different S/N. A spectral range $[3940\,\mathrm{\AA},7000\,\mathrm{\AA}]$ was chosen to match the minimum wavelength coverage of standard high-resolution spectrographs. Regions in that interval that are generally affected by telluric features were excluded. The spectral resolving power was set to $R = 45\,000$, which is very close to the resolution of the Echelle spectra which are analyzed in this study.

Table~\ref{table:uncertainties} lists the results of this exercise for ten exemplary cases. The input parameters and, thus, the global minimum were recovered with excellent accuracy after only a run-time of few minutes on a standard $3.1$\,GHz single-core processor and independent of the choice of the starting parameters within the grid, which shows that our method is fast and reliable.

It is important to stress here that all mock spectra were constructed from complete {\sc Surface} models, which, in contrast to the fitting function, treat every microscopic line blend correctly by simultaneously computing lines of all chemical species under consideration. Additionally, models off the grid points were chosen to check that our mesh is sufficiently spaced for the linear interpolation scheme applied. Because the differences of input and output values in Table~\ref{table:uncertainties} are often covered by the very small statistical uncertainties (see Sect.~\ref{subsection:uncertainties}) that result from the high S/N assigned to the mock spectra, we conclude that inaccuracies introduced by simplifications in our approach are negligible.

In a second step, three mock composite spectra were created with the help of Eq.~(\ref{eq:composite}). The parameters chosen here are motivated by real SB2 systems and anticipate the results presented in Sect.~\ref{sec:analysis}. They cover a sharp-lined, well-separated and, thus, easy to analyze system and a very difficult configuration with heavily blended spectral features. Similar to the previous tests, most of the input parameters are recovered with very high precision or at least within the derived uncertainties, as seen in Table~\ref{table:uncertainties_binary}. In particular, the degree of accuracy in the inferred parameters of both components of the extremely blended composite spectrum is astonishing, hence, making us quite confident that our method is also highly suitable for investigations of SB2 systems.

Although our method is able to model individual abundances if necessary, we generally prefer to assume an identical chemical composition for the two components within the binary system. In this way, the number of free parameters and, consequently, the numerical complexity of the problem is significantly reduced. In cases where the parameters of the secondary component are only poorly constrained due to their little impact on the composite spectrum, it is even necessary to impose these constraints, which compensate for the lack of spectral indicators to derive reasonable atmospheric parameters. Note that the assumption of an equal chemical composition is well justified for SB2 systems containing B-type or late O-type stars. On the one hand, the components of SB2 systems are in general similar regarding to their masses (Otherwise, the flux contribution of the fainter companion would not be visible in the spectrum.), ages (the whole system formed at once), and pristine chemical composition (both components formed from the same building material). On the other hand, processes causing chemical anomalies are rare among B- and late O-type stars and primarily affect helium \citep{helium_weak_rich}. Since chemical peculiarities are possibly even less frequent in detached binary systems \citep{pavlovski_southworth}, elemental abundances are expected to evolve in the same way in both components. To estimate the influence of this approximation on the spectral analysis, Table~\ref{table:uncertainties_binary} lists the results obtained from fitting the three mock composite spectra with adjustable abundances and an equal chemical composition. Even for the system with different individual abundances, the results derived by assuming an identical chemical composition are very satisfying. In particular, this is with respect to the primary component, which dominates the spectrum and in this way also the estimates for the system abundances and their respective confidence limits. As a consequence, the actual abundances of the secondary component may sometimes lie outside of the uncertainty intervals determined for the binary system as a whole. Nevertheless, it is obvious that the decision of whether or not to use separate abundances during the fitting process depends on the individual object and has to be checked, for example, a posteriori by inspecting the final match of the model to the observation.
\begin{table*}
\tiny
\setlength{\tabcolsep}{0.15cm}
\renewcommand{\arraystretch}{1.26}
\caption{\label{table:uncertainties} Comparison of input parameters (\textit{``In'' row}) and corresponding parameters obtained from fits (\textit{``Out'' row}) for ten exemplary mock spectra computed from complete {\sc Surface} models, which simultaneously account for all lines considered in the fitting function.}
\centering
\begin{tabular}{lrrrrrrrrrrrrrrrrrrr}
\hline\hline
 & S/N & $T_{\mathrm{eff}}$ & $\log(g)$ & $\varv_{\mathrm{rad}}$ & $\varv\,\sin(i)$ & $\zeta$ & $\xi$ & & \multicolumn{11}{c}{$\log(n(x))$} \\ 
\cline{5-8} \cline{10-20}
& & (K) & (cgs) & \multicolumn{4}{c}{$(\mathrm{km\,s^{-1}})$} & & He & C & N & O & Ne & Mg & Al & Si & S & Ar & Fe\\
\hline
In & $300$ & $15000$ & $3.750$ & $-18.0$ & $15.0$ & $7.0$ & $2.00$ & & $-1.06$ & $-3.70$ & $-4.30$ & $-3.20$ & $-4.00$ & $-4.70$ & $-5.60$ & $-4.60$ & $-4.80$ & $-5.60$ & $-4.60$ \\
Out & $305$ & $15000$ & $3.750$ & $-18.0$ & $14.8$ & $8.0$ & $2.10$ & & $-1.06$ & $-3.71$ & $-4.28$ & $-3.20$ & $-4.01$ & $-4.70$ & $-5.60$ & $-4.61$ & $-4.81$ & $-5.68$ & $-4.61$ \\
Stat. & $^{+3}_{-3}$ & $^{+20}_{-20}$ & $^{+0.003}_{-0.004}$ & $^{+0.1}_{-0.1}$ & $^{+0.1}_{-0.1}$ & $^{+0.2}_{-0.1}$ & $^{+0.04}_{-0.03}$ & & $^{+0.01}_{-0.01}$ & $^{+0.02}_{-0.02}$ & $^{+0.04}_{-0.04}$ & $^{+0.02}_{-0.02}$ & $^{+0.02}_{-0.02}$ & $^{+0.02}_{-0.02}$ & $^{+0.02}_{-0.02}$ & $^{+0.02}_{-0.02}$ & $^{+0.01}_{-0.01}$ & $^{+0.07}_{-0.08}$ & $^{+0.01}_{-0.01}$ \\
Sys. & \ldots & $^{+300}_{-300}$ & $^{+0.100}_{-0.100}$ & $^{+0.1}_{-0.1}$ & $^{+0.2}_{-0.1}$ & $^{+0.7}_{-1.2}$ & $^{+0.79}_{-1.09}$ & & $^{+0.15}_{-0.09}$ & $^{+0.07}_{-0.09}$ & $^{+0.05}_{-0.08}$ & $^{+0.04}_{-0.08}$ & $^{+0.04}_{-0.02}$ & $^{+0.06}_{-0.07}$ & $^{+0.02}_{-0.02}$ & $^{+0.12}_{-0.18}$ & $^{+0.04}_{-0.04}$ & $^{+0.08}_{-0.14}$ & $^{+0.06}_{-0.10}$ \\
Start & \ldots & $19000$ & $3.700$ & $-15.0$ & $10.0$ & $10.0$ & $3.00$ & & $-0.85$ & $-3.60$ & $-4.20$ & $-3.20$ & $-4.00$ & $-4.60$ & $-5.80$ & $-4.40$ & $-4.90$ & $-5.60$ & $-4.60$ \\
\hline
In & $300$ & $15000$ & $4.250$ & $36.0$ & $17.0$ & $16.0$ & $2.00$ & & $-1.06$ & $-3.70$ & $-4.30$ & $-3.20$ & $-4.00$ & $-4.70$ & $-5.60$ & $-4.60$ & $-4.80$ & $-5.60$ & $-4.60$ \\
Out & $295$ & $14860$ & $4.215$ & $36.0$ & $17.0$ & $16.1$ & $1.90$ & & $-1.04$ & $-3.68$ & $-4.32$ & $-3.22$ & $-4.00$ & $-4.72$ & $-5.60$ & $-4.61$ & $-4.80$ & $-5.50$ & $-4.65$ \\
Stat. & $^{+2}_{-2}$ & $^{+30}_{-20}$ & $^{+0.008}_{-0.006}$ & $^{+0.1}_{-0.1}$ & $^{+0.1}_{-0.1}$ & $^{+0.3}_{-0.2}$ & $^{+0.07}_{-0.03}$ & & $^{+0.01}_{-0.01}$ & $^{+0.02}_{-0.01}$ & $^{+0.07}_{-0.07}$ & $^{+0.01}_{-0.01}$ & $^{+0.02}_{-0.02}$ & $^{+0.02}_{-0.01}$ & $^{+0.02}_{-0.02}$ & $^{+0.02}_{-0.01}$ & $^{+0.01}_{-0.01}$ & $^{+0.10}_{-0.11}$ & $^{+0.01}_{-0.01}$ \\
Sys. & \ldots & $^{+300}_{-300}$ & $^{+0.100}_{-0.100}$ & $^{+0.1}_{-0.1}$ & $^{+0.1}_{-0.1}$ & $^{+0.2}_{-0.3}$ & $^{+0.62}_{-0.86}$ & & $^{+0.15}_{-0.12}$ & $^{+0.07}_{-0.08}$ & $^{+0.06}_{-0.09}$ & $^{+0.07}_{-0.07}$ & $^{+0.06}_{-0.04}$ & $^{+0.06}_{-0.06}$ & $^{+0.02}_{-0.02}$ & $^{+0.11}_{-0.11}$ & $^{+0.04}_{-0.04}$ & $^{+0.08}_{-0.14}$ & $^{+0.08}_{-0.10}$ \\
Start & \ldots & $19000$ & $3.700$ & $35.0$ & $10.0$ & $10.0$ & $3.00$ & & $-0.85$ & $-3.60$ & $-4.20$ & $-3.20$ & $-4.00$ & $-4.60$ & $-5.80$ & $-4.40$ & $-4.90$ & $-5.60$ & $-4.60$ \\
\hline       
In & $100$ & $20000$ & $3.750$ & $19.0$ & $25.0$ & $18.0$ & $2.0$ & & $-1.06$ & $-3.70$ & $-4.30$ & $-3.20$ & $-4.00$ & $-4.70$ & $-5.60$ & $-4.60$ & $-4.80$ & $-5.60$ & $-4.60$ \\
Out & $100$ & $19960$ & $3.734$ & $18.9$ & $23.5$ & $22.0$ & $1.94$ & & $-1.05$ & $-3.68$ & $-4.28$ & $-3.19$ & $-4.02$ & $-4.71$ & $-5.56$ & $-4.57$ & $-4.78$ & $-5.59$ & $-4.59$ \\
Stat. & $^{+1}_{-1}$ & $^{+120}_{-120}$ & $^{+0.015}_{-0.017}$ & $^{+0.3}_{-0.3}$ & $^{+1.1}_{-1.1}$ & $^{+1.7}_{-0.9}$ & $^{+0.3}_{-0.7}$ & & $^{+0.02}_{-0.02}$ & $^{+0.04}_{-0.03}$ & $^{+0.04}_{-0.04}$ & $^{+0.04}_{-0.03}$ & $^{+0.04}_{-0.04}$ & $^{+0.06}_{-0.05}$ & $^{+0.05}_{-0.04}$ & $^{+0.07}_{-0.05}$ & $^{+0.03}_{-0.03}$ & $^{+0.06}_{-0.08}$ & $^{+0.05}_{-0.04}$ \\
Sys. & \ldots & $^{+400}_{-400}$ & $^{+0.100}_{-0.100}$ & $^{+0.1}_{-0.1}$ & $^{+0.2}_{-0.2}$ & $^{+0.7}_{-0.7}$ & $^{+0.8}_{-1.7}$ & & $^{+0.09}_{-0.07}$ & $^{+0.08}_{-0.06}$ & $^{+0.08}_{-0.07}$ & $^{+0.11}_{-0.11}$ & $^{+0.04}_{-0.04}$ & $^{+0.11}_{-0.09}$ & $^{+0.08}_{-0.07}$ & $^{+0.14}_{-0.12}$ & $^{+0.05}_{-0.05}$ & $^{+0.05}_{-0.05}$ & $^{+0.05}_{-0.05}$ \\
Start & \ldots & $19000$ & $3.700$ & $10.0$ & $10.0$ & $10.0$ & $3.0$ & & $-0.85$ & $-3.40$ & $-4.10$ & $-3.10$ & $-4.20$ & $-4.80$ & $-5.60$ & $-4.20$ & $-4.70$ & $-5.70$ & $-4.50$ \\
\hline
In & $250$ & $20000$ & $4.250$ & $25.0$ & $5.0$ & $26.0$ & $2.00$ & & $-1.06$ & $-3.70$ & $-4.30$ & $-3.20$ & $-4.00$ & $-4.70$ & $-5.60$ & $-4.60$ & $-4.80$ & $-5.60$ & $-4.60$ \\
Out & $254$ & $19990$ & $4.251$ & $25.0$ & $0.0$ & $26.9$ & $2.26$ & & $-1.07$ & $-3.73$ & $-4.31$ & $-3.22$ & $-4.02$ & $-4.72$ & $-5.58$ & $-4.64$ & $-4.81$ & $-5.61$ & $-4.60$ \\
Stat. & $^{+2}_{-2}$ & $^{+20}_{-20}$ & $^{+0.003}_{-0.004}$ & $^{+0.1}_{-0.1}$ & $^{+1.8}_{-0.0}$ & $^{+0.1}_{-0.1}$ & $^{+0.12}_{-0.07}$ & & $^{+0.01}_{-0.01}$ & $^{+0.02}_{-0.02}$ & $^{+0.02}_{-0.02}$ & $^{+0.02}_{-0.02}$ & $^{+0.02}_{-0.02}$ & $^{+0.02}_{-0.02}$ & $^{+0.02}_{-0.02}$ & $^{+0.02}_{-0.02}$ & $^{+0.01}_{-0.01}$ & $^{+0.03}_{-0.04}$ & $^{+0.02}_{-0.02}$ \\
Sys. & \ldots & $^{+400}_{-400}$ & $^{+0.100}_{-0.100}$ & $^{+0.1}_{-0.1}$ & $^{+3.4}_{-0.0}$ & $^{+0.1}_{-0.1}$ & $^{+1.04}_{-1.66}$ & & $^{+0.08}_{-0.06}$ & $^{+0.09}_{-0.07}$ & $^{+0.07}_{-0.07}$ & $^{+0.11}_{-0.12}$ & $^{+0.04}_{-0.03}$ & $^{+0.12}_{-0.13}$ & $^{+0.06}_{-0.06}$ & $^{+0.13}_{-0.11}$ & $^{+0.06}_{-0.05}$ & $^{+0.05}_{-0.05}$ & $^{+0.01}_{-0.03}$ \\
Start & \ldots & $17000$ & $4.000$ & $20.0$ & $10.0$ & $10.0$ & $3.00$ & & $-0.85$ & $-3.40$ & $-4.10$ & $-3.10$ & $-3.80$ & $-4.40$ & $-5.90$ & $-4.70$ & $-4.80$ & $-5.50$ & $-4.70$ \\
\hline
In & $400$ & $25000$ & $3.750$ & $-25.0$ & $40.0$ & $28.0$ & $2.00$ & & $-1.06$ & $-3.70$ & $-4.30$ & $-3.20$ & $-4.00$ & $-4.70$ & $-5.60$ & $-4.60$ & $-4.80$ & $-5.60$ & $-4.60$ \\
Out & $396$ & $24930$ & $3.738$ & $-25.0$ & $39.7$ & $28.5$ & $2.02$ & & $-1.07$ & $-3.72$ & $-4.31$ & $-3.21$ & $-4.00$ & $-4.70$ & $-5.61$ & $-4.61$ & $-4.81$ & $-5.63$ & $-4.61$ \\
Stat. & $^{+2}_{-2}$ & $^{+20}_{-20}$ & $^{+0.002}_{-0.003}$ & $^{+0.1}_{-0.1}$ & $^{+0.1}_{-0.1}$ & $^{+0.1}_{-0.3}$ & $^{+0.04}_{-0.04}$ & & $^{+0.01}_{-0.01}$ & $^{+0.01}_{-0.01}$ & $^{+0.01}_{-0.02}$ & $^{+0.01}_{-0.01}$ & $^{+0.02}_{-0.02}$ & $^{+0.02}_{-0.02}$ & $^{+0.01}_{-0.01}$ & $^{+0.01}_{-0.01}$ & $^{+0.01}_{-0.01}$ & $^{+0.04}_{-0.04}$ & $^{+0.01}_{-0.01}$ \\
Sys. & \ldots & $^{+500}_{-500}$ & $^{+0.100}_{-0.100}$ & $^{+0.1}_{-0.1}$ & $^{+0.3}_{-0.1}$ & $^{+1.3}_{-1.5}$ & $^{+0.71}_{-0.02}$ & & $^{+0.06}_{-0.07}$ & $^{+0.01}_{-0.06}$ & $^{+0.03}_{-0.03}$ & $^{+0.07}_{-0.08}$ & $^{+0.01}_{-0.02}$ & $^{+0.05}_{-0.08}$ & $^{+0.04}_{-0.06}$ & $^{+0.02}_{-0.06}$ & $^{+0.02}_{-0.03}$ & $^{+0.10}_{-0.12}$ & $^{+0.03}_{-0.03}$ \\
Start & \ldots & $23500$ & $3.800$ & $-20.0$ & $10.0$ & $10.0$ & $8.00$ & & $-1.10$ & $-3.50$ & $-4.10$ & $-3.30$ & $-4.10$ & $-4.50$ & $-6.00$ & $-4.20$ & $-4.70$ & $-5.40$ & $-4.30$ \\
\hline
In & $275$ & $25000$ & $4.250$ & $25.0$ & $70.0$ & $0.0$ & $2.00$ & & $-1.06$ & $-3.70$ & $-4.30$ & $-3.20$ & $-4.00$ & $-4.70$ & $-5.60$ & $-4.60$ & $-4.80$ & $-5.60$ & $-4.60$ \\
Out & $274$ & $25010$ & $4.259$ & $25.0$ & $69.8$ & $3.6$ & $2.00$ & & $-1.05$ & $-3.71$ & $-4.29$ & $-3.21$ & $-3.99$ & $-4.68$ & $-5.60$ & $-4.61$ & $-4.82$ & $-5.59$ & $-4.61$ \\
Stat. & $^{+1}_{-1}$ & $^{+30}_{-30}$ & $^{+0.003}_{-0.003}$ & $^{+0.2}_{-0.1}$ & $^{+0.1}_{-0.1}$ & $^{+3.7}_{-2.8}$ & $^{+0.06}_{-0.11}$ & & $^{+0.01}_{-0.01}$ & $^{+0.02}_{-0.02}$ & $^{+0.01}_{-0.01}$ & $^{+0.01}_{-0.01}$ & $^{+0.03}_{-0.03}$ & $^{+0.03}_{-0.03}$ & $^{+0.02}_{-0.02}$ & $^{+0.02}_{-0.02}$ & $^{+0.02}_{-0.02}$ & $^{+0.05}_{-0.07}$ & $^{+0.02}_{-0.02}$ \\
Sys. & \ldots & $^{+500}_{-500}$ & $^{+0.100}_{-0.100}$ & $^{+0.1}_{-0.1}$ & $^{+0.5}_{-0.1}$ & $^{+3.5}_{-3.6}$ & $^{+1.41}_{-2.00}$ & & $^{+0.06}_{-0.06}$ & $^{+0.08}_{-0.10}$ & $^{+0.03}_{-0.03}$ & $^{+0.12}_{-0.11}$ & $^{+0.02}_{-0.04}$ & $^{+0.09}_{-0.10}$ & $^{+0.04}_{-0.04}$ & $^{+0.12}_{-0.11}$ & $^{+0.07}_{-0.06}$ & $^{+0.12}_{-0.15}$ & $^{+0.05}_{-0.04}$ \\
Start & \ldots & $19000$ & $3.700$ & $30.0$ & $10.0$ & $10.0$ & $3.00$ & & $-0.85$ & $-3.80$ & $-4.40$ & $-3.40$ & $-4.20$ & $-4.80$ & $-5.90$ & $-4.50$ & $-4.70$ & $-5.80$ & $-4.90$ \\
\hline
In & $125$ & $30000$ & $4.250$ & $-15.0$ & $0.0$ & $18.0$ & $2.00$ & & $-1.06$ & $-3.70$ & $-4.30$ & $-3.20$ & $-4.00$ & $-4.70$ & $-5.60$ & $-4.60$ & $-4.80$ & $-5.60$ & $-4.60$ \\
Out & $120$ & $30110$ & $4.259$ & $-15.0$ & $5.8$ & $16.4$ & $1.84$ & & $-1.04$ & $-3.72$ & $-4.31$ & $-3.21$ & $-4.02$ & $-4.69$ & $-5.61$ & $-4.61$ & $-4.81$ & $-5.47$ & $-4.60$ \\
Stat. & $^{+2}_{-2}$ & $^{+40}_{-60}$ & $^{+0.007}_{-0.013}$ & $^{+0.1}_{-0.1}$ & $^{+0.4}_{-0.5}$ & $^{+0.4}_{-0.4}$ & $^{+0.17}_{-0.14}$ & & $^{+0.02}_{-0.02}$ & $^{+0.03}_{-0.02}$ & $^{+0.02}_{-0.02}$ & $^{+0.01}_{-0.02}$ & $^{+0.03}_{-0.03}$ & $^{+0.04}_{-0.04}$ & $^{+0.03}_{-0.03}$ & $^{+0.02}_{-0.02}$ & $^{+0.03}_{-0.03}$ & $^{+0.15}_{-0.24}$ & $^{+0.02}_{-0.02}$ \\
Sys. & \ldots & $^{+610}_{-610}$ & $^{+0.100}_{-0.100}$ & $^{+0.1}_{-0.1}$ & $^{+1.1}_{-1.2}$ & $^{+0.1}_{-0.1}$ & $^{+0.36}_{-0.32}$ & & $^{+0.03}_{-0.03}$ & $^{+0.05}_{-0.06}$ & $^{+0.05}_{-0.05}$ & $^{+0.05}_{-0.03}$ & $^{+0.03}_{-0.02}$ & $^{+0.04}_{-0.04}$ & $^{+0.06}_{-0.06}$ & $^{+0.03}_{-0.03}$ & $^{+0.07}_{-0.06}$ & $^{+0.15}_{-0.17}$ & $^{+0.08}_{-0.08}$ \\
Start & \ldots & $32000$ & $4.400$ & $-15.0$ & $1.0$ & $0.0$ & $7.00$ & & $-1.15$ & $-3.70$ & $-4.10$ & $-3.10$ & $-4.20$ & $-4.70$ & $-5.90$ & $-4.70$ & $-4.60$ & $-5.40$ & $-4.80$ \\
\hline
\multicolumn{20}{l}{Mock spectrum as a proxy to the observed spectrum of object \#1 in Table~\ref{table:program_stars}:}\\
In & $250$ & $23880$ & $4.127$ & $23.0$ & $5.0$ & $4.0$ & $2.00$ & & $-0.99$ & $-3.73$ & $-4.30$ & $-3.29$ & $-4.00$ & $-4.57$ & $-5.79$ & $-4.66$ & $-4.88$ & $-5.49$ & $-4.71$ \\
Out & $232$ & $23800$ & $4.109$ & $23.0$ & $4.7$ & $4.0$ & $2.02$ & & $-0.99$ & $-3.74$ & $-4.30$ & $-3.30$ & $-4.02$ & $-4.57$ & $-5.80$ & $-4.66$ & $-4.90$ & $-5.51$ & $-4.73$ \\
Stat. & $^{+2}_{-2}$ & $^{+50}_{-50}$ & $^{+0.006}_{-0.007}$ & $^{+0.1}_{-0.1}$ & $^{+0.5}_{-2.9}$ & $^{+2.2}_{-0.4}$ & $^{+0.08}_{-0.11}$ & & $^{+0.01}_{-0.01}$ & $^{+0.01}_{-0.01}$ & $^{+0.01}_{-0.01}$ & $^{+0.01}_{-0.01}$ & $^{+0.02}_{-0.02}$ & $^{+0.02}_{-0.02}$ & $^{+0.01}_{-0.01}$ & $^{+0.02}_{-0.02}$ & $^{+0.01}_{-0.02}$ & $^{+0.02}_{-0.02}$ & $^{+0.01}_{-0.01}$ \\
Sys. & \ldots & $^{+480}_{-480}$ & $^{+0.100}_{-0.100}$ & $^{+0.1}_{-0.1}$ & $^{+0.4}_{-0.6}$ & $^{+1.3}_{-1.5}$ & $^{+0.43}_{-0.90}$ & & $^{+0.05}_{-0.05}$ & $^{+0.04}_{-0.03}$ & $^{+0.03}_{-0.02}$ & $^{+0.08}_{-0.08}$ & $^{+0.04}_{-0.04}$ & $^{+0.08}_{-0.07}$ & $^{+0.01}_{-0.02}$ & $^{+0.06}_{-0.05}$ & $^{+0.02}_{-0.03}$ & $^{+0.06}_{-0.06}$ & $^{+0.02}_{-0.02}$ \\
Start & \ldots & $19000$ & $3.700$ & $20.0$ & $10.0$ & $10.0$ & $3.00$ & & $-0.90$ & $-3.50$ & $-4.30$ & $-3.30$ & $-4.20$ & $-4.50$ & $-5.80$ & $-4.50$ & $-4.90$ & $-5.50$ & $-4.70$ \\
\hline
\multicolumn{20}{l}{Mock spectrum as a proxy to the observed spectrum of object \#2 in Table~\ref{table:program_stars}:}\\
In & $250$ & $19250$ & $4.052$ & $31.0$ & $7.0$ & $17.0$ & $2.00$ & & $-1.00$ & $-3.64$ & $-4.23$ & $-3.21$ & $-4.06$ & $-4.60$ & $-5.71$ & $-4.48$ & $-4.89$ & $-5.57$ & $-4.63$ \\
Out & $242$ & $19240$ & $4.058$ & $31.0$ & $4.8$ & $18.0$ & $2.07$ & & $-1.00$ & $-3.64$ & $-4.23$ & $-3.22$ & $-4.06$ & $-4.62$ & $-5.70$ & $-4.51$ & $-4.89$ & $-5.57$ & $-4.65$ \\
Stat. & $^{+2}_{-2}$ & $^{+20}_{-20}$ & $^{+0.005}_{-0.005}$ & $^{+0.1}_{-0.1}$ & $^{+0.4}_{-0.5}$ & $^{+0.1}_{-0.1}$ & $^{+0.05}_{-0.07}$ & & $^{+0.01}_{-0.01}$ & $^{+0.01}_{-0.02}$ & $^{+0.02}_{-0.02}$ & $^{+0.02}_{-0.01}$ & $^{+0.02}_{-0.02}$ & $^{+0.02}_{-0.02}$ & $^{+0.02}_{-0.02}$ & $^{+0.02}_{-0.02}$ & $^{+0.01}_{-0.01}$ & $^{+0.03}_{-0.03}$ & $^{+0.02}_{-0.02}$ \\
Sys. & \ldots & $^{+390}_{-390}$ & $^{+0.100}_{-0.100}$ & $^{+0.1}_{-0.1}$ & $^{+0.9}_{-3.2}$ & $^{+0.1}_{-0.1}$ & $^{+0.84}_{-1.43}$ & & $^{+0.08}_{-0.08}$ & $^{+0.08}_{-0.09}$ & $^{+0.08}_{-0.08}$ & $^{+0.11}_{-0.11}$ & $^{+0.03}_{-0.03}$ & $^{+0.10}_{-0.11}$ & $^{+0.05}_{-0.06}$ & $^{+0.13}_{-0.12}$ & $^{+0.04}_{-0.05}$ & $^{+0.05}_{-0.05}$ & $^{+0.01}_{-0.03}$ \\
Start & \ldots & $17000$ & $3.700$ & $20.0$ & $10.0$ & $10.0$ & $3.00$ & & $-0.85$ & $-3.50$ & $-3.90$ & $-3.40$ & $-3.80$ & $-4.40$ & $-5.40$ & $-4.00$ & $-4.70$ & $-5.40$ & $-4.80$ \\
\hline
\multicolumn{20}{l}{Mock spectrum as a proxy to the observed spectrum of object \#3 in Table~\ref{table:program_stars}:}\\
In & $200$ & $29210$ & $4.284$ & $30.0$ & $31.0$ & $0.0$ & $3.20$ & & $-1.05$ & $-3.71$ & $-4.13$ & $-3.40$ & $-4.01$ & $-4.58$ & $-5.73$ & $-4.66$ & $-4.97$ & $-5.83$ & $-4.62$ \\
Out & $197$ & $29190$ & $4.273$ & $30.1$ & $30.8$ & $0.0$ & $3.12$ & & $-1.04$ & $-3.74$ & $-4.13$ & $-3.41$ & $-4.04$ & $-4.63$ & $-5.76$ & $-4.66$ & $-5.00$ & $-5.87$ & $-4.62$ \\
Stat. & $^{+1}_{-1}$ & $^{+30}_{-30}$ & $^{+0.005}_{-0.005}$ & $^{+0.1}_{-0.1}$ & $^{+0.1}_{-0.1}$ & $^{+1.8}_{-0.0}$ & $^{+0.08}_{-0.08}$ & & $^{+0.01}_{-0.01}$ & $^{+0.02}_{-0.02}$ & $^{+0.01}_{-0.01}$ & $^{+0.01}_{-0.01}$ & $^{+0.03}_{-0.03}$ & $^{+0.03}_{-0.03}$ & $^{+0.03}_{-0.03}$ & $^{+0.02}_{-0.02}$ & $^{+0.03}_{-0.02}$ & $^{+0.21}_{-0.13}$ & $^{+0.02}_{-0.02}$ \\
Sys. & \ldots & $^{+590}_{-590}$ & $^{+0.100}_{-0.100}$ & $^{+0.1}_{-0.1}$ & $^{+0.1}_{-0.1}$ & $^{+4.6}_{-0.0}$ & $^{+0.50}_{-0.57}$ & & $^{+0.04}_{-0.03}$ & $^{+0.03}_{-0.04}$ & $^{+0.05}_{-0.05}$ & $^{+0.04}_{-0.03}$ & $^{+0.04}_{-0.03}$ & $^{+0.05}_{-0.05}$ & $^{+0.06}_{-0.05}$ & $^{+0.04}_{-0.04}$ & $^{+0.06}_{-0.05}$ & $^{+0.32}_{-0.13}$ & $^{+0.09}_{-0.07}$ \\
Start & \ldots & $31000$ & $4.300$ & $20.0$ & $10.0$ & $10.0$ & $7.00$ & & $-1.05$ & $-3.60$ & $-4.20$ & $-3.20$ & $-4.10$ & $-4.60$ & $-5.80$ & $-4.40$ & $-4.90$ & $-5.60$ & $-4.60$ \\
\hline
\end{tabular}
\tablefoot{Starting parameters for the fitting algorithm are given as well (\textit{``Start'' row}). See Sect.~\ref{subsection:performance} for details. The S/N estimates in the \textit{``Out'' row} are based on the method outlined in Sect.~\ref{subsection:noise_estimation}. The abundance $n(x)$ is given as fractional particle number of species $x$ with respect to all elements. Statistical uncertainties (\textit{``Stat.'' row}) correspond to $\Delta \chi^2 = 6.63$ and are 99\%-confidence limits. Systematic uncertainties (\textit{``Sys.'' row}) cover only the effects induced by additional variations of $2\%$ in $T_{\mathrm{eff}}$ and $0.1\,\mathrm{dex}$ in $\log(g)$ (see Sect.~\ref{subsection:uncertainties} for details) and are formally taken to be 99\%-confidence limits.}
\end{table*}
\begin{table*}
\begin{center}
\tiny
\setlength{\tabcolsep}{0.115cm}
\renewcommand{\arraystretch}{1.3}
\caption{\label{table:uncertainties_binary} Same as Table~\ref{table:uncertainties} but for three exemplary mock composite spectra.}
\begin{tabular}{lrrrrrrrrrrrrrrrrrrr}
\hline\hline
& S/N & $T_{\mathrm{eff}}$ & $\log(g)$ & $\varv_{\mathrm{rad}}$ & $\varv\,\sin(i)$ & $\zeta$ & $\xi$ & $A_{\mathrm{eff,s}}/A_{\mathrm{eff,p}}$ & \multicolumn{11}{c}{$\log(n(x))$} \\ 
\cline{5-8} \cline{10-20}
& & (K) & (cgs) & \multicolumn{4}{c}{$(\mathrm{km\,s^{-1}})$} & & He & C & N & O & Ne & Mg & Al & Si & S & Ar & Fe\\
\hline
\multicolumn{20}{l}{Mock spectrum as a proxy to the observed spectrum of object \#5 in Table~\ref{table:program_stars} (sharp and well-separated features):}\\
In p & $220$ & $16680$ & $4.098$ & $-84.7$ & $7.9$ & $11.4$ & $2.10$ & \ldots & $-0.96$ & $-3.55$ & $-4.16$ & $-3.25$ & $-4.04$ & $-4.73$ & $-5.86$ & $-4.45$ & $-4.91$ & $-5.58$ & $-4.66$ \\
Out f & $215$ & $16680$ & $4.098$ & $-84.7$ & $6.9$ & $12.2$ & $2.17$ & \ldots & $-0.97$ & $-3.56$ & $-4.19$ & $-3.25$ & $-4.03$ & $-4.74$ & $-5.91$ & $-4.47$ & $-4.91$ & $-5.60$ & $-4.68$ \\
Stat. & $^{+1}_{-1}$ & $^{+80}_{-90}$ & $^{+0.023}_{-0.020}$ & $^{+0.1}_{-0.1}$ & $^{+1.4}_{-1.1}$ & $^{+0.7}_{-1.3}$ & $^{+0.14}_{-0.16}$ & \ldots & $^{+0.02}_{-0.03}$ & $^{+0.03}_{-0.03}$ & $^{+0.04}_{-0.04}$ & $^{+0.02}_{-0.03}$ & $^{+0.02}_{-0.02}$ & $^{+0.03}_{-0.03}$ & $^{+0.04}_{-0.06}$ & $^{+0.03}_{-0.03}$ & $^{+0.02}_{-0.02}$ & $^{+0.07}_{-0.10}$ & $^{+0.02}_{-0.03}$ \\
Sys. & \ldots & $^{+340}_{-340}$ & $^{+0.100}_{-0.100}$ & $^{+0.1}_{-0.1}$ & $^{+0.7}_{-0.6}$ & $^{+0.8}_{-0.8}$ & $^{+0.27}_{-0.29}$ & \ldots & $^{+0.06}_{-0.07}$ & $  ^{+0.07}_{-0.07}$ & $^{+0.06}_{-0.07}$ & $^{+0.02}_{-0.02}$ & $^{+0.02}_{-0.02}$ & $^{+0.03}_{-0.03}$ & $^{+0.02}_{-0.03}$ & $^{+0.02}_{-0.03}$ & $^{+0.02}_{-0.03}$ & $^{+0.05}_{-0.06}$ & $^{+0.06}_{-0.06}$ \\
Out i & \ldots & $16660$ & $4.096$ & $-84.7$ & $6.6$ & $12.4$ & $2.20$ & \ldots & $-0.97$ & $-3.56$ & $-4.18$ & $-3.26$ & $-4.03$ & $-4.75$ & $-5.90$ & $-4.47$ & $-4.91$ & $-5.60$ & $-4.68$ \\
Stat. & \ldots & $^{+70}_{-70}$ & $^{+0.017}_{-0.022}$ & $^{+0.1}_{-0.1}$ & $^{+1.4}_{-1.0}$ & $^{+0.9}_{-1.3}$ & $^{+0.13}_{-0.15}$ & \ldots & $^{+0.02}_{-0.02}$ & $ ^{+0.03}_{-0.03}$ & $^{+0.03}_{-0.04}$ & $^{+0.02}_{-0.02}$ & $^{+0.02}_{-0.02}$ & $^{+0.03}_{-0.03}$ & $ ^{+0.04}_{-0.05}$ & $^{+0.03}_{-0.03}$ & $^{+0.02}_{-0.02}$ & $^{+0.07}_{-0.11}$ & $^{+0.02}_{-0.03}$ \\
Sys. & \ldots & $^{+340}_{-340}$ & $^{+0.100}_{-0.100}$ & $^{+0.1}_{-0.1}$ & $^{+0.6}_{-0.5}$ & $^{+0.4}_{-0.8}$ & $^{+0.27}_{-0.26}$ & \ldots & $^{+0.05}_{-0.06}$ & $^{+0.06}_{-0.06}$ & $^{+0.05}_{-0.06}$ & $^{+0.01}_{-0.01}$ & $^{+0.01}_{-0.01}$ & $^{+0.03}_{-0.03}$ & $^{+0.02}_{-0.02}$ & $^{+0.02}_{-0.02}$ & $^{+0.02}_{-0.02}$ & $^{+0.04}_{-0.06}$ & $^{+0.05}_{-0.05}$ \\
Start & \ldots & $15000$ & $4.100$ & $-80.0$ & $10.0$ & $10.0$ & $3.00$ & \ldots & $-0.85$ & $-3.60$ & $-4.20$ & $-3.20$ & $-4.00$ & $-4.60$ & $-5.80$ & $-4.40$ & $-4.90$ & $-5.60$ & $-4.60$ \\
In s & \ldots & $13490$ & $4.274$ & $125.0$ & $28.3$ & $15.6$ & $0.79$ & $0.642$ & $-0.96$ & $-3.55$ & $-4.16$ & $-3.25$ & $-4.04$ & $-4.73$ & $-5.86$ & $-4.45$ & $-4.91$ & $-5.58$ & $-4.66$ \\
Out f & \ldots & $13200$ & $4.210$ & $125.3$ & $ 29.3$ & $14.2$ & $0.30$ & $0.662$ & $-0.91$ & $-3.46$ & $-4.21$ & $-3.29$ & $-4.12$ & $-4.79$ & $-5.73$ & $-4.50$ & $-4.97$ & \ldots & $-4.76$  \\
Stat. & \ldots & $^{+200}_{-250}$ & $^{+0.050}_{-0.070}$ & $^{+0.5}_{-0.5}$ & $^{+0.8}_{-1.3}$ & $^{+1.7}_{-2.9}$ & $^{+0.33}_{-0.30}$ & $^{+0.020}_{-0.018}$ & $^{+0.10}_{-0.07}$ & $^{+0.13}_{-0.12}$ & $^{+0.26}_{-0.52}$ & $^{+0.06}_{-0.08}$ & $^{+0.13}_{-0.14}$ & $^{+0.06}_{-0.06}$ & $^{+0.13}_{-0.16}$ & $^{+0.05}_{-0.06}$ & $^{+0.11}_{-0.10}$ & \ldots & $^{+0.07}_{-0.09}$ \\
Sys. & \ldots & $^{+260}_{-360}$ & $^{+0.100}_{-0.100}$ & $^{+0.2}_{-0.2}$ & $^{+0.2}_{-0.6}$ & $^{+0.9}_{-1.2}$ & $^{+0.15}_{-0.22}$ & $^{+0.036}_{-0.033}$ & $^{+0.19}_{-0.17}$ & $^{+0.12}_{-0.12}$ & $^{+0.01}_{-0.06}$ & $^{+0.04}_{-0.07}$ & $^{+0.10}_{-0.11}$ & $^{+0.03}_{-0.04}$ & $^{+0.03}_{-0.04}$ & $^{+0.02}_{-0.02}$ & $^{+0.10}_{-0.09}$ & \ldots & $^{+0.09}_{-0.13}$ \\
Out i & \ldots & $13360$ & $4.258$ & $125.3$ & $28.7$ & $15.1$ & $0.15$ & $0.650$ & \ldots & \ldots & \ldots & \ldots & \ldots & \ldots & \ldots & \ldots & \ldots & \ldots & \ldots  \\
Stat. & \ldots & $^{+80}_{-90}$ & $^{+0.028}_{-0.029}$ & $^{+0.5}_{-0.5}$ & $^{+1.3}_{-1.1}$ & $^{+2.1}_{-2.5}$ & $^{+0.34}_{-0.15}$ & $^{+0.015}_{-0.014}$ & \ldots & \ldots & \ldots & \ldots & \ldots & \ldots & \ldots & \ldots & \ldots & \ldots & \ldots  \\
Sys. & \ldots & $^{+270}_{-270}$ & $^{+0.100}_{-0.113}$ & $^{+0.2}_{-0.3}$ & $^{+0.3}_{-0.3}$ & $^{+1.6}_{-1.3}$ & $^{+0.29}_{-0.15}$ & $^{+0.031}_{-0.032}$ & \ldots & \ldots & \ldots & \ldots & \ldots & \ldots & \ldots & \ldots & \ldots & \ldots & \ldots  \\
Start & \ldots & $15000$ & $4.100$ & $120.0$ & $10.0$ & $10.0$ & $3.00$ & $0.800$ & $-0.85$ & $-3.60$ & $-4.20$ & $-3.20$ & $-4.00$ & $-4.60$ & $-5.80$ & $-4.40$ & $-4.90$ & $-5.60$ & $-4.60$ \\
\hline
\multicolumn{20}{l}{Mock spectrum as a proxy to the observed spectrum of object \#6 in Table~\ref{table:program_stars} (extremely blended features):}\\
In p & $350$ & $20600$ & $3.485$ & $-11.2$ & $54.2$ & $9.4$ & $6.04$ & \ldots & $-1.02$ & $-3.79$ & $-4.38$ & $-3.39$ & $-4.02$ & $-4.74$ & $-5.87$ & $-4.66$ & $-4.99$ & $-5.59$ & $-4.79$ \\
Out f & $350$ & $20740$ & $3.502$ & $-11.3$ & $54.4$ & $1.9$ & $5.80$ & \ldots & $-1.02$ & $-3.76$ & $-4.37$ & $-3.39$ & $-4.03$ & $-4.74$ & $-5.87$ & $-4.68$ & $-5.02$ & $-5.57$ & $-4.76$ \\
Stat. & $^{+2}_{-2}$ & $^{+30}_{-20}$ & $^{+0.005}_{-0.003}$ & $^{+0.2}_{-0.2}$ & $^{+0.2}_{-0.2}$ & $^{+4.4}_{-1.9}$ & $^{+0.07}_{-0.12}$ & \ldots & $^{+0.02}_{-0.03}$ & $^{+0.01}_{-0.01}$ & $^{+0.02}_{-0.01}$ & $^{+0.02}_{-0.02}$ & $^{+0.01}_{-0.03}$ & $^{+0.04}_{-0.02}$ & $^{+0.01}_{-0.02}$ & $^{+0.02}_{-0.01}$ & $^{+0.01}_{-0.04}$ & $^{+0.03}_{-0.04}$ & $^{+0.01}_{-0.01}$ \\
Sys. & \ldots & $^{+420}_{-420}$ & $^{+0.100}_{-0.100}$ & $^{+0.1}_{-0.1}$ & $^{+0.3}_{-0.1}$ & $^{+3.1}_{-1.9}$ & $^{+0.24}_{-0.49}$ & \ldots & $^{+0.03}_{-0.04}$ & $^{+0.02}_{-0.02}$ & $^{+0.05}_{-0.05}$ & $^{+0.09}_{-0.07}$ & $^{+0.03}_{-0.04}$ & $^{+0.05}_{-0.05}$ & $^{+0.03}_{-0.03}$ & $^{+0.03}_{-0.03}$ & $^{+0.02}_{-0.05}$ & $^{+0.03}_{-0.04}$ & $^{+0.04}_{-0.04}$ \\
Out i & \ldots & $20790$ & $3.506$ & $-11.3$ & $54.4$ & $3.1$ & $5.74$ & \ldots & $-1.01$ & $-3.79$ & $-4.38$ & $-3.41$ & $-4.03$ & $-4.74$ & $-5.87$ & $-4.68$ & $-5.01$ & $-5.59$ & $-4.79$ \\
Stat. & \ldots & $^{+30}_{-30}$ & $^{+0.004}_{-0.004}$ & $^{+0.1}_{-0.1}$ & $^{+0.1}_{-0.1}$ & $^{+0.1}_{-3.1}$ & $^{+0.05}_{-0.02}$ & \ldots & $^{+0.01}_{-0.01}$ & $^{+0.01}_{-0.01}$ & $^{+0.01}_{-0.01}$ & $^{+0.01}_{-0.01}$ & $^{+0.02}_{-0.02}$ & $^{+0.02}_{-0.02}$ & $^{+0.02}_{-0.01}$ & $^{+0.02}_{-0.02}$ & $^{+0.02}_{-0.02}$ & $^{+0.02}_{-0.03}$ & $^{+0.02}_{-0.02}$ \\
Sys. & \ldots & $^{+420}_{-420}$ & $^{+0.100}_{-0.100}$ & $^{+0.1}_{-0.1}$ & $^{+0.4}_{-0.1}$ & $^{+2.8}_{-3.1}$ & $^{+0.46}_{-0.46}$ & \ldots & $^{+0.01}_{-0.01}$ & $^{+0.01}_{-0.01}$ & $^{+0.02}_{-0.02}$ & $^{+0.04}_{-0.04}$ & $^{+0.01}_{-0.01}$ & $^{+0.02}_{-0.01}$ & $^{+0.02}_{-0.02}$ & $^{+0.02}_{-0.02}$ & $^{+0.01}_{-0.01}$ & $^{+0.02}_{-0.02}$ & $ ^{+0.02}_{-0.02}$ \\
Start & \ldots & $20000$ & $3.500$ & $-10.0$ & $50.0$ & $10.0$ & $8.00$ & \ldots & $-1.05$ & $-3.60$ & $-4.20$ & $-3.20$ & $-4.00$ & $-4.60$ & $-5.80$ & $-4.40$ & $-4.90$ & $-5.60$ & $-4.60$ \\
In s & \ldots & $18610$ & $3.227$ & $-9.1$ & $134.0$ & $59.5$ & $2.90$ & $0.936$ & $-1.02$ & $-3.79$ & $-4.38$ & $-3.39$ & $-4.02$ & $-4.74$ & $-5.87$ & $-4.66$ & $-4.99$ & $-5.59$ & $-4.79$ \\
Out f & \ldots & $18520$ & $3.200$ & $-9.6$ & $118.0$ & $87.0$ & $3.82$ & $1.085$ & $-1.00$ & $-3.89$ & $-4.40$ & $-3.46$ & $-4.05$ & $-4.75$ & $-5.86$ & $-4.68$ & $ -5.01$ & $-5.68$ & $-4.89$ \\
Stat. & \ldots & $^{+60}_{-40}$ & $^{+0.003}_{-0.002}$ & $^{+0.4}_{-0.7}$ & $^{+1.6}_{-0.5}$ & $^{+0.5}_{-3.0}$ & $^{+0.16}_{-0.12}$ & $^{+0.008}_{-0.005}$ & $^{+0.02}_{-0.02}$ & $^{+0.04}_{-0.04}$ & $^{+0.03}_{-0.05}$ & $^{+0.04}_{-0.05}$ & $^{+0.04}_{-0.02}$ & $ ^{+0.05}_{-0.05}$ & $^{+0.03}_{-0.06}$ & $^{+0.03}_{-0.04}$ & $^{+0.03}_{-0.03}$ & $^{+0.06}_{-0.05}$ & $^{+0.04}_{-0.05}$ \\
Sys.  & \ldots & $^{+580}_{-500}$ & $^{+0.124}_{-0.124}$ & $^{+0.3}_{-0.1}$ & $^{+1.3}_{-0.8}$ & $^{+2.0}_{-1.6}$ & $^{+0.52}_{-0.36}$ & $^{+0.069}_{-0.131}$ & $^{+0.06}_{-0.03}$ & $^{+0.07}_{-0.06}$ & $^{+0.09}_{-0.10}$ & $^{+0.11}_{-0.13}$ & $^{+0.06}_{-0.03}$ & $^{+0.10}_{-0.06}$ & $^{+0.06}_{-0.05}$ & $^{+0.04}_{-0.04}$ & $^{+0.07}_{-0.03}$ & $^{+0.04}_{-0.04}$ & $^{+0.06}_{-0.07}$ \\
Out i & \ldots & $18410$ & $3.197$ & $-9.9$ & $120.3$ & $84.9$ & $3.71$ & $1.066$ & \ldots & \ldots & \ldots & \ldots & \ldots & \ldots & \ldots & \ldots & \ldots & \ldots & \ldots  \\
Stat. & \ldots & $^{+70}_{-20}$ & $^{+0.010}_{-0.005}$ & $^{+0.7}_{-0.7}$ & $^{+4.8}_{-0.5}$ & $^{+3.0}_{-3.3}$ & $^{+0.15}_{-0.09}$ & $^{+0.015}_{-0.008}$ & \ldots & \ldots & \ldots & \ldots & \ldots & \ldots & \ldots & \ldots & \ldots & \ldots & \ldots \\
Sys. & \ldots & $^{+410}_{-390}$ & $^{+0.119}_{-0.111}$ & $^{+0.2}_{-0.2}$ & $^{+0.7}_{-0.6}$ & $^{+1.4}_{-1.5}$ & $^{+0.62}_{-0.71}$ & $^{+0.054}_{-0.061}$ & \ldots & \ldots & \ldots & \ldots & \ldots & \ldots & \ldots & \ldots & \ldots & \ldots & \ldots \\
Start & \ldots & $20000$ & $3.500$ & $-10.0$ & $100.0$ & $50.0$ & $8.00$ & $1.000$ & $-1.05$ & $-3.60$ & $-4.20$ & $-3.20$ & $-4.00$ & $-4.60$ & $-5.80$ & $-4.40$ & $-4.90$ & $-5.60$ & $-4.60$ \\
\hline
\multicolumn{20}{l}{Mock spectrum as a proxy to the observed spectrum of object \#4 b in Table~\ref{table:program_stars} (but with individual metal abundances for the two components):}\\
In p & $340$ & $29710$ & $3.669$ & $104.0$ & $23.7$ & $41.2$ & $14.92$ & \ldots & $-1.17$ & $-3.79$ & $-4.38$ & $-3.39$ & $-4.02$ & $-4.74$ & $-5.87$ & $-4.66$ & $-4.99$ & $-5.59$ & $-4.79$ \\
Out f & $353$ & $29730$ & $3.678$ & $103.9$ & $24.9$ & $39.3$ & $15.02$ & \ldots & $-1.17$ & $-3.81$ & $-4.39$ & $-3.41$ & $-4.03$ & $-4.75$ & $-5.88$ & $-4.68$ & $-5.00$ & \ldots & $-4.86$ \\
Stat. & $^{+3}_{-2}$ & $^{+20}_{-40}$ & $^{+0.003}_{-0.003}$ & $^{+0.1}_{-0.1}$ & $^{+0.3}_{-0.3}$ & $^{+0.1}_{-0.1}$ & $^{+0.12}_{-0.09}$ & \ldots & $^{+0.01}_{-0.01}$ & $^{+0.01}_{-0.01}$ & $^{+0.01}_{-0.01}$ & $^{+0.01}_{-0.01}$ & $^{+0.02}_{-0.02}$ & $^{+0.02}_{-0.02}$ & $^{+0.03}_{-0.03}$ & $^{+0.01}_{-0.01}$ & $^{+0.02}_{-0.02}$ & \ldots & $^{+0.05}_{-0.04}$ \\
Sys. & \ldots & $^{+600}_{-600}$ & $ ^{+0.100}_{-0.100}$ & $^{+0.1}_{-0.1}$ & $^{+0.9}_{-1.0}$ & $^{+0.3}_{-0.1}$ & $^{+0.48}_{-0.61}$ & \ldots & $^{+0.04}_{-0.04}$ & $^{+0.02}_{-0.01}$ & $^{+0.04}_{-0.04}$ & $^{+0.05}_{-0.05}$ & $^{+0.02}_{-0.02}$ & $^{+0.03}_{-0.03}$ & $^{+0.04}_{-0.03}$ & $^{+0.04}_{-0.04}$ & $^{+0.07}_{-0.06}$ & \ldots & $^{+0.07}_{-0.06}$ \\
Out i & \ldots & $29700$ & $3.673$ & $103.9$ & $24.7$ & $39.6$ & $14.98$ & \ldots & $ -1.17$ & $-3.81$ & $-4.39$ & $-3.41$ & $-4.04$ & $-4.74$ & $-5.86$ & $-4.68$ & $-5.00$ & \ldots & $-4.80$ \\
Stat. & \ldots & $^{+10}_{-20}$ & $^{+0.004}_{-0.003}$ & $^{+0.1}_{-0.1}$ & $^{+0.2}_{-0.1}$ & $^{+0.1}_{-0.1}$ & $^{+0.08}_{-0.05}$ & \ldots & $^{+0.01}_{-0.01}$ & $^{+0.01}_{-0.01}$ & $^{+0.01}_{-0.01}$ & $^{+0.01}_{-0.01}$ & $^{+0.02}_{-0.02}$ & $^{+0.02}_{-0.02}$ & $^{+0.03}_{-0.03}$ & $^{+0.01}_{-0.01}$ & $^{+0.02}_{-0.02}$ & \ldots  & $^{+0.04}_{-0.04}$ \\
Sys. & \ldots & $^{+600}_{-600}$ & $^{+0.100}_{-0.100}$ & $^{+0.1}_{-0.1}$ & $^{+0.8}_{-1.2}$ & $^{+0.3}_{-0.1}$ & $^{+0.50}_{-0.63}$ & \ldots & $^{+0.04}_{-0.03}$ & $^{+0.02}_{-0.02}$ & $^{+0.03}_{-0.04}$ & $^{+0.05}_{-0.05}$ & $^{+0.02}_{-0.02}$ & $^{+0.03}_{-0.03}$ & $^{+0.03}_{-0.03}$ & $^{+0.04}_{-0.04}$ & $^{+0.07}_{-0.06}$ & \ldots & $^{+0.03}_{-0.02}$ \\
Start & \ldots & $31000$ & $3.800$ & $100.0$ & $40.0$ & $10.0$ & $9.00$ & \ldots & $-1.05$ & $-3.60$ & $-4.20$ & $-3.20$ & $-4.10$ & $-4.60$ & $-5.80$ & $-4.40$ & $-4.90$ & $-5.60$ & $-4.60$ \\
In s & \ldots & $28070$ & $4.343$ & $-110.9$ & $35.5$ & $62.6$ & $6.04$ & $0.218$ & $-1.17$ & $-3.76$ & $-4.50$ & $-3.59$ & $-4.07$ & $-4.61$ & $-5.70$ & $-4.71$ & $-4.94$ & $-5.60$ & $-4.69$ \\
Out f & \ldots & $27740$ & $4.253$ & $-110.2$ & $28.2$ & $67.7$ & $6.50$ & $0.221$ & $-1.23$ & $-3.86$ & $-4.50$ & $-3.64$ & $-4.28$ & $-4.45$ & $-5.73$ & $-4.85$ & $-4.96$ & \ldots & $-4.64$ \\
Stat. & \ldots & $^{+120}_{-{\color{white}0}80}$ & $^{+0.019}_{-0.019}$ & $^{+1.0}_{-0.9}$ & $^{+3.0}_{-3.3}$ & $^{+3.0}_{-5.9}$ & $^{+0.70}_{-0.40}$ & $^{+0.001}_{-0.002}$ & $^{+0.05}_{-0.04}$ & $^{+0.08}_{-0.07}$ & $^{+0.05}_{-0.05}$ & $^{+0.03}_{-0.03}$ & $^{+0.20}_{-0.18}$ & $^{+0.31}_{-0.60}$ & $^{+0.09}_{-0.09}$ & $^{+0.06}_{-0.06}$ & $^{+0.08}_{-0.08}$ & \ldots & $^{+0.05}_{-0.05}$ \\
Sys. & \ldots & $^{+710}_{-880}$ & $^{+0.334}_{-0.523}$ & $^{+1.3}_{-0.6}$ & $^{+4.0}_{-2.5}$ & $^{+0.6}_{-0.2}$ & $^{+1.10}_{-2.00}$ & $^{+0.022}_{-0.012}$ & $^{+0.14}_{-0.11}$ & $^{+0.06}_{-0.05}$ & $^{+0.05}_{-0.07}$ & $^{+0.08}_{-0.03}$ & $^{+0.04}_{-0.03}$ & $^{+0.17}_{-0.26}$ & $^{+0.06}_{-0.05}$ & $^{+0.07}_{-0.03}$ & $^{+0.04}_{-0.03}$ & \ldots & $^{+0.11}_{-0.07}$ \\
Out i & \ldots & $26870$ & $4.220$ & $-110.7$ & $33.5$ & $65.0$ & $4.74$ & $0.231$ & \ldots & \ldots & \ldots & \ldots & \ldots & \ldots & \ldots & \ldots & \ldots & \ldots & \ldots \\
Stat. & \ldots & $^{+120}_{-140}$ & $^{+0.015}_{-0.018}$ & $^{+0.8}_{-1.0}$ & $^{+2.5}_{-2.7}$ & $^{+2.3}_{-2.9}$ & $^{+0.29}_{-0.30}$ & $^{+0.001}_{-0.002}$ & \ldots & \ldots & \ldots & \ldots & \ldots & \ldots & \ldots & \ldots & \ldots & \ldots & \ldots  \\
Sys. & \ldots & $^{+870}_{-850}$ & $^{+0.366}_{-0.526}$ & $^{+1.3}_{-0.9}$ & $^{+3.8}_{-2.7}$ & $^{+0.6}_{-0.5}$ & $^{+0.74}_{-1.76}$ & $^{+0.016}_{-0.010}$ & \ldots & \ldots & \ldots & \ldots & \ldots & \ldots & \ldots & \ldots & \ldots & \ldots & \ldots  \\
Start & \ldots & $23500$ & $3.800$ & $-100.0$ & $30.0$ & $10.0$ & $8.00$ & $0.300$ & $-1.05$ & $-3.60$ & $-4.20$ & $-3.20$ & $-4.10$ & $-4.60$ & $-5.80$ & $-4.40$ & $-4.90$ & $-5.60$ & $-4.60$ \\
\hline
\end{tabular}
\tablefoot{Same as Table~\ref{table:uncertainties}. The letter ``p'' denotes the primary and ``s'' the secondary component. The letter ``f'' indicates that all abundances were allowed to vary freely during the fitting process, whereas ``i'' denotes the assumption of an identical chemical composition of both components. Argon lines are not visible for all temperatures.}
\end{center}
\end{table*}
%
\subsection{Discussion of statistical and systematic uncertainties}\label{subsection:uncertainties}
\begin{figure*}
\centering
\includegraphics[width=0.49\textwidth]{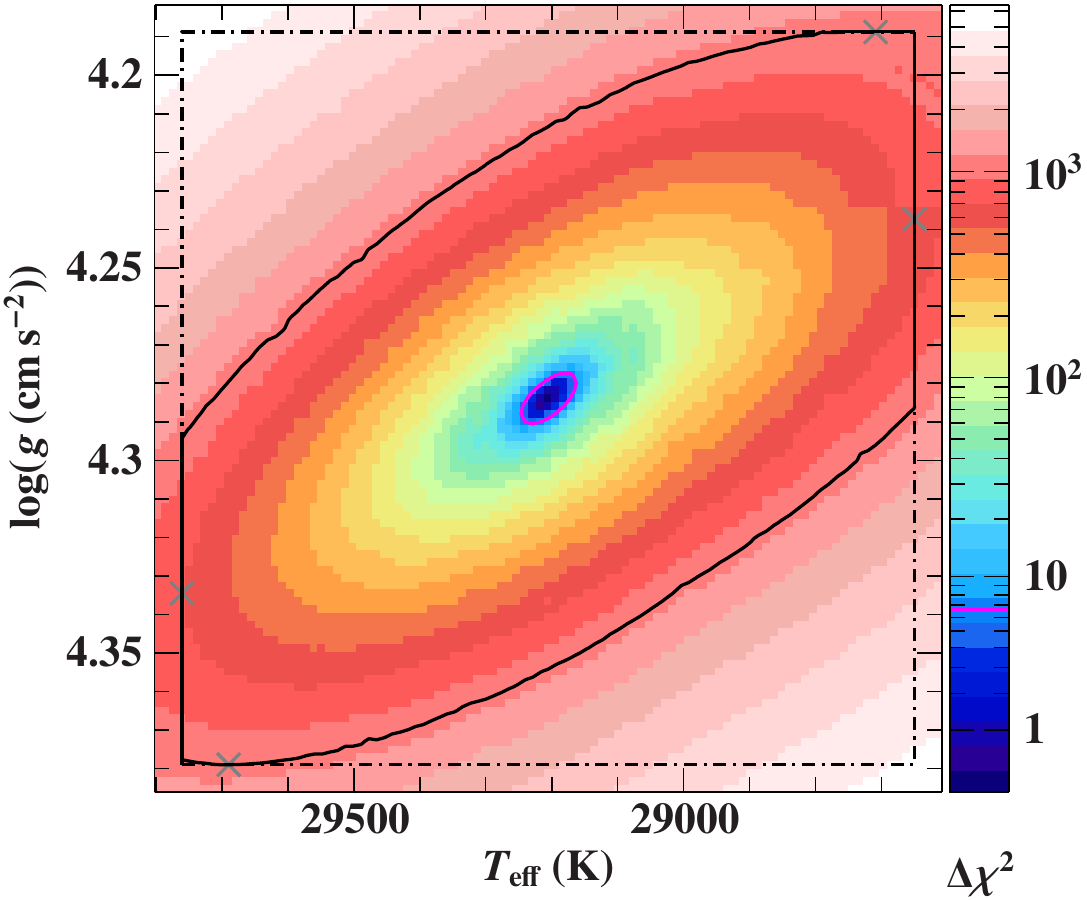}
\includegraphics[width=0.49\textwidth]{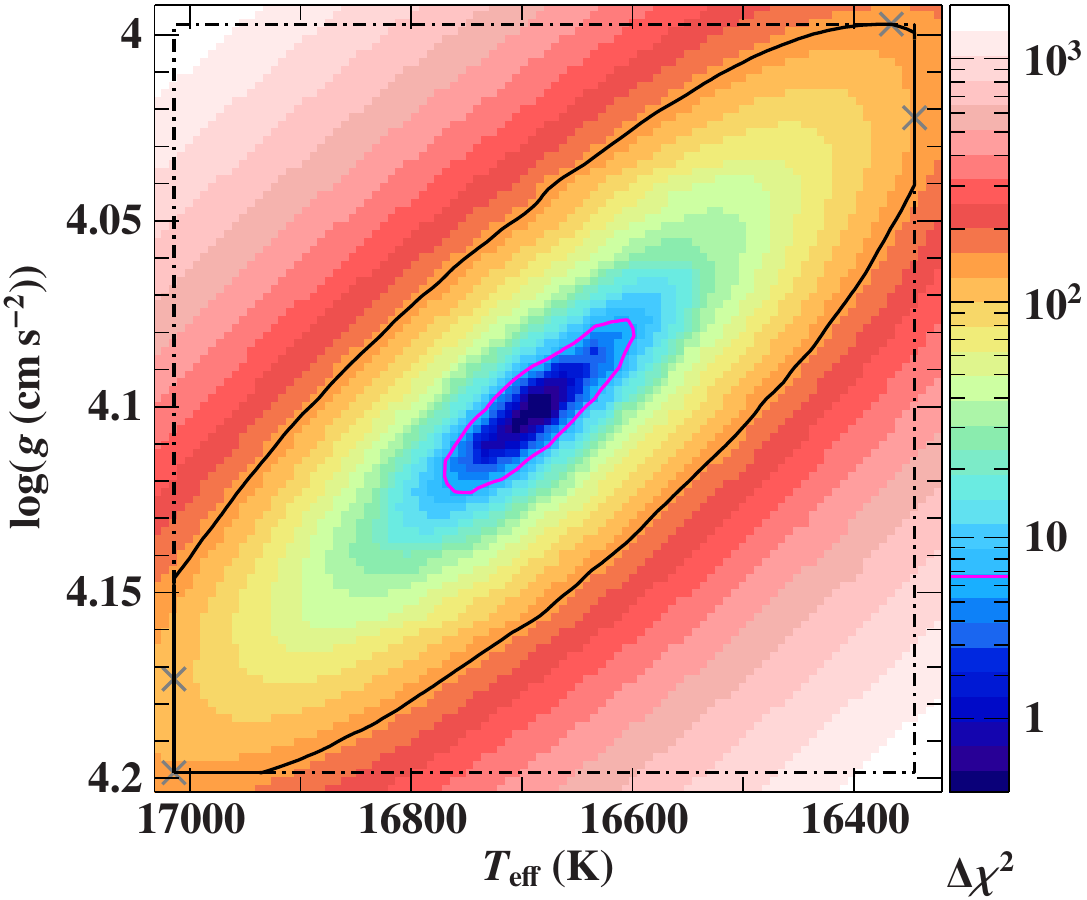}
\caption{Examples of a color coded $\Delta \chi^2$ map as a function of effective temperature and surface gravity for the single star HD\,37042 (\textit{left}) and for the primary component of the SB2 system HD\,119109 (\textit{right}). The magenta line is the $\Delta \chi^2 = 6.63$ contour line, therefore, indicating the statistical (single parameter) 99\%-confidence interval for abscissa and ordinate. The four corners of the black dashed-dotted rectangle are defined by the four combinations that result from adding or subtracting the respective total uncertainty, which is a quadratic sum of statistical and systematic uncertainty, to each coordinate of the best fit location. The point of minimum $\Delta \chi^2$ on each edge of the rectangle is marked by a gray cross. The solid black line surrounds the region within the rectangle with $\Delta \chi^2$ values lower or equal the maximum of the four $\Delta \chi^2$ values given by the gray crosses. In this way, areas within the rectangle where the models fit the observation worst are excluded, while it is ensured that each edge of the rectangle contributes at least one point to the solid line at the same time. This construction is our approach to combine statistical and systematic uncertainties.}
\label{fig:teff_logg_confmap_chisqr}
\end{figure*}
The accuracy of spectral analyses is generally limited by the quality of the obtained data and the ability of the model to reproduce the observation. As shown in this subsection, shortcomings in the model, which may be due to inaccurate atomic data or deficient line broadening theory, are the main obstacles to overcome to perform more precise investigations.

Statistical uncertainties result from the noise in the observed spectrum and can be deduced from the $\chi^2$ statistics in the standard way: starting from the best fit with a reduced $\chi^2$ of about one\footnote{\label{footnote:chisqr}This condition is generally not met because there are always some lines that our models still cannot reproduce on the small scales given by the high S/N of the available observations. In that case, the $\delta_i$ values corresponding to these lines are increased until their $\chi_i$ values (see Eq.~(\ref{eq:chisqr})) approach $\pm 1$ at the best fit eventually yielding a reduced $\chi^2$ of about $1$.}, the parameter under consideration is increased/decreased, while all remaining parameters are fitted, until a certain increment $\Delta \chi^2$ from the minimum $\chi^2$ is reached \citep[for details, see][]{bevrob}. Here, each $\Delta \chi^2$ corresponds to a confidence level; for example, $\Delta \chi^2 = 6.63$ is equivalent to the 99\%-confidence interval (see the magenta line in Fig.~\ref{fig:teff_logg_confmap_chisqr} for an illustration). The resulting uncertainties are, of course, only trustworthy if the $\delta_i$ of Eq.~(\ref{eq:chisqr}) are reasonably estimated. The method outlined in Sect.~\ref{subsection:noise_estimation} can do so, as shown by the tests with mock spectra with known noise level (see Tables~\ref{table:uncertainties} and \ref{table:uncertainties_binary}). Moreover, those tests, which use the same models as the fitting routine and, thus, exclude all sources of systematic errors apart from  microscopic line blends, give an estimate of the statistical uncertainties that can be expected in real data with a similar S/N.

Systematic uncertainties are much harder to cope with. Sources of systematic errors occur almost everywhere in the course of the analysis \citep[see the discussion in][]{nieva_iterative}. At the same time, their effects are by no means trivial, and it is extremely difficult and sometimes even impossible to quantify them. In particular, this is true for atomic data (such as energy levels, oscillator strengths, and photo-ionization cross sections), which affect individual spectral lines and the atmospheric structure. Monte Carlo simulations in the style of \citet{sigut,sigut2} offer the possibility of estimating the effects on spectra caused by variations in these input data. However, a thorough error analysis has to take all sources of systematic errors into consideration at the same time to account for correlations as well, which is an unfeasible task.

Our analysis strategy is designed to keep systematic uncertainties as small as possible. For instance, an inaccurate local continuum definition can introduce considerable uncertainties to the determination of metal abundances, especially in fast-rotating stars or low-resolution spectra where metal line blends lower the actual continuum. In our routine, these effects are allowed by re-normalizing the observed spectrum with the help of the synthetic ones. Here, the latter are used to properly locate the continuum regions, which are sufficiently frequent in optical spectra of early-type stars. For these, a correction factor is obtained by dividing the (smoothed) observed data with the model data. Interpolating this factor to the whole wavelength grid gives the local continuum correction term for all spectral lines. For this approach to work, a high degree of completeness in terms of modeled lines is necessary, which is verified by high-resolution, high S/N spectra of slow rotators, as seen in Figs.~\ref{fig:spectra_1}--\ref{fig:spectra_9} (available online only).

Another crucial part of our strategy is that we are simultaneously fitting the maximum useful range of the optical spectrum. In this way, parameters are determined not just from one or two spectral indicators but from all available ones. As the systematic errors of the individual indicators are typically independent of each other, which can be exemplified by ionization equilibria of different metals or oscillator strengths of various multiplets, there is a good chance that their effects on the parameter determination average out because some lines systematically give higher and others lower abundances, thus reducing the impact of systematics.

To crudely estimate the systematic uncertainties, we start from the assumption that they mainly appear as inaccuracies in the determination of effective temperature and surface gravity. From our extensive experience with the applied synthetic spectra, we find it realistic but conservative to assign errors of $\pm 2\%$ in $T_{\mathrm{eff}}$ and $\pm 0.1\,\mathrm{dex}$ in $\log(g)$. The ranges given by these errors are formally treated as 99\%-confidence intervals. The precision in fixing the microturbulence and the abundances of the chemical elements is then estimated from propagating the errors in $T_{\mathrm{eff}}$ and $\log(g)$. Here, a fit of all remaining parameters is performed for each pixel (that is for each combination of temperature and surface gravity) surrounded by the solid black line in Fig.~\ref{fig:teff_logg_confmap_chisqr}. In the case of a binary system, this procedure is carried out for the secondary component as well. The resulting (combined) ranges of parameter values are then taken to be 99\%-confidence intervals. This approach is valid as long as uncertainties induced by variations in $T_{\mathrm{eff}}$ and $\log(g)$ dominate other sources of systematic errors. While this is likely to be true for $\xi$ and $n(x)$, it is clearly not the case for $\varv_{\mathrm{rad}}$, $\varv\,\sin(i)$, and $\zeta$. Determination of radial velocities is generally limited by the accuracy of the wavelength calibration and ranges between $0.1$-$2\,\mathrm{km\,s^{-1}}$ for common spectrographs. Projected rotational velocity and macroturbulence are incorporated via convolution with corresponding profile functions. Because of simplifications (for example in the treatment of limb-darkening or the assumption of radial-tangential macroturbulence) during the derivation of the latter \citep[see][]{gray}, their validity may be limited to a few $\mathrm{km\,s^{-1}}$.

The comparison of statistical and systematic uncertainties, as listed in Table~\ref{table:uncertainties}, shows that our method's total uncertainty is dominated by systematic effects down to at least a S/N of $100$. However, this by no means implies that high S/N data are an unnecessary luxury. They are indispensable to detect weak features, such as contributions from a faint companion star, which would otherwise be hidden by noise. Moreover, shortcomings in the models are much more likely to remain unrecognized in low S/N spectra. This is particularly  true if they can be partly compensated by tuning some fitting parameters, which, in turn, would cause erroneous results. Instead, this comparison shows that the accuracy of the presented spectral analysis technique is currently limited by modeling and not by observation.
\section{Analysis}\label{sec:analysis}
\subsection{The program stars}\label{subsection:star_sample}
\begin{table}
\begin{center}
\small
\setlength{\tabcolsep}{0.19cm}
\caption{\label{table:program_stars} Program stars: ID, spectroscopy, photometry.}
\begin{tabular}{lrrrrr}
\hline\hline
\# & Star & Instrument & S/N & Photometry & Remark\\
\hline
1   & \object{HD\,35299}  & {\sc Fies}\tablefootmark{a} & 250 & (1),     (3), (4), (5) & single\\
2   & \object{HD\,35912}  & {\sc Fies}\tablefootmark{a} & 240 & (1),     (3), (4), (5) & single\\
3   & \object{HD\,37042}  & {\sc Fies}\tablefootmark{a} & 205 &     (2), (3), (4)      & single\\
4 a & \object{HD\,75821}  & F{\sc eros}                 & 350 & (1),     (3), (4), (5) & SB2\\
4 b & HD\,75821  & F{\sc eros}                 & 340 & (1),     (3), (4), (5) & SB2\\
4 c & HD\,75821  & F{\sc eros}                 & 340 & (1),     (3), (4), (5) & SB2\\
5   & \object{HD\,119109} & {\sc Feros}                 & 220 & (1),          (4), (5) & SB2\\
6   & \object{HD\,213420} & {\sc Fies}                  & 350 & (1),     (3), (4), (5) & SB2\\
\hline
\end{tabular}
\tablefoot{The fourth column is the mean S/N of the spectrum obtained with one of the two high-resolution spectrographs {\sc Fies} \citep[$\lambda / \Delta \lambda = 45\,000$,][]{fies} or {\sc Feros} \citep[$\lambda / \Delta \lambda = 48\,000$,][]{feros}.
\tablefoottext{a}{Spectra have been taken from the IACOB database \citep{IACOB} by courtesy of S.~Sim{\'o}n-D{\'{\i}}az and were first presented and analyzed in \citet{orion_fies}.}}
\tablebib{(1)~\citet{mermilliod}; (2)~\citet{ducati}; (3)~\citet{hauck_mermilliod}; (4)~\citet{tycho}; (5)~\citet{vanleeuwen}.}
\end{center}
\end{table}
To illustrate the capabilities of our new method, we re-analyzed three well-studied early-type stars from the Orion region and performed a spectral analysis of three SB2 systems based on very high quality single epoch spectra. For one of the binaries, two additional spectra covering different orbital phases were also investigated. Table~\ref{table:program_stars} lists the relevant information about the program stars. We focused on targets which cover a wide range of effective temperatures, have moderate projected stellar rotations, and show a variety of binary configurations.
\subsection{Atmospheric parameters and abundances}\label{subsection:atmospheric_parameters}
\begin{table*}
\begin{center}
\tiny
\setlength{\tabcolsep}{0.13cm}
\renewcommand{\arraystretch}{1.3}
\caption{\label{table:atmospheric_parameters} Atmospheric parameters and abundances of the program stars.}
\begin{tabular}{lrrrrrrrrrrrrrrrrrr}
\hline\hline
\# & $T_{\mathrm{eff}}$ & $\log(g)$ & $\varv_{\mathrm{rad}}$ & $\varv\,\sin(i)$ & $\zeta$ & $\xi$ & $A_{\mathrm{eff,s}}/A_{\mathrm{eff,p}}$ & \multicolumn{11}{c}{$\log(n(x))$} \\ 
\cline{4-7} \cline{9-19}
& (K) & (cgs) & \multicolumn{4}{c}{$(\mathrm{km\,s^{-1}})$} & & He & C & N & O & Ne & Mg & Al & Si & S & Ar & Fe\\
\hline
1  & $23880$ & $4.127$ &   $23.0$ &   $5.2$ &  $4.4$ &  $2.02$ &  \ldots & $-0.98$ & $-3.73$ & $-4.30$ & $-3.29$ & $-4.00$ & $-4.56$ & $-5.79$ & $-4.66$ & $-4.88$ & $-5.49$ & $-4.71$ \\ Stat. &   $^{+70}_{-70}$ & $^{+0.008}_{-0.008}$ & $^{+0.1}_{-0.1}$ & $^{+0.7}_{-2.2}$ & $^{+2.0}_{-0.7}$ & $^{+0.11}_{-0.23}$ &               \ldots & $^{+0.01}_{-0.02}$ & $^{+0.02}_{-0.02}$ & $^{+0.01}_{-0.01}$ & $^{+0.02}_{-0.02}$ & $^{+0.02}_{-0.02}$ & $^{+0.02}_{-0.03}$ & $^{+0.01}_{-0.02}$ & $^{+0.02}_{-0.02}$ & $^{+0.02}_{-0.01}$ & $^{+0.02}_{-0.02}$ & $^{+0.02}_{-0.02}$ \\ Sys. &  $^{+480}_{-480}$ & $^{+0.100}_{-0.100}$ &   $^{+0.1}_{-0.1}$ &  $^{+0.5}_{-0.3}$ &   $^{+0.8}_{-1.5}$ & $^{+0.68}_{-1.39}$ &               \ldots & $^{+0.04}_{-0.06}$ & $^{+0.02}_{-0.02}$ & $^{+0.03}_{-0.02}$ & $^{+0.09}_{-0.08}$ & $^{+0.04}_{-0.04}$ & $^{+0.10}_{-0.10}$ & $^{+0.03}_{-0.03}$ & $^{+0.07}_{-0.06}$ & $^{+0.06}_{-0.06}$ & $^{+0.07}_{-0.06}$ & $^{+0.03}_{-0.02}$ \\
1\tablefootmark{a} & $24000$   & $4.20$     & \ldots & $8$     & \ldots  & $0$     & \ldots & $-1.06$ & $-3.67$    & $-4.23$    & $-3.32$    & $-3.92$    & $-4.43$    & \ldots  & $-4.54$    & \ldots  & \ldots  & $-4.55$ \\ 
                   & $\pm 200$ & $\pm 0.08$ & \ldots & $\pm 1$ & \ldots  & $\pm 1$ & \ldots & \ldots  & $\pm 0.07$ & $\pm 0.07$ & $\pm 0.07$ & $\pm 0.09$ & $\pm 0.07$ & \ldots  & $\pm 0.08$ & \ldots  & \ldots  & $\pm 0.10$  \\
\hline
2  & $19250$ & $4.052$ &   $30.7$ &   $7.3$ & $17.1$ &  $1.99$ &  \ldots & $-1.00$ & $-3.64$ & $-4.23$ & $-3.21$ & $-4.06$ & $-4.60$ & $-5.71$ & $-4.48$ & $-4.89$ & $-5.57$ & $-4.63$ \\ Stat. &   $^{+60}_{-50}$ & $^{+0.007}_{-0.007}$ & $^{+0.2}_{-0.1}$ & $^{+0.4}_{-0.5}$ & $^{+0.5}_{-1.1}$ & $^{+0.14}_{-0.26}$ &               \ldots & $^{+0.01}_{-0.02}$ & $^{+0.03}_{-0.02}$ & $^{+0.02}_{-0.02}$ & $^{+0.02}_{-0.02}$ & $^{+0.02}_{-0.02}$ & $^{+0.02}_{-0.03}$ & $^{+0.03}_{-0.02}$ & $^{+0.03}_{-0.03}$ & $^{+0.02}_{-0.01}$ & $^{+0.03}_{-0.03}$ & $^{+0.02}_{-0.02}$ \\ Sys. &  $^{+390}_{-390}$ & $^{+0.100}_{-0.100}$ &   $^{+0.1}_{-0.1}$ &  $^{+0.7}_{-1.8}$ &   $^{+0.2}_{-0.1}$ & $^{+1.08}_{-1.56}$ &               \ldots & $^{+0.08}_{-0.09}$ & $^{+0.07}_{-0.07}$ & $^{+0.08}_{-0.08}$ & $^{+0.10}_{-0.11}$ & $^{+0.03}_{-0.03}$ & $^{+0.11}_{-0.13}$ & $^{+0.06}_{-0.07}$ & $^{+0.13}_{-0.14}$ & $^{+0.05}_{-0.05}$ & $^{+0.04}_{-0.05}$ & $^{+0.01}_{-0.03}$ \\
2\tablefootmark{a} & $19000$   & $4.00$     & \ldots & $15$    & $8$     & $2$     & \ldots & $-1.06$ & $-3.71$    & $-4.28$    & $-3.25$    & $-3.99$    & $-4.54$    & \ldots  & $-4.56$    & \ldots  & \ldots  & $-4.52$ \\ 
                   & $\pm 300$ & $\pm 0.10$ & \ldots & $\pm 1$ & $\pm 1$ & $\pm 1$ & \ldots & \ldots  & $\pm 0.09$ & $\pm 0.07$ & $\pm 0.09$ & $\pm 0.11$ & $\pm 0.05$ & \ldots  & $\pm 0.07$ & \ldots  & \ldots  & $\pm 0.08$ \\
\hline
3  & $29210$ & $4.284$ &   $29.9$ &  $30.9$ &  $0.0$ &  $3.22$ &  \ldots & $-1.04$ & $-3.71$ & $-4.13$ & $-3.40$ & $-4.01$ & $-4.58$ & $-5.73$ & $-4.66$ & $-4.97$ &  \ldots & $-4.62$ \\ Stat. &   $^{+30}_{-50}$ & $^{+0.006}_{-0.007}$ & $^{+0.2}_{-0.1}$ & $^{+0.1}_{-0.1}$ & $^{+1.1}_{-0.0}$ & $^{+0.10}_{-0.11}$ &               \ldots & $^{+0.01}_{-0.02}$ & $^{+0.02}_{-0.02}$ & $^{+0.01}_{-0.01}$ & $^{+0.02}_{-0.01}$ & $^{+0.03}_{-0.03}$ & $^{+0.02}_{-0.03}$ & $^{+0.03}_{-0.02}$ & $^{+0.03}_{-0.01}$ & $^{+0.03}_{-0.03}$ &             \ldots & $^{+0.02}_{-0.02}$ \\ Sys. &  $^{+580}_{-590}$ & $^{+0.100}_{-0.100}$ &   $^{+0.1}_{-0.1}$ &  $^{+0.2}_{-0.1}$ &   $^{+2.6}_{-0.0}$ & $^{+0.47}_{-0.59}$ &               \ldots & $^{+0.04}_{-0.04}$ & $^{+0.01}_{-0.03}$ & $^{+0.05}_{-0.05}$ & $^{+0.04}_{-0.02}$ & $^{+0.02}_{-0.02}$ & $^{+0.05}_{-0.05}$ & $^{+0.05}_{-0.05}$ & $^{+0.04}_{-0.03}$ & $^{+0.07}_{-0.05}$ &             \ldots & $^{+0.08}_{-0.06}$ \\
3\tablefootmark{a} & $29300$   & $4.30$     & \ldots & $30$    & $10$    & $2$     & \ldots & $-1.06$ & $-3.71$    & $-4.00$    & $-3.29$    & $-3.91$    & $-4.40$    & \ldots  & $-4.49$    & \ldots  & \ldots  & $-4.50$ \\ 
                   & $\pm 300$ & $\pm 0.09$ & \ldots & $\pm 2$ & $\pm 3$ & $\pm 1$ & \ldots & \ldots  & $\pm 0.11$ & $\pm 0.08$ & $\pm 0.08$ & $\pm 0.09$ & \ldots     & \ldots  & $\pm 0.03$ & \ldots  & \ldots  & $\pm 0.09$ \\
\hline
4p a & $29420$ & $3.620$ &   $17.7$ &  $24.0$ & $37.4$ & $13.58$ &  \ldots & $-1.10$ & $-3.81$ & $-4.47$ & $-3.60$ & $-4.10$ & $-4.60$ & $-5.68$ & $-4.70$ & $-4.91$ &  \ldots & $-4.59$ \\ Stat. &   $^{+20}_{-20}$ & $^{+0.003}_{-0.003}$ & $^{+0.2}_{-0.2}$ & $^{+0.7}_{-0.4}$ & $^{+0.2}_{-1.3}$ & $^{+0.12}_{-0.11}$ &               \ldots & $^{+0.01}_{-0.01}$ & $^{+0.02}_{-0.01}$ & $^{+0.02}_{-0.01}$ & $^{+0.01}_{-0.01}$ & $^{+0.01}_{-0.02}$ & $^{+0.02}_{-0.01}$ & $^{+0.02}_{-0.01}$ & $^{+0.01}_{-0.01}$ & $^{+0.02}_{-0.03}$ &             \ldots & $^{+0.03}_{-0.02}$ \\ Sys. &  $^{+590}_{-590}$ & $^{+0.100}_{-0.100}$ &   $^{+1.5}_{-2.0}$ &  $^{+2.9}_{-5.0}$ &   $^{+0.5}_{-0.1}$ & $^{+0.75}_{-1.04}$ &               \ldots & $^{+0.01}_{-0.02}$ & $^{+0.02}_{-0.01}$ & $^{+0.04}_{-0.02}$ & $^{+0.02}_{-0.03}$ & $^{+0.01}_{-0.02}$ & $^{+0.02}_{-0.01}$ & $^{+0.03}_{-0.02}$ & $^{+0.04}_{-0.04}$ & $^{+0.04}_{-0.05}$ &             \ldots & $^{+0.05}_{-0.04}$ \\
4s a & $32900$ & $4.687$ &   $56.6$ &   $0.0$ & $69.1$ & $14.34$ & $0.218$ &  \ldots &  \ldots &  \ldots &  \ldots &  \ldots &  \ldots &  \ldots &  \ldots &  \ldots &  \ldots &  \ldots \\ Stat. &   $^{+80}_{-80}$ & $^{+0.012}_{-0.014}$ & $^{+0.9}_{-0.6}$ & $^{+7.8}_{-0.0}$ & $^{+1.4}_{-1.1}$ & $^{+0.72}_{-0.53}$ & $^{+0.002}_{-0.007}$ &             \ldots &             \ldots &             \ldots &             \ldots &             \ldots &             \ldots &             \ldots &             \ldots &             \ldots &             \ldots &             \ldots \\ Sys. & $^{+{\color{white}0}910}_{-1770}$ & $^{+\geq0.063}_{-\enspace0.305}$ & $^{+14.8}_{-18.3}$ & $^{+34.8}_{-{\color{white}0}0.0}$ &   $^{+0.8}_{-0.2}$ & $^{+1.66}_{-2.61}$ & $^{+0.314}_{-0.095}$ &             \ldots &             \ldots &             \ldots &             \ldots &             \ldots &             \ldots &             \ldots &             \ldots &             \ldots &             \ldots &             \ldots \\
\hline
4p b & $29710$ & $3.669$ &  $104.0$ &  $23.7$ & $41.2$ & $14.92$ &  \ldots & $-1.17$ & $-3.76$ & $-4.50$ & $-3.59$ & $-4.07$ & $-4.61$ & $-5.70$ & $-4.71$ & $-4.94$ &  \ldots & $-4.69$ \\ Stat. &   $^{+40}_{-40}$ & $^{+0.005}_{-0.004}$ & $^{+0.2}_{-0.1}$ & $^{+0.5}_{-0.5}$ & $^{+0.4}_{-0.6}$ & $^{+0.12}_{-0.12}$ &               \ldots & $^{+0.01}_{-0.01}$ & $^{+0.02}_{-0.01}$ & $^{+0.02}_{-0.02}$ & $^{+0.01}_{-0.01}$ & $^{+0.02}_{-0.01}$ & $^{+0.02}_{-0.02}$ & $^{+0.02}_{-0.02}$ & $^{+0.01}_{-0.01}$ & $^{+0.02}_{-0.03}$ &             \ldots & $^{+0.03}_{-0.03}$ \\ Sys. &  $^{+600}_{-600}$ & $^{+0.100}_{-0.100}$ &   $^{+0.1}_{-0.1}$ &  $^{+1.2}_{-1.5}$ &   $^{+0.1}_{-0.1}$ & $^{+0.27}_{-0.24}$ &               \ldots & $^{+0.03}_{-0.03}$ & $^{+0.02}_{-0.01}$ & $^{+0.06}_{-0.05}$ & $^{+0.05}_{-0.05}$ & $^{+0.02}_{-0.01}$ & $^{+0.03}_{-0.03}$ & $^{+0.04}_{-0.04}$ & $^{+0.05}_{-0.05}$ & $^{+0.07}_{-0.07}$ &             \ldots & $^{+0.05}_{-0.06}$ \\
4s b & $28070$ & $4.343$ & $-110.9$ &  $35.5$ & $62.6$ &  $6.04$ & $0.218$ &  \ldots &  \ldots &  \ldots &  \ldots &  \ldots &  \ldots &  \ldots &  \ldots &  \ldots &  \ldots &  \ldots \\ Stat. & $^{+140}_{-170}$ & $^{+0.016}_{-0.021}$ & $^{+0.8}_{-1.0}$ & $^{+2.7}_{-2.8}$ & $^{+3.5}_{-3.4}$ & $^{+0.23}_{-0.35}$ & $^{+0.003}_{-0.002}$ &             \ldots &             \ldots &             \ldots &             \ldots &             \ldots &             \ldots &             \ldots &             \ldots &             \ldots &             \ldots &             \ldots \\ Sys. &  $^{+610}_{-870}$ & $^{+0.310}_{-0.434}$ &   $^{+0.6}_{-0.4}$ &  $^{+2.6}_{-1.9}$ &   $^{+0.4}_{-0.6}$ & $^{+0.78}_{-0.80}$ & $^{+0.014}_{-0.011}$ &             \ldots &             \ldots &             \ldots &             \ldots &             \ldots &             \ldots &             \ldots &             \ldots &             \ldots &             \ldots &             \ldots \\
\hline
4p c & $29630$ & $3.622$ &  $-17.2$ &  $22.4$ & $37.2$ & $14.68$ &  \ldots & $-1.17$ & $-3.78$ & $-4.57$ & $-3.58$ & $-4.00$ & $-4.60$ & $-5.71$ & $-4.70$ & $-4.92$ &  \ldots & $-4.65$ \\ Stat. &   $^{+20}_{-20}$ & $^{+0.003}_{-0.003}$ & $^{+0.2}_{-0.1}$ & $^{+0.4}_{-0.5}$ & $^{+0.2}_{-0.1}$ & $^{+0.12}_{-0.12}$ &               \ldots & $^{+0.01}_{-0.01}$ & $^{+0.01}_{-0.02}$ & $^{+0.02}_{-0.02}$ & $^{+0.01}_{-0.01}$ & $^{+0.01}_{-0.02}$ & $^{+0.02}_{-0.01}$ & $^{+0.02}_{-0.03}$ & $^{+0.01}_{-0.01}$ & $^{+0.03}_{-0.02}$ &             \ldots & $^{+0.03}_{-0.03}$ \\ Sys. &  $^{+600}_{-600}$ & $^{+0.100}_{-0.100}$ &   $^{+0.5}_{-1.0}$ &  $^{+1.4}_{-8.9}$ &   $^{+1.1}_{-0.1}$ & $^{+0.48}_{-0.19}$ &               \ldots & $^{+0.03}_{-0.02}$ & $^{+0.01}_{-0.02}$ & $^{+0.03}_{-0.03}$ & $^{+0.05}_{-0.04}$ & $^{+0.02}_{-0.02}$ & $^{+0.03}_{-0.02}$ & $^{+0.03}_{-0.04}$ & $^{+0.05}_{-0.05}$ & $^{+0.06}_{-0.05}$ &             \ldots & $^{+0.05}_{-0.04}$ \\
4s c & $30630$ & $4.750$ &   $83.5$ &  $68.9$ & $38.3$ &  $9.73$ & $0.228$ &  \ldots &  \ldots &  \ldots &  \ldots &  \ldots &  \ldots &  \ldots &  \ldots &  \ldots &  \ldots &  \ldots \\ Stat. & $^{+140}_{-150}$ & $^{+ \ldots}_{-0.013}$ & $^{+1.0}_{-1.0}$ & $^{+0.1}_{-2.8}$ & $^{+4.1}_{-4.2}$ & $^{+0.50}_{-0.47}$ & $^{+0.003}_{-0.003}$ &             \ldots &             \ldots &             \ldots &             \ldots &             \ldots &             \ldots &             \ldots &             \ldots &             \ldots &             \ldots &             \ldots \\ Sys. & $^{+{\color{white}0}780}_{-1130}$ & $^{+ \ldots}_{-0.358}$ &  $^{+{\color{white}0}4.4}_{-36.2}$ & $^{+42.1}_{-{\color{white}0}0.8}$ & $^{+37.7}_{-22.1}$ & $^{+0.97}_{-0.73}$ & $^{+0.204}_{-0.034}$ &             \ldots &             \ldots &             \ldots &             \ldots &             \ldots &             \ldots &             \ldots &             \ldots &             \ldots &             \ldots &             \ldots \\
\hline
5p & $16680$ & $4.098$ &  $-84.7$ &   $7.9$ & $11.4$ &  $2.10$ &  \ldots & $-0.96$ & $-3.55$ & $-4.16$ & $-3.25$ & $-4.04$ & $-4.73$ & $-5.86$ & $-4.45$ & $-4.91$ & $-5.58$ & $-4.66$ \\ Stat. &  $^{+100}_{-{\color{white}0}90}$ & $^{+0.027}_{-0.021}$ & $^{+0.1}_{-0.2}$ & $^{+1.9}_{-1.3}$ & $^{+1.0}_{-2.2}$ & $^{+0.17}_{-0.19}$ &               \ldots & $^{+0.02}_{-0.02}$ & $^{+0.03}_{-0.03}$ & $^{+0.03}_{-0.04}$ & $^{+0.02}_{-0.03}$ & $^{+0.02}_{-0.03}$ & $^{+0.04}_{-0.03}$ & $^{+0.05}_{-0.04}$ & $^{+0.03}_{-0.03}$ & $^{+0.02}_{-0.03}$ & $^{+0.06}_{-0.08}$ & $^{+0.03}_{-0.03}$ \\ Sys. &  $^{+330}_{-340}$ & $^{+0.100}_{-0.100}$ &   $^{+0.1}_{-0.1}$ &  $^{+1.2}_{-0.2}$ &   $^{+0.5}_{-1.7}$ & $^{+0.25}_{-0.29}$ &               \ldots & $^{+0.06}_{-0.06}$ & $^{+0.06}_{-0.06}$ & $^{+0.06}_{-0.06}$ & $^{+0.01}_{-0.02}$ & $^{+0.01}_{-0.02}$ & $^{+0.04}_{-0.03}$ & $^{+0.03}_{-0.03}$ & $^{+0.01}_{-0.03}$ & $^{+0.02}_{-0.03}$ & $^{+0.04}_{-0.05}$ & $^{+0.04}_{-0.05}$ \\
5s & $13490$ & $4.274$ &  $125.0$ &  $28.3$ & $15.6$ &  $0.79$ & $0.642$ &  \ldots &  \ldots &  \ldots &  \ldots &  \ldots &  \ldots &  \ldots &  \ldots &  \ldots &  \ldots &  \ldots \\ Stat. &   $^{+90}_{-80}$ & $^{+0.030}_{-0.025}$ & $^{+0.5}_{-0.4}$ & $^{+1.1}_{-1.5}$ & $^{+3.1}_{-2.0}$ & $^{+0.32}_{-0.31}$ & $^{+0.015}_{-0.013}$ &             \ldots &             \ldots &             \ldots &             \ldots &             \ldots &             \ldots &             \ldots &             \ldots &             \ldots &             \ldots &             \ldots \\ Sys. &  $^{+270}_{-280}$ & $^{+0.100}_{-0.102}$ &   $^{+0.1}_{-0.2}$ &  $^{+0.2}_{-0.3}$ &   $^{+1.6}_{-0.7}$ & $^{+0.23}_{-0.23}$ & $^{+0.027}_{-0.028}$ &             \ldots &             \ldots &             \ldots &             \ldots &             \ldots &             \ldots &             \ldots &             \ldots &             \ldots &             \ldots &             \ldots \\
\hline
6p & $20590$ & $3.485$ &  $-11.2$ &  $54.2$ &  $9.4$ &  $6.04$ &  \ldots & $-1.02$ & $-3.79$ & $-4.38$ & $-3.39$ & $-4.02$ & $-4.74$ & $-5.87$ & $-4.66$ & $-4.98$ & $-5.59$ & $-4.78$ \\ Stat. &   $^{+30}_{-40}$ & $^{+0.004}_{-0.005}$ & $^{+0.3}_{-0.2}$ & $^{+0.1}_{-0.1}$ & $^{+0.1}_{-2.2}$ & $^{+0.09}_{-0.08}$ &               \ldots & $^{+0.01}_{-0.02}$ & $^{+0.02}_{-0.02}$ & $^{+0.01}_{-0.02}$ & $^{+0.02}_{-0.01}$ & $^{+0.02}_{-0.02}$ & $^{+0.02}_{-0.03}$ & $^{+0.02}_{-0.02}$ & $^{+0.02}_{-0.02}$ & $^{+0.01}_{-0.02}$ & $^{+0.02}_{-0.03}$ & $^{+0.02}_{-0.02}$ \\ Sys. &  $^{+420}_{-410}$ & $^{+0.100}_{-0.100}$ &   $^{+0.1}_{-0.1}$ &  $^{+0.3}_{-0.1}$ &   $^{+1.7}_{-3.2}$ & $^{+0.60}_{-0.61}$ &               \ldots & $^{+0.01}_{-0.02}$ & $^{+0.02}_{-0.01}$ & $^{+0.02}_{-0.03}$ & $^{+0.06}_{-0.04}$ & $^{+0.02}_{-0.01}$ & $^{+0.02}_{-0.02}$ & $^{+0.02}_{-0.03}$ & $^{+0.03}_{-0.03}$ & $^{+0.01}_{-0.03}$ & $^{+0.01}_{-0.03}$ & $^{+0.02}_{-0.02}$ \\
6s & $18610$ & $3.227$ &   $-9.1$ & $134.0$ & $59.5$ &  $2.90$ & $0.936$ &  \ldots &  \ldots &  \ldots &  \ldots &  \ldots &  \ldots &  \ldots &  \ldots &  \ldots &  \ldots &  \ldots \\ Stat. &   $^{+50}_{-70}$ & $^{+0.004}_{-0.006}$ & $^{+1.1}_{-1.0}$ & $^{+0.1}_{-0.6}$ & $^{+4.3}_{-4.7}$ & $^{+0.18}_{-0.18}$ & $^{+0.014}_{-0.015}$ &             \ldots &             \ldots &             \ldots &             \ldots &             \ldots &             \ldots &             \ldots &             \ldots &             \ldots &             \ldots &             \ldots \\ Sys. &  $^{+460}_{-370}$ & $^{+0.148}_{-0.130}$ &   $^{+0.3}_{-0.2}$ &  $^{+0.5}_{-0.8}$ &   $^{+2.2}_{-1.8}$ & $^{+0.96}_{-0.82}$ & $^{+0.069}_{-0.060}$ &             \ldots &             \ldots &             \ldots &             \ldots &             \ldots &             \ldots &             \ldots &             \ldots &             \ldots &             \ldots &             \ldots \\

\hline
\sun\tablefootmark{b} & & & & & & & & $-1.06$ & $-3.57$ & $-4.17$ & $-3.31$ & $-4.07$ & $-4.40$ & $-5.55$ & $-4.49$ & $-4.88$ & $-5.60$ & $-4.50$ \\
                     & & & & & & & & $^{+0.01}_{-0.01}$ & $^{+0.05}_{-0.05}$ & $^{+0.05}_{-0.05}$ & $^{+0.05}_{-0.05}$ & $^{+0.10}_{-0.10}$ & $^{+0.04}_{-0.04}$ & $^{+0.03}_{-0.03}$ & $^{+0.04}_{-0.04}$ & $^{+0.03}_{-0.03}$ & $^{+0.13}_{-0.13}$ & $^{+0.04}_{-0.04}$ \\
\hline
\end{tabular}
\tablefoot{Same as Table~\ref{table:uncertainties}. Numbering according to Table~\ref{table:program_stars}. Argon lines are not visible for all temperatures. Owing to the assumption of a homo\-geneous chemical composition, abundances of the secondary components ``s'' are tied to the ones of the primaries ``p'' during the analysis. \tablefoottext{a}{Values and uncertainties from \citet{orion_composition} or in the case of oxygen and silicon from \citet{orion_fies}.}\tablefoottext{b}{Protosolar nebula values and uncertainties from \citet{sun}.} }
\end{center}
\end{table*}
Atmospheric parameters and abundances are determined by fitting synthetic to observed spectra, as outlined in detail in Sect.~\ref{sec:setting_up}. Comparisons of final, best-fitting models with observations are shown for a large portion of the used spectral range in Figs.~\ref{fig:spectra_1}--\ref{fig:spectra_9} for the Orion stars and in Figs.~\ref{fig:binary_spectra_1}--\ref{fig:binary_spectra_9} (available online only) for the SB2 systems. The overall match of metal lines is almost perfect for the cooler stars of the sample and still very good for the hotter ones where our model atoms begin to be partially incomplete because they were not optimized for this temperature regime. This is particularly true with respect to \ion{O}{ii} lines. Consequently, more regions have to be excluded from the analysis for higher effective temperatures due to (blends with) missing spectral lines. 

Note that we generally exclude several \ion{He}{i} lines from the analysis owing to recurrent issues with their detailed spectral line shapes, which is apparent only in high-resolution spectra with high S/N. This specifically affects the diffuse \ion{He}{I} lines that show small but perceptible systematic deficiencies in their forbidden components, which can be attributed to shortcomings in their line broadening theory. These issues are independent of $T_{\mathrm{eff}}$ and $\log(g)$ and cannot be resolved, even if these lines are fitted individually. With a sufficient number of alternative, highly trustworthy \ion{He}{I} lines present in the optical spectral range, we generally refrain from fitting the diffuse lines to avoid a possible source of systematic error and use them instead as a consistency check. Nevertheless, given the fact that the synthetic profiles of these lines match the observed ones quite well, except for the forbidden components (see Figs.~\ref{fig:spectra_1}--\ref{fig:spectra_9}), this decision is probably too restrictive. The diffuse \ion{He}{I} lines could therefore be considered for spectral fitting as well with resulting changes in parameters that are well below the stated systematic uncertainties.

Furthermore, we ignore the temperature-sensitive cores of the Balmer lines during the fitting process by excluding those parts of these lines where their normalized flux is smaller than a cutoff, which is typically chosen to be $0.8$. The reason for this is that they are formed in the outer stellar atmosphere, where deviations from the LTE stratification are more pronounced \citep[see][]{hybrid2,hybrid3} and where the assumption of hydrostatic equilibrium also becomes less and less valid in the accelerating (weak) stellar wind. Moreover, by simultaneously fitting the entire useful spectral range, there are enough other indicators for $T_{\mathrm{eff}}$ such as (multiple) ionization equilibria available so that it is sufficient to use the Balmer line cores as a consistency check (Figures~\ref{fig:spectra_1}--\ref{fig:spectra_9} and Figs.~\ref{fig:binary_spectra_1}--\ref{fig:binary_spectra_9} show how well this works.). As a side product, we also reduce the otherwise overwhelming influence of these lines on the parameter determination. 

Table~\ref{table:atmospheric_parameters} lists the atmospheric parameters and abundances of the program stars. Instead of the ``classical'' notation for the abundance, $\log(x/\mathrm{H})+12$, we have chosen $n(x)$, which is the fractional particle number of species $x$ with respect to \emph{all} elements. The motivation for this is that the helium abundance is variable from star to star, which, in turn, causes the hydrogen abundance to vary since hydrogen and helium abundances are coupled via the fixed number of total particles. As a consequence, the quantity $\log(x/\mathrm{H})$ can change even if $x$ stays constant. For a better comparison of metal abundances in stars with different helium content, we therefore prefer the notation that gives abundances relative to all elements.
\onlfig{
\setcounter{subfig}{1}
\renewcommand\thefigure{\arabic{figure}\alph{subfig}}
\begin{figure*}
\centering
\includegraphics[height=1\textwidth, angle=-90]{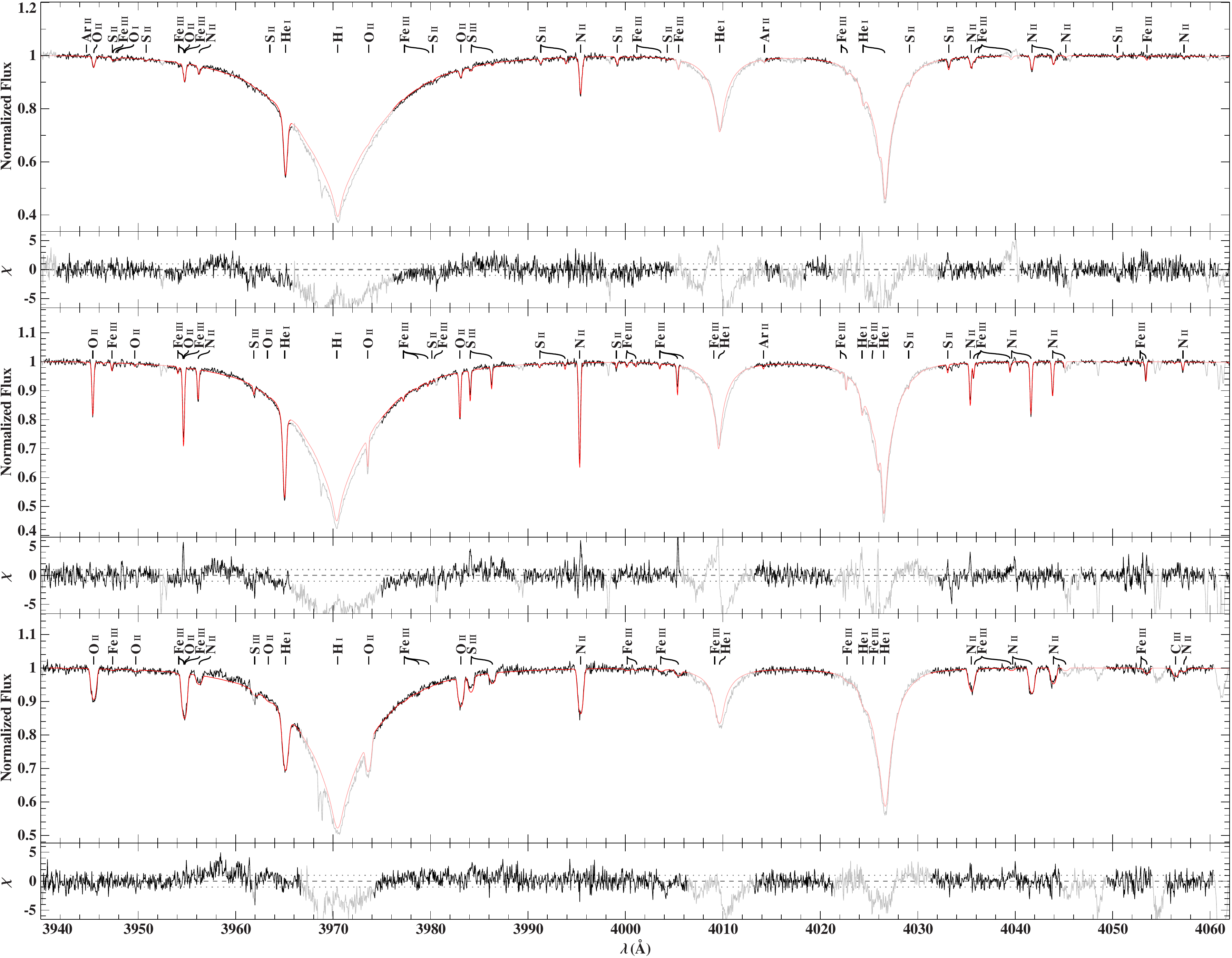}
\caption{Comparison of best-fitting model spectrum (red line) with re-normalized observation (black line) for the stars, HD\,35912 (\textit{right}), HD\,35299 (\textit{middle}), and HD\,37042 (\textit{left}), in the spectral range $\lambda\lambda$\,3940--4062\,\AA. Light colors mark regions that have been excluded from fitting due to the presence of features that are not (properly) included in our models. For the sake of clarity, only the strongest out of all lines that have been used in the analysis are labeled. The residuals $\chi$ are defined by the bracket in Eq.~(\ref{eq:chisqr}).}
\label{fig:spectra_1}
\addtocounter{subfig}{1}
\end{figure*}
}
\addtocounter{figure}{-1}
\onlfig{
\begin{figure*}
\centering
\includegraphics[height=1\textwidth, angle=-90]{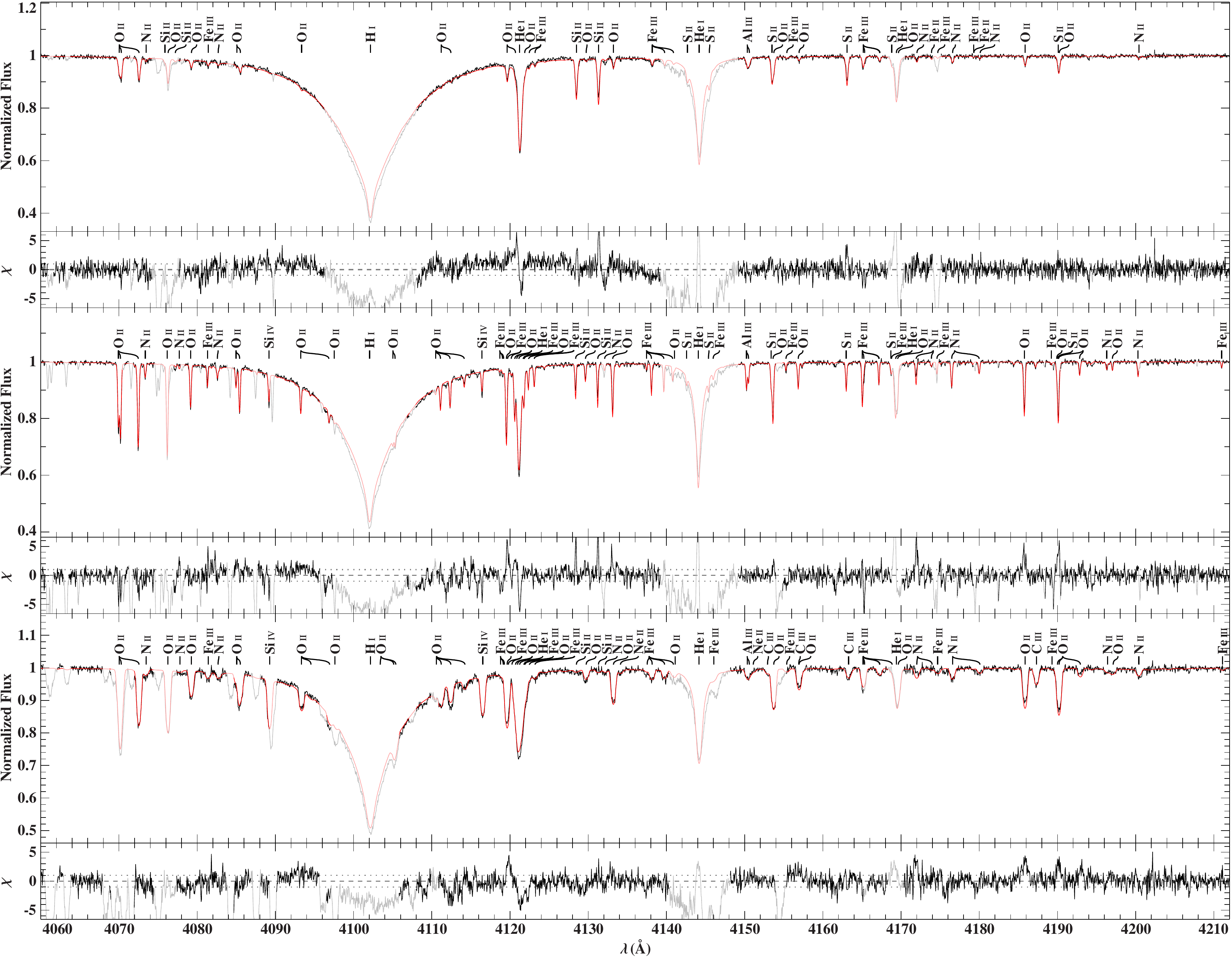}
\caption{Same as Fig.~\ref{fig:spectra_1} in the spectral range $\lambda\lambda$\,4060--4212\,\AA.}
\label{fig:spectra_2}
\addtocounter{subfig}{1}
\end{figure*}
}
\addtocounter{figure}{-1}
\onlfig{
\begin{figure*}
\centering
\includegraphics[height=1\textwidth, angle=-90]{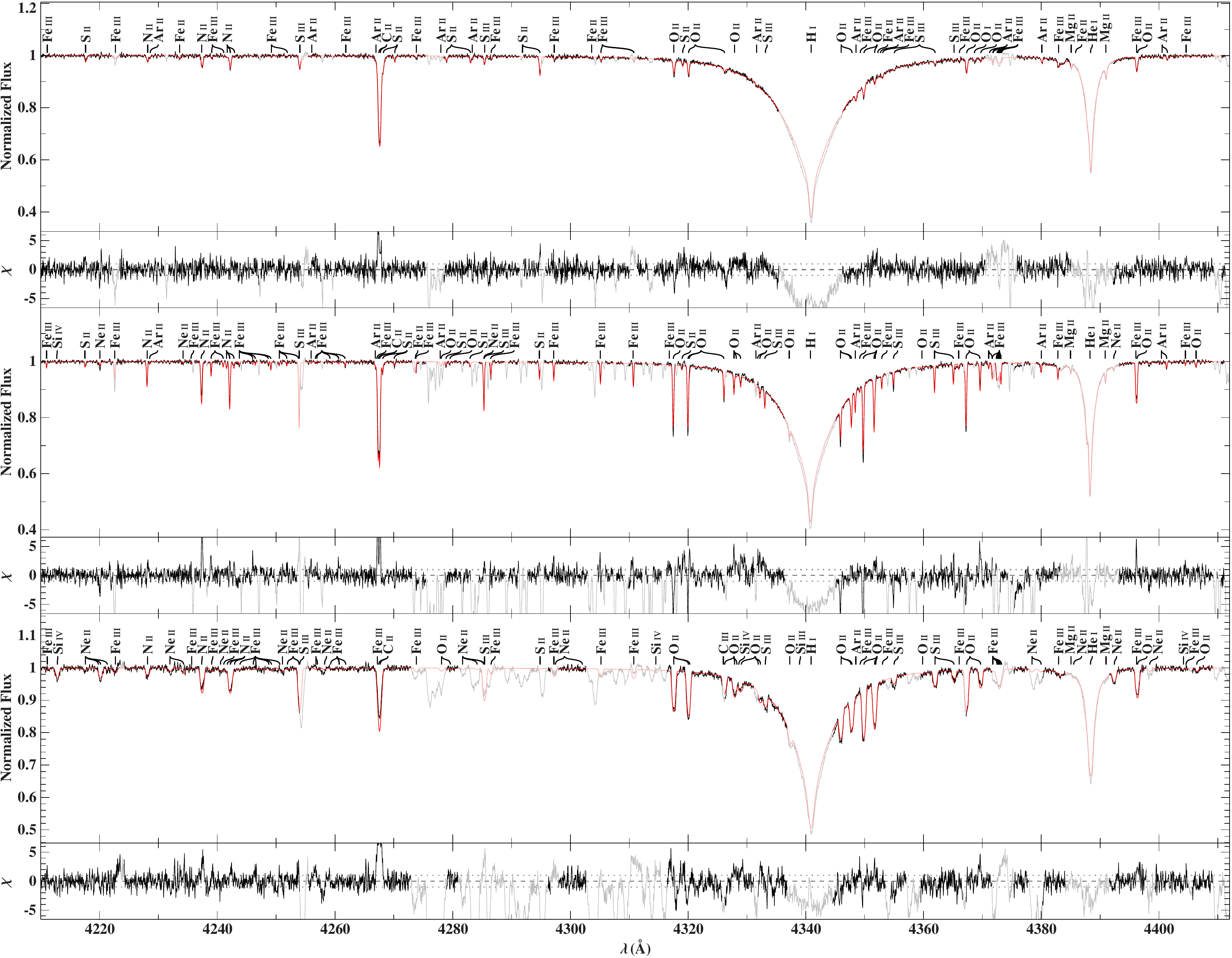}
\caption{Same as Fig.~\ref{fig:spectra_1} in the spectral range $\lambda\lambda$\,4210--4412\,\AA.}
\label{fig:spectra_3}
\addtocounter{subfig}{1}
\end{figure*}
}
\addtocounter{figure}{-1}
\onlfig{
\begin{figure*}
\centering
\includegraphics[height=1\textwidth, angle=-90]{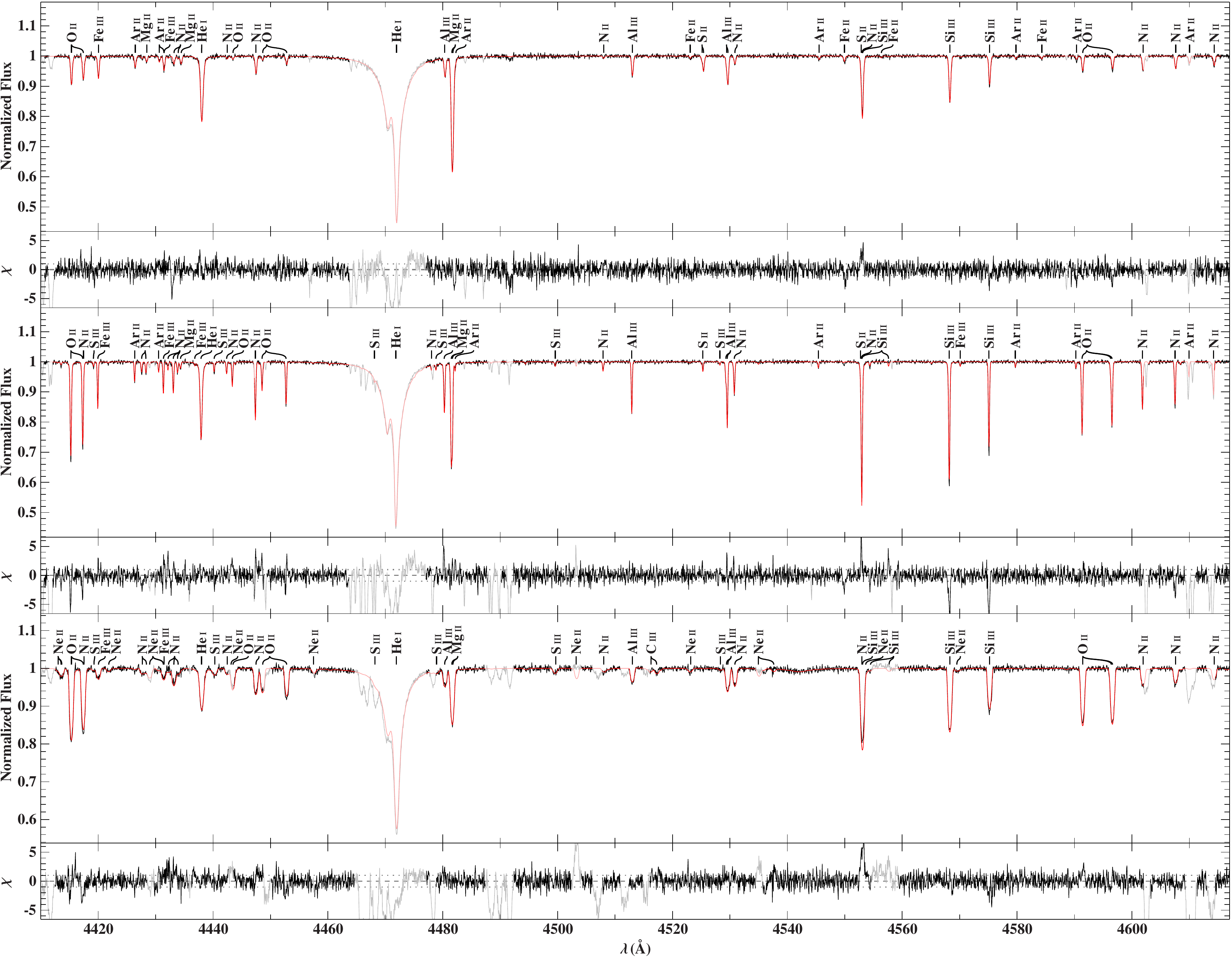}
\caption{Same as Fig.~\ref{fig:spectra_1} in the spectral range $\lambda\lambda$\,4410--4617\,\AA.}
\label{fig:spectra_4}
\addtocounter{subfig}{1}
\end{figure*}
}
\addtocounter{figure}{-1}
\onlfig{
\begin{figure*}
\centering
\includegraphics[height=1\textwidth, angle=-90]{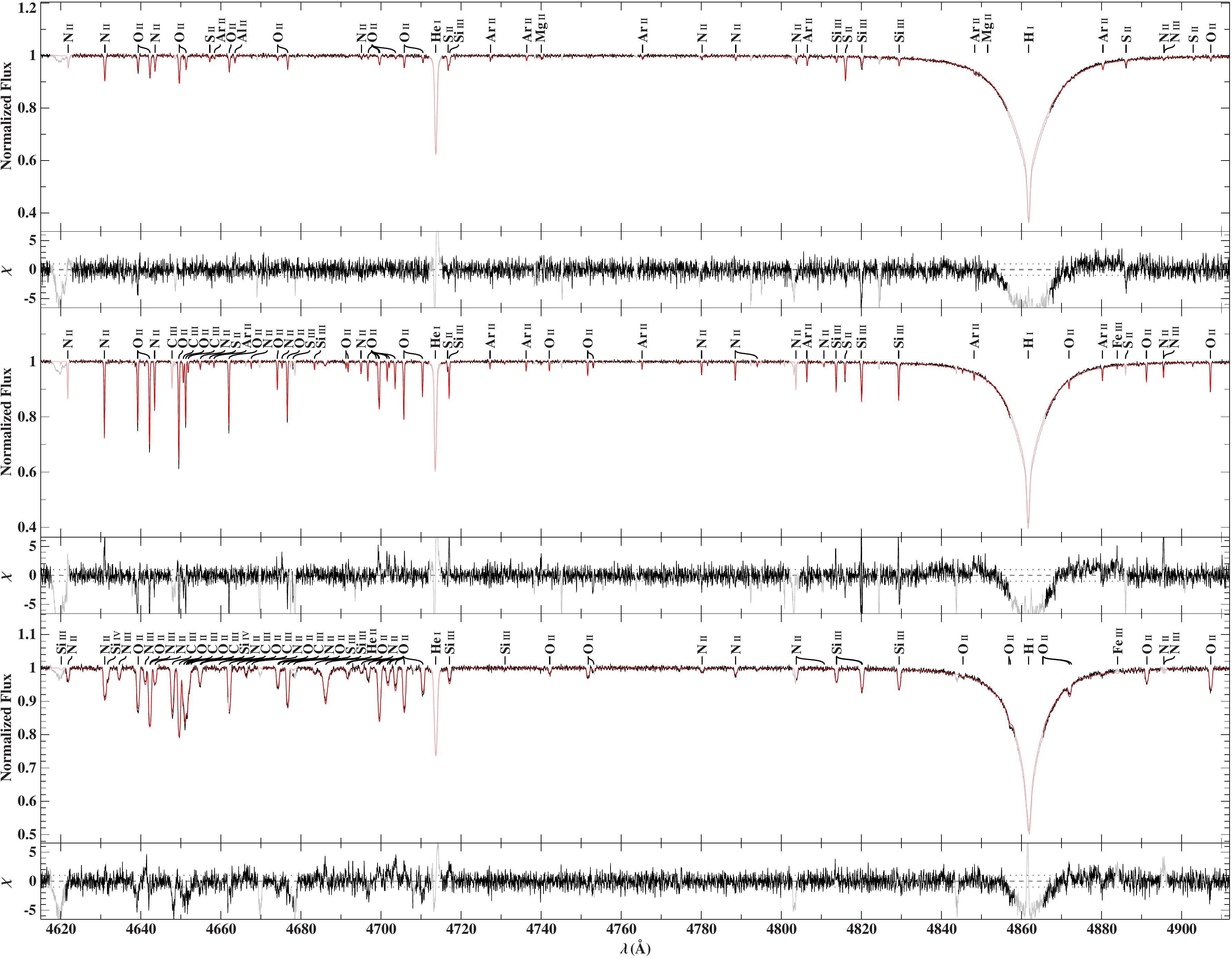}
\caption{Same as Fig.~\ref{fig:spectra_1} in the spectral range $\lambda\lambda$\,4615--4912\,\AA.}
\label{fig:spectra_5}
\addtocounter{subfig}{1}
\end{figure*}
}
\addtocounter{figure}{-1}
\onlfig{
\begin{figure*}
\centering
\includegraphics[height=1\textwidth, angle=-90]{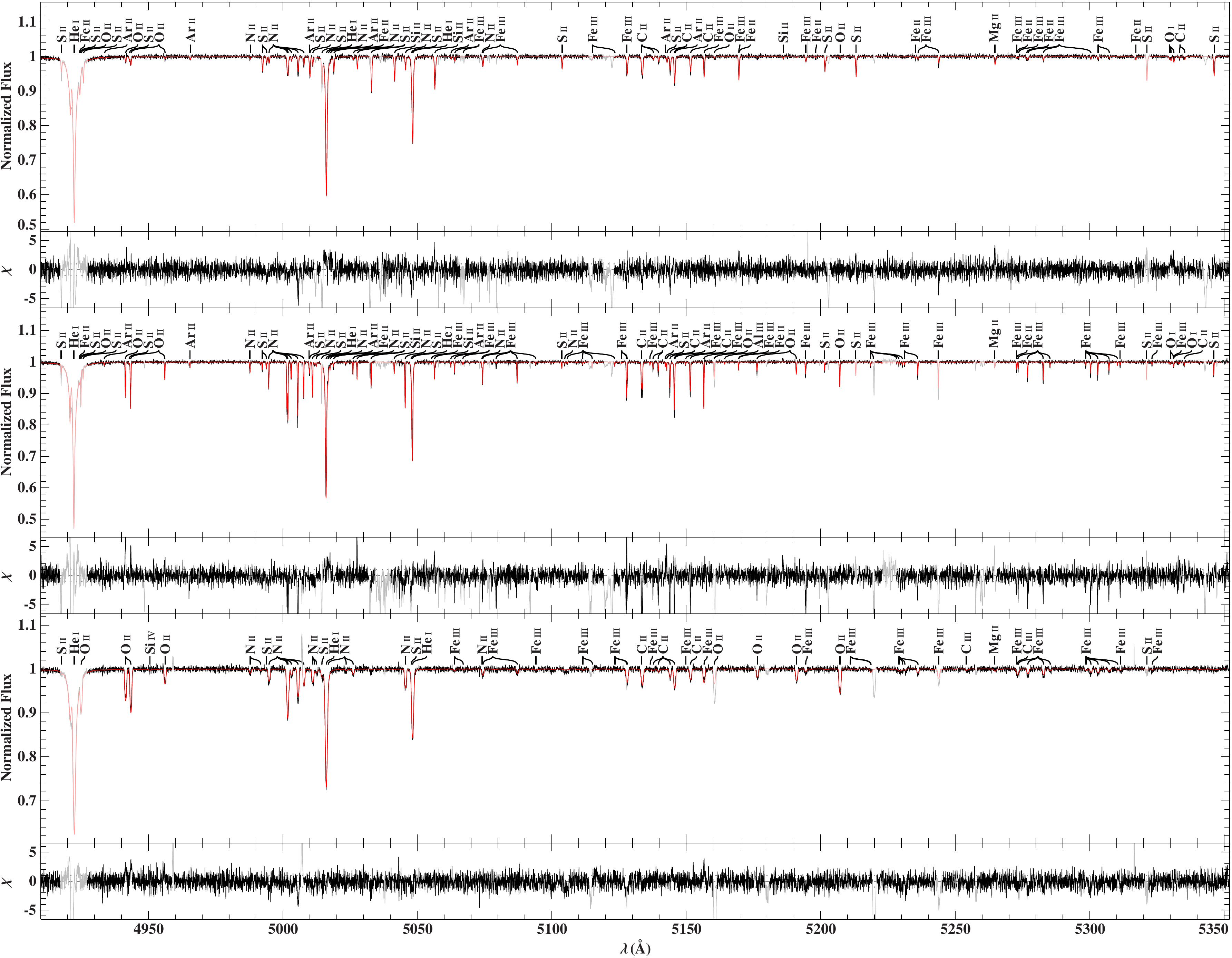}
\caption{Same as Fig.~\ref{fig:spectra_1} in the spectral range $\lambda\lambda$\,4910--5352\,\AA.}
\label{fig:spectra_6}
\addtocounter{subfig}{1}
\end{figure*}
}
\addtocounter{figure}{-1}
\onlfig{
\begin{figure*}
\centering
\includegraphics[height=1\textwidth, angle=-90]{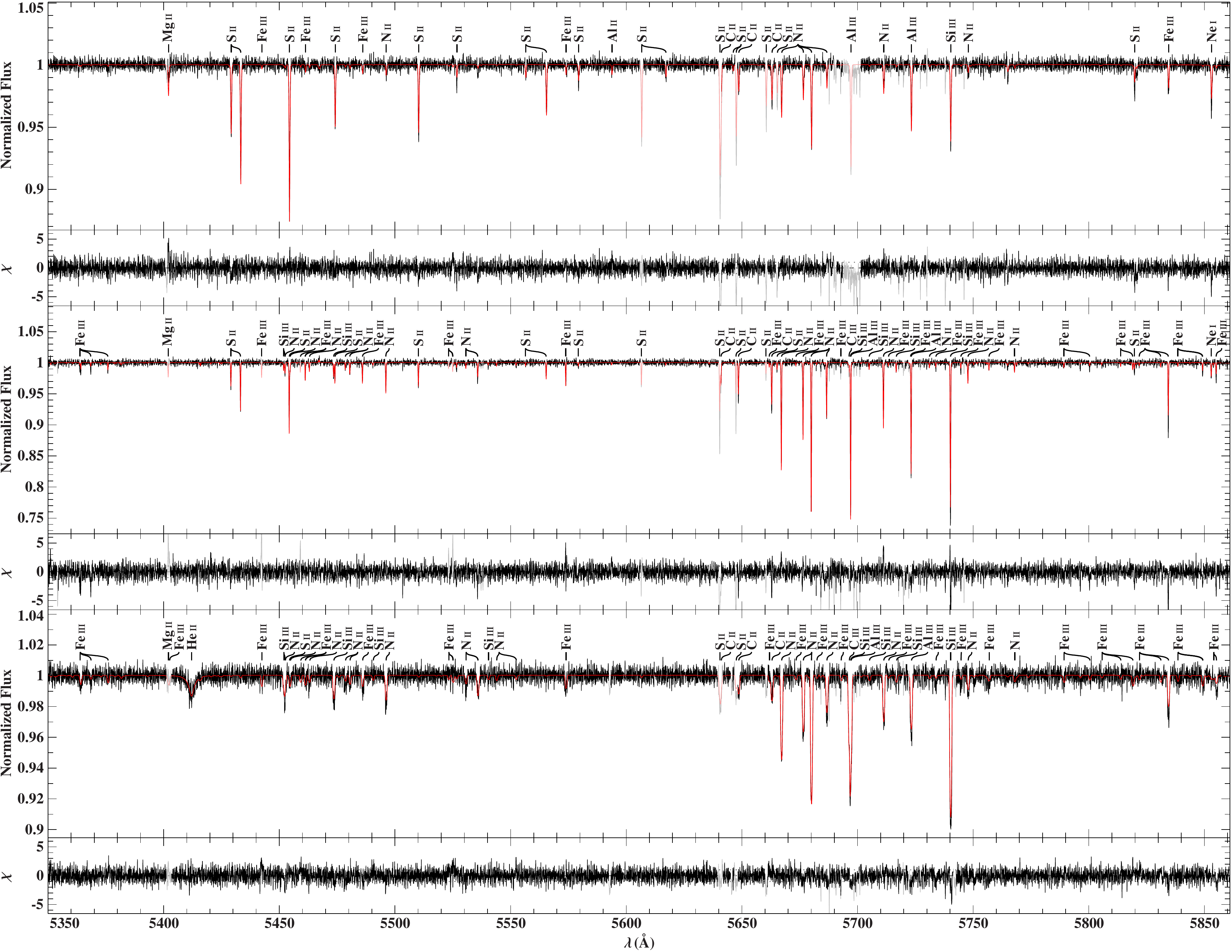}
\caption{Same as Fig.~\ref{fig:spectra_1} in the spectral range $\lambda\lambda$\,5350--5861\,\AA. Note the comparatively small ordinate scales.}
\label{fig:spectra_7}
\addtocounter{subfig}{1}
\end{figure*}
}
\addtocounter{figure}{-1}
\onlfig{
\begin{figure*}
\centering
\includegraphics[height=1\textwidth, angle=-90]{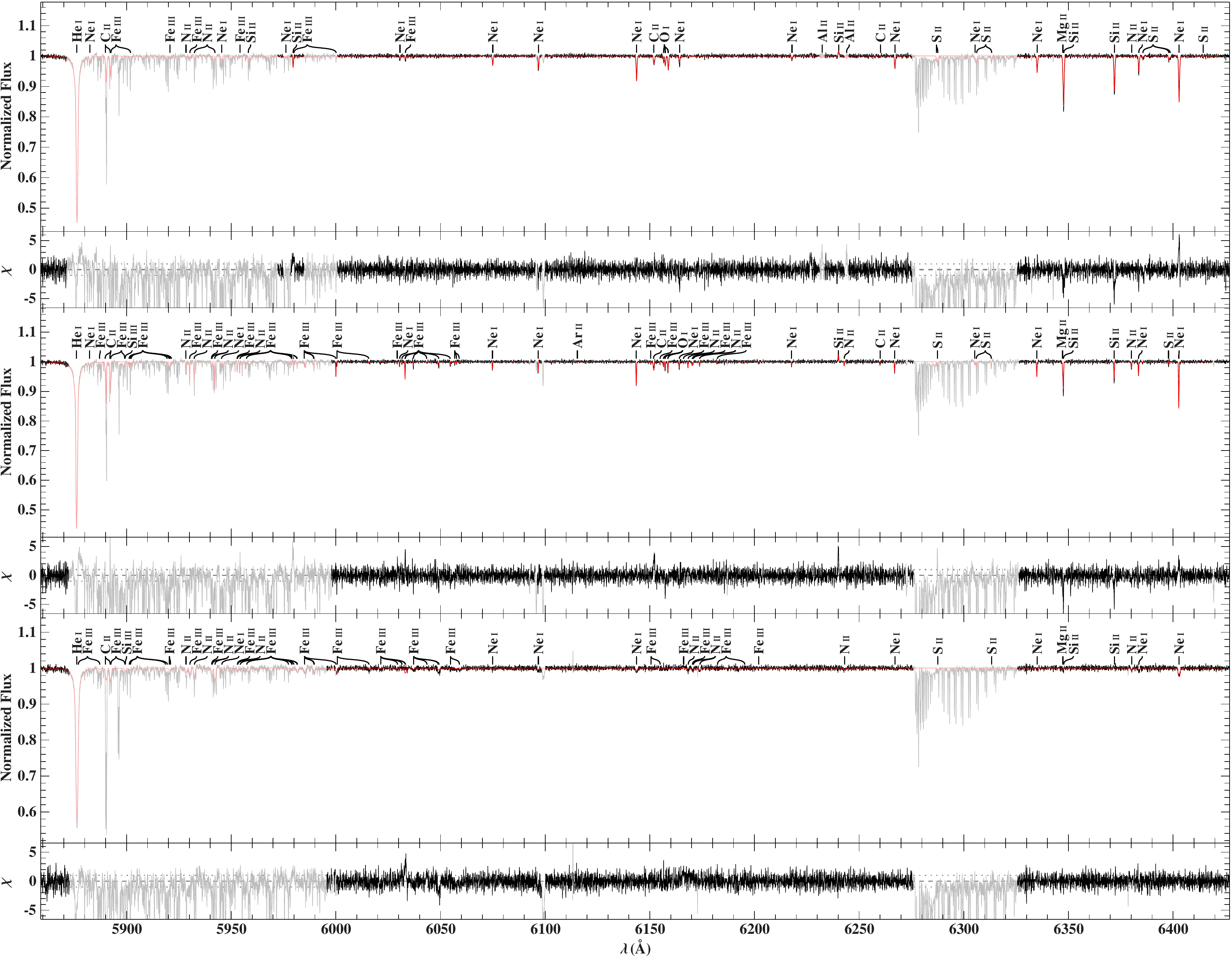}
\caption{Same as Fig.~\ref{fig:spectra_1} in the spectral range $\lambda\lambda$\,5859--6427\,\AA. Strong contamination with telluric lines.}
\label{fig:spectra_8}
\addtocounter{subfig}{1}
\end{figure*}
}
\addtocounter{figure}{-1}
\onlfig{
\begin{figure*}
\centering
\includegraphics[height=1\textwidth, angle=-90]{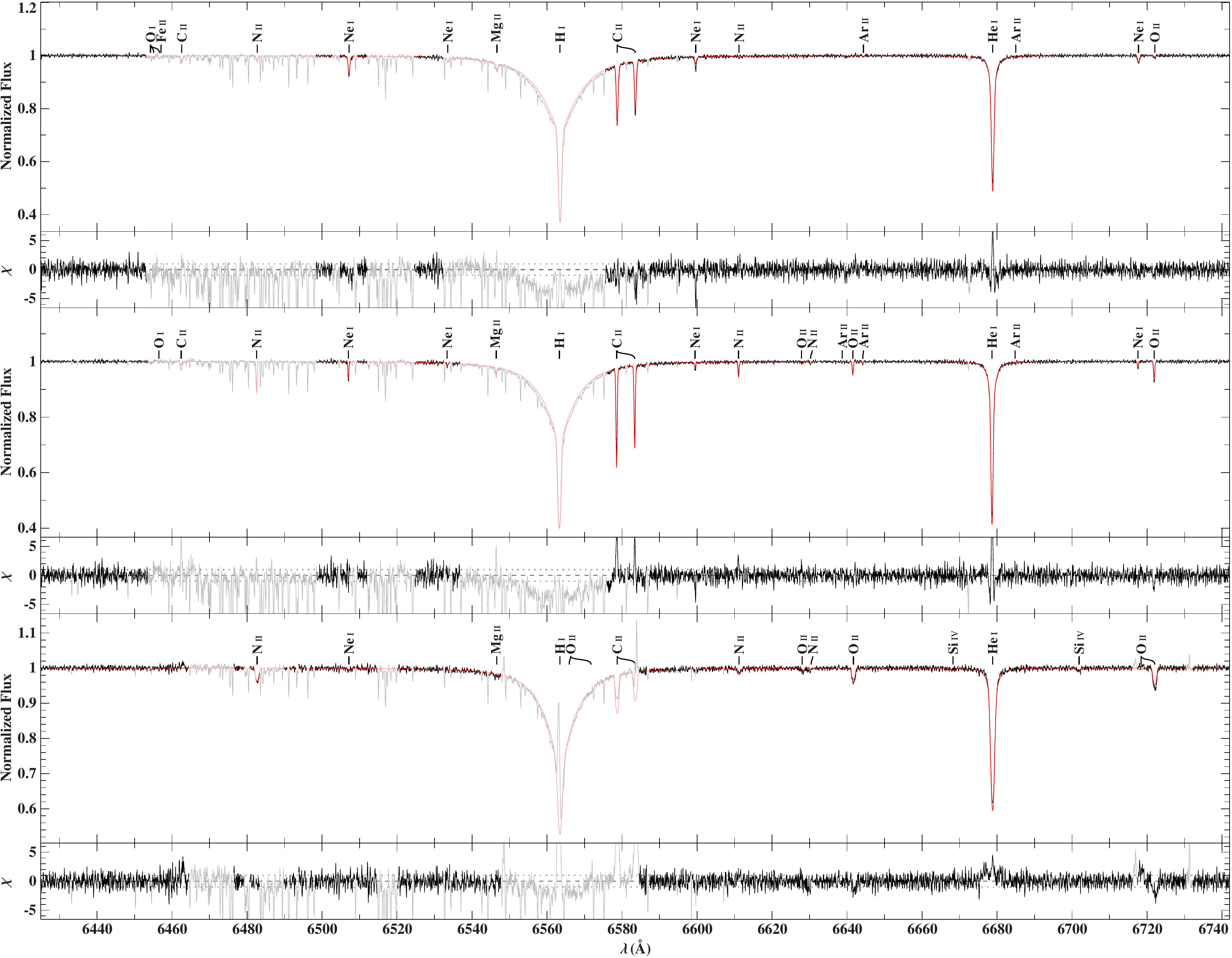}
\caption{Same as Fig.~\ref{fig:spectra_1} in the spectral range $\lambda\lambda$\,6425--6742\,\AA. Strong contamination with telluric lines. Nebula emission lines visible in HD\,37042's spectrum (\textit{left}).}
\label{fig:spectra_9}
\end{figure*}
\renewcommand{\thefigure}{\arabic{figure}}
\setcounter{subfig}{1}
}
\onlfig{
\setcounter{subfig}{1}
\renewcommand\thefigure{\arabic{figure}\alph{subfig}}
\begin{figure*}
\centering
\includegraphics[height=1\textwidth, angle=-90]{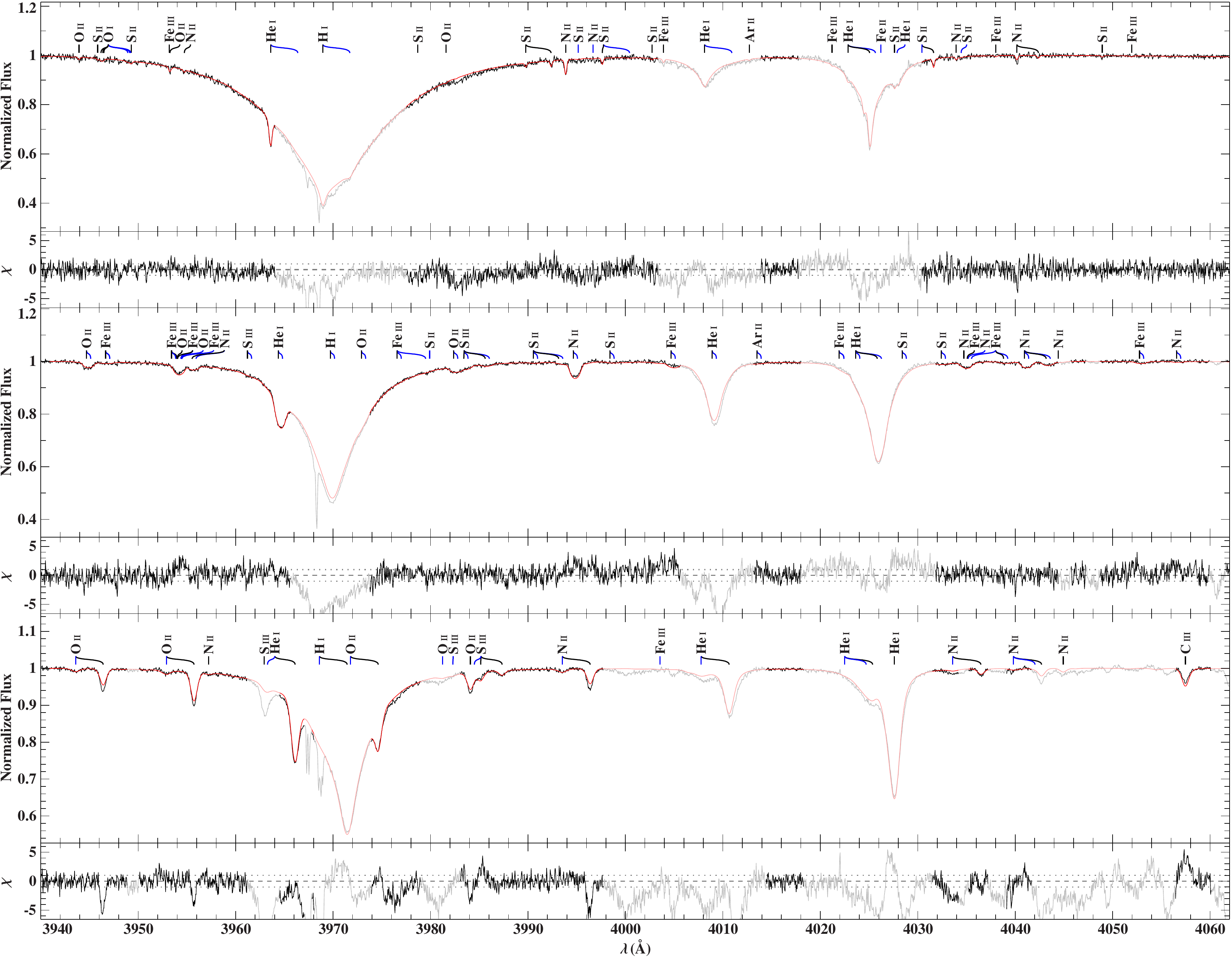}
\caption{Comparison of best-fitting model spectrum (red line) with re-normalized observation (black line) for the objects, HD\,119109 (\textit{right}), HD\,213420 (\textit{middle}), and HD\,75821 (\textit{left}), in the spectral range $\lambda\lambda$\,3940--4062\,\AA. Light colors mark regions that have been excluded from the fitting due to the presence of features that are not (properly) included in our models. For the sake of clarity, only the strongest out of all lines that have been used in the analysis are labeled. Blue connection lines mark contributions of the secondary component. The residuals $\chi$ are defined by the bracket in Eq.~(\ref{eq:chisqr}).}
\label{fig:binary_spectra_1}
\addtocounter{subfig}{1}
\end{figure*}
}
\addtocounter{figure}{-1}
\onlfig{
\begin{figure*}
\centering
\includegraphics[height=1\textwidth, angle=-90]{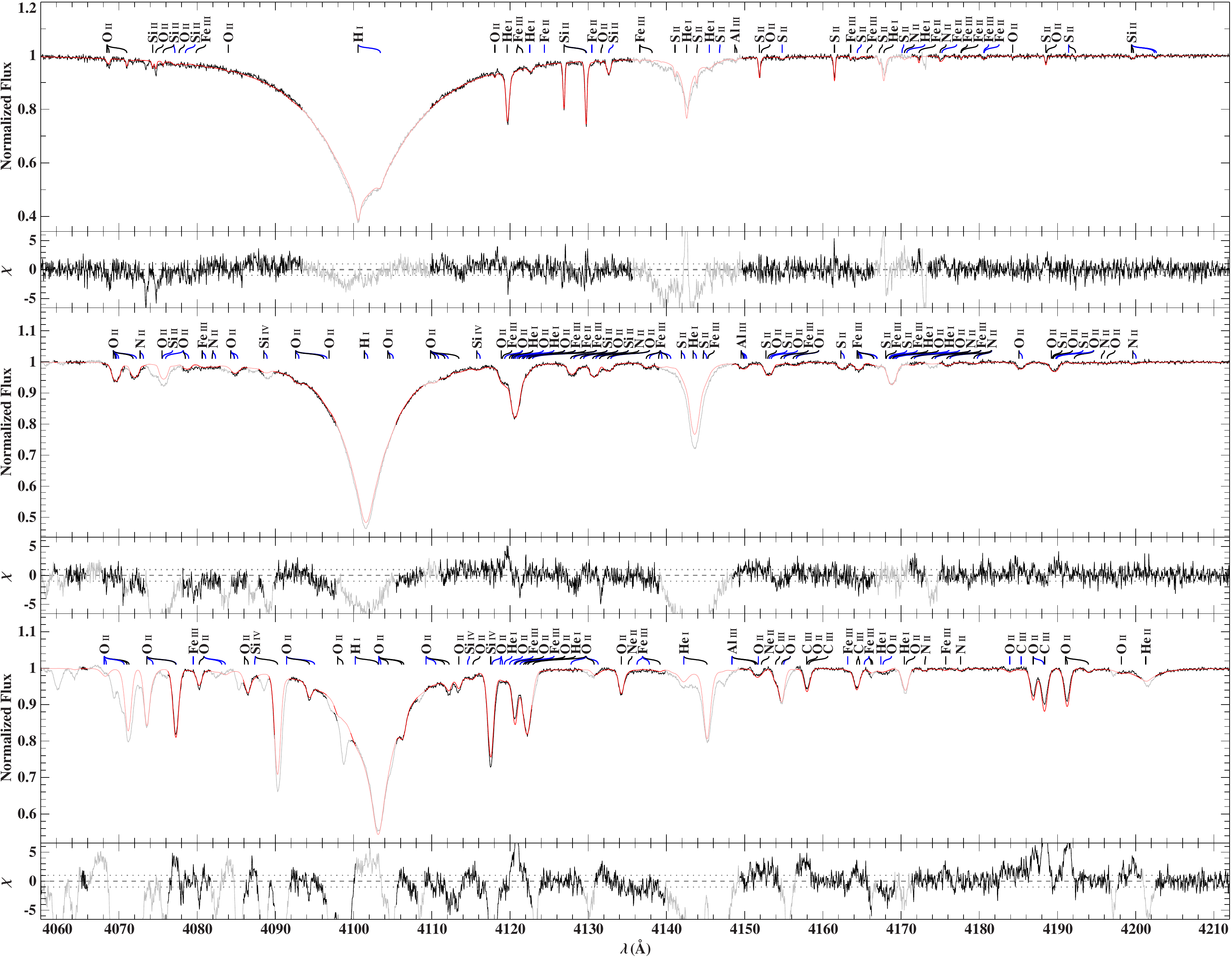}
\caption{Same as Fig.~\ref{fig:binary_spectra_1} in the spectral range $\lambda\lambda$\,4060--4212\,\AA.}
\label{fig:binary_spectra_2}
\addtocounter{subfig}{1}
\end{figure*}
}
\addtocounter{figure}{-1}
\onlfig{
\begin{figure*}
\centering
\includegraphics[height=1\textwidth, angle=-90]{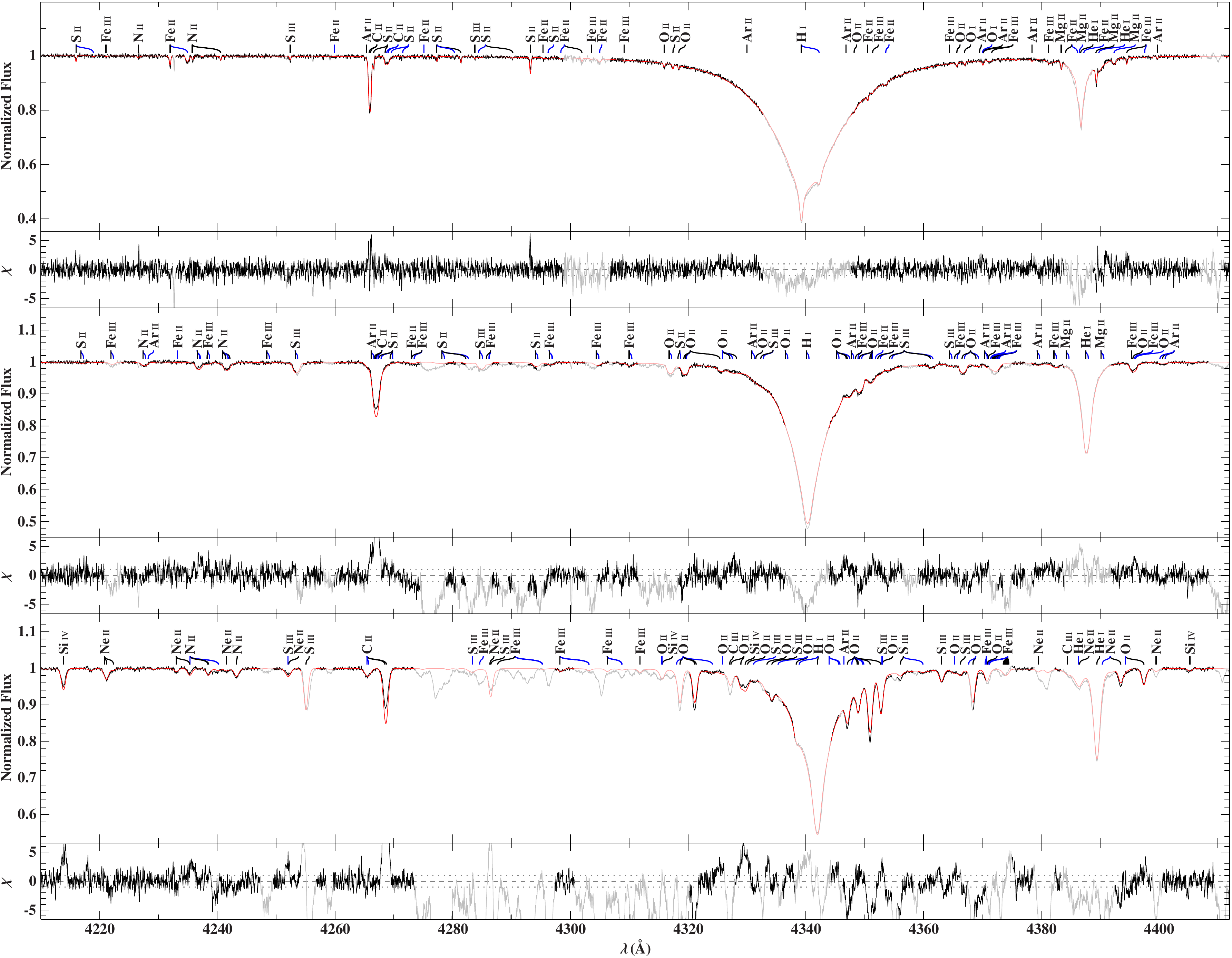}
\caption{Same as Fig.~\ref{fig:binary_spectra_1} in the spectral range $\lambda\lambda$\,4210--4412\,\AA.}
\label{fig:binary_spectra_3}
\addtocounter{subfig}{1}
\end{figure*}
}
\addtocounter{figure}{-1}
\onlfig{
\begin{figure*}
\centering
\includegraphics[height=1\textwidth, angle=-90]{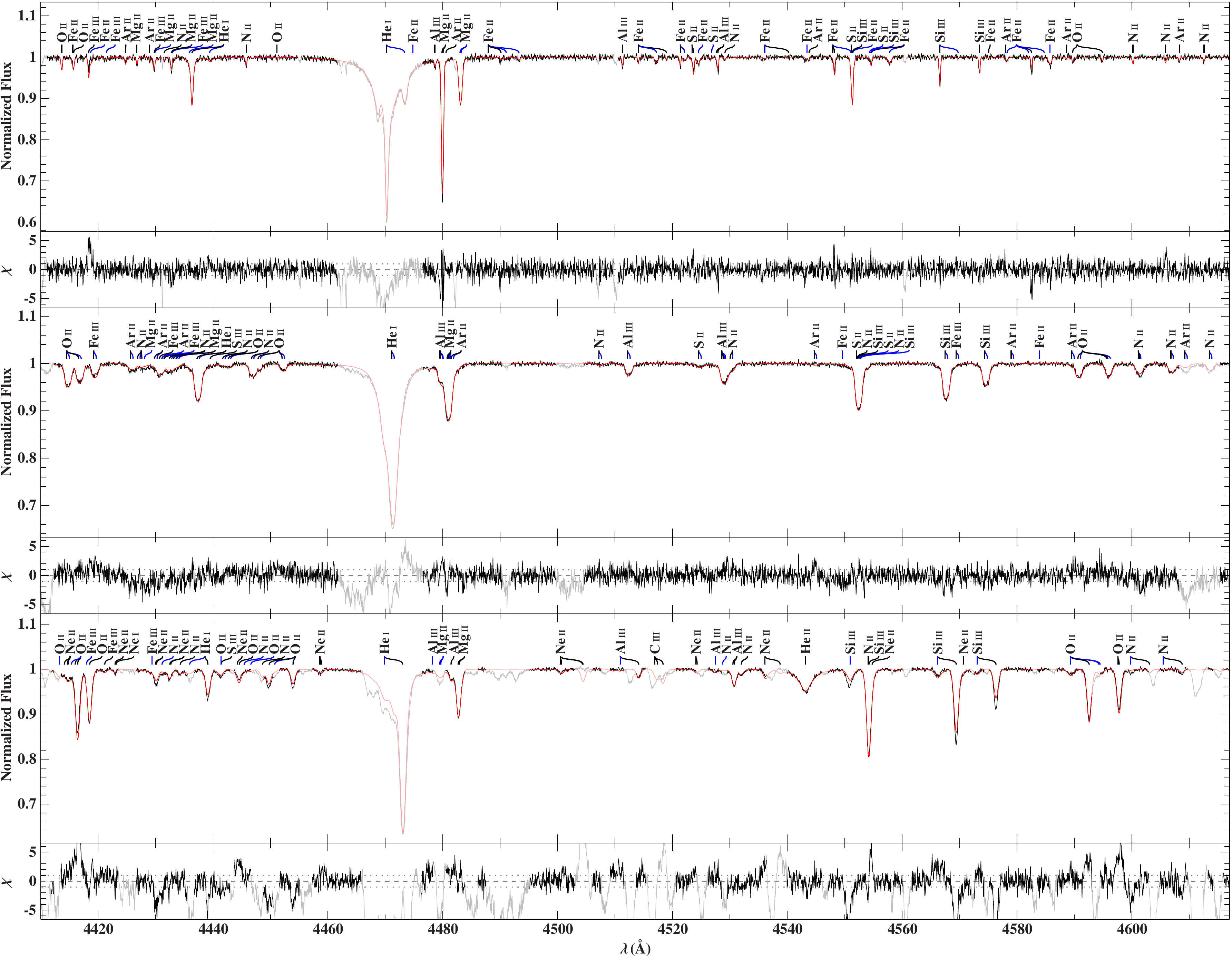}
\caption{Same as Fig.~\ref{fig:binary_spectra_1} in the spectral range $\lambda\lambda$\,4410--4617\,\AA.}
\label{fig:binary_spectra_4}
\addtocounter{subfig}{1}
\end{figure*}
}
\addtocounter{figure}{-1}
\onlfig{
\begin{figure*}
\centering
\includegraphics[height=1\textwidth, angle=-90]{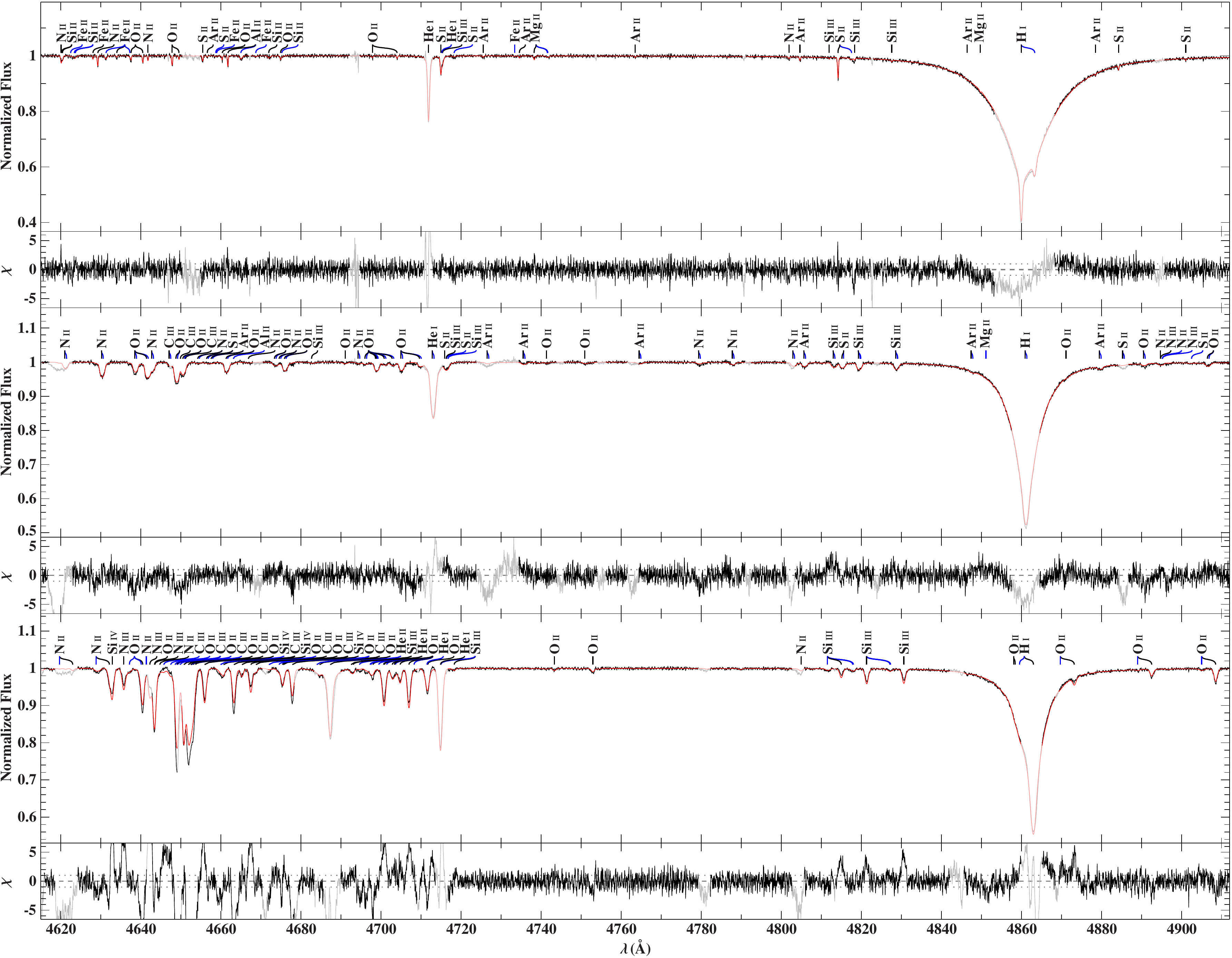}
\caption{Same as Fig.~\ref{fig:binary_spectra_1} in the spectral range $\lambda\lambda$\,4615--4912\,\AA.}
\label{fig:binary_spectra_5}
\addtocounter{subfig}{1}
\end{figure*}
}
\addtocounter{figure}{-1}
\onlfig{
\begin{figure*}
\centering
\includegraphics[height=1\textwidth, angle=-90]{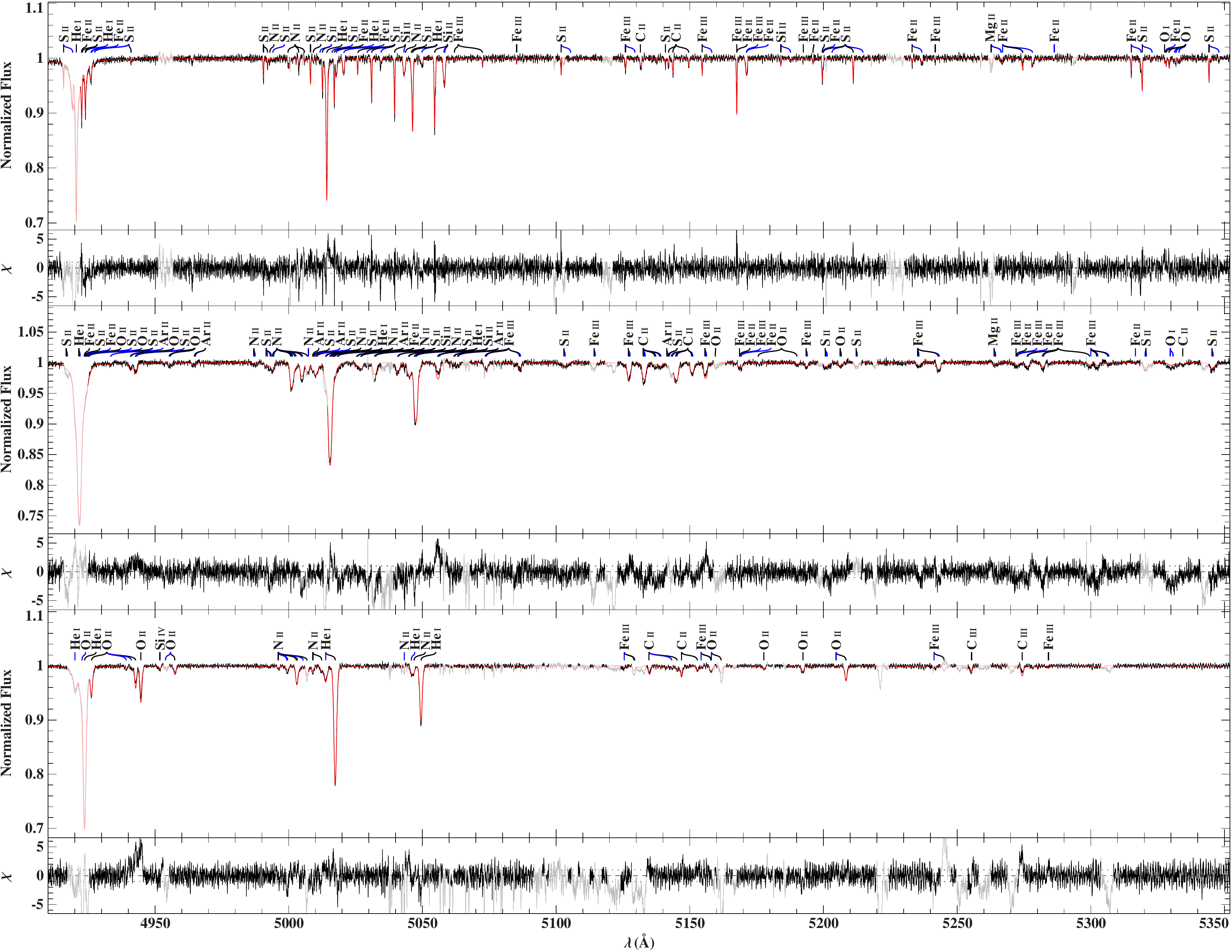}
\caption{Same as Fig.~\ref{fig:binary_spectra_1} in the spectral range $\lambda\lambda$\,4910--5352\,\AA.}
\label{fig:binary_spectra_6}
\addtocounter{subfig}{1}
\end{figure*}
}
\addtocounter{figure}{-1}
\onlfig{
\begin{figure*}
\centering
\includegraphics[height=1\textwidth, angle=-90]{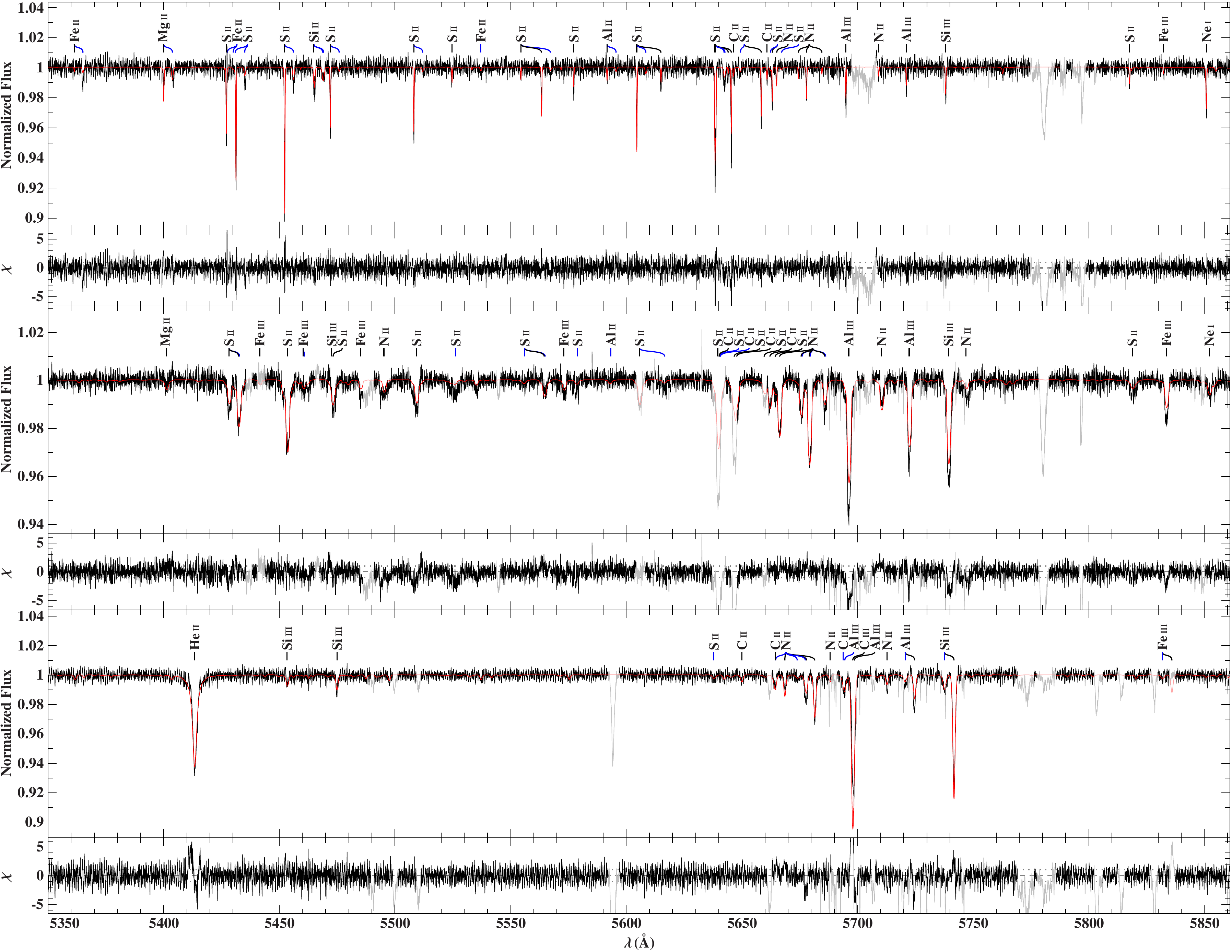}
\caption{Same as Fig.~\ref{fig:binary_spectra_1} in the spectral range $\lambda\lambda$\,5350--5861\,\AA. Note the comparatively small ordinate scales.}
\label{fig:binary_spectra_7}
\addtocounter{subfig}{1}
\end{figure*}
}
\addtocounter{figure}{-1}
\onlfig{
\begin{figure*}
\centering
\includegraphics[height=1\textwidth, angle=-90]{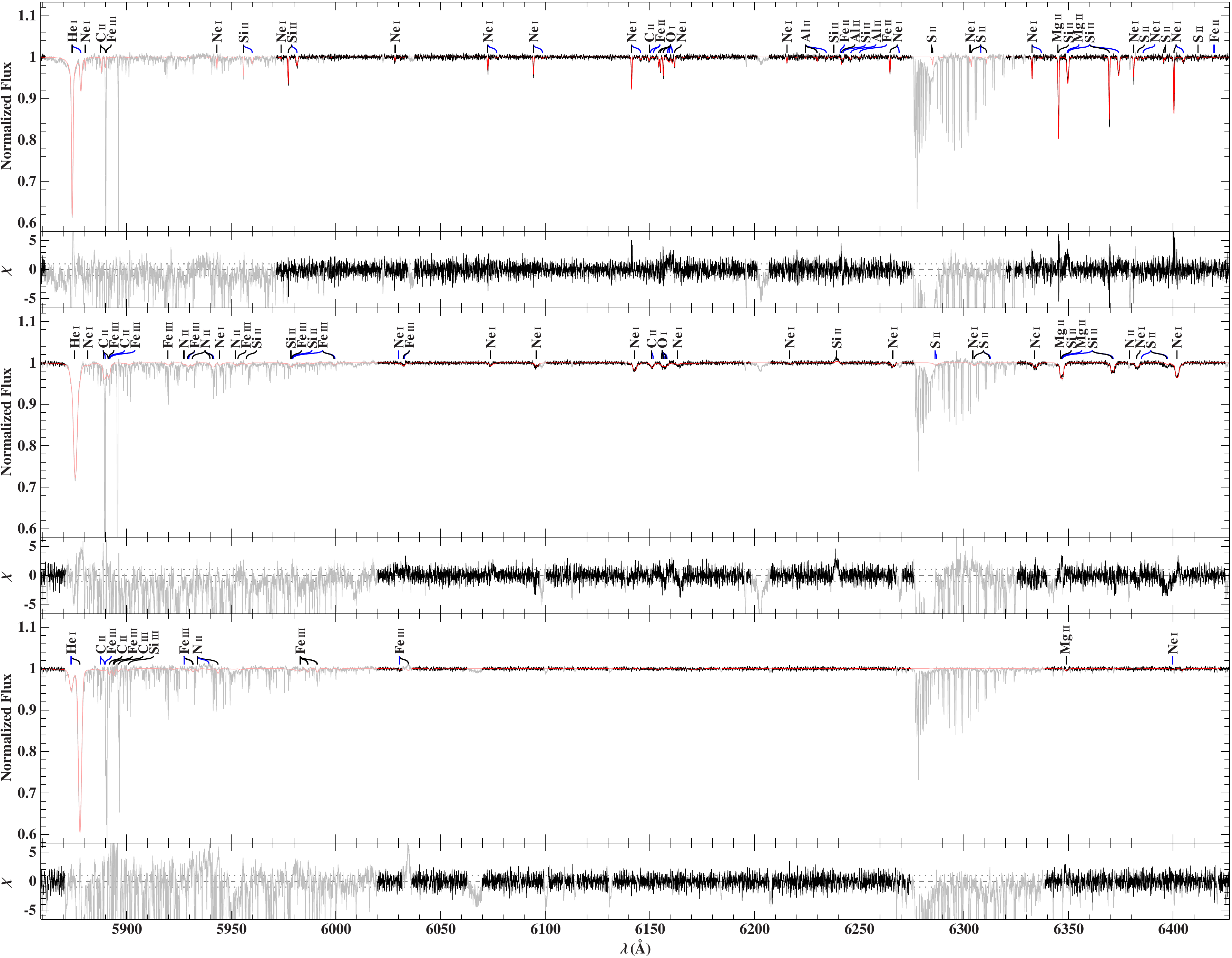}
\caption{Same as Fig.~\ref{fig:binary_spectra_1} in the spectral range $\lambda\lambda$\,5859--6427\,\AA. Strong contamination with telluric lines.}
\label{fig:binary_spectra_8}
\addtocounter{subfig}{1}
\end{figure*}
}
\addtocounter{figure}{-1}
\onlfig{
\begin{figure*}
\centering
\includegraphics[height=1\textwidth, angle=-90]{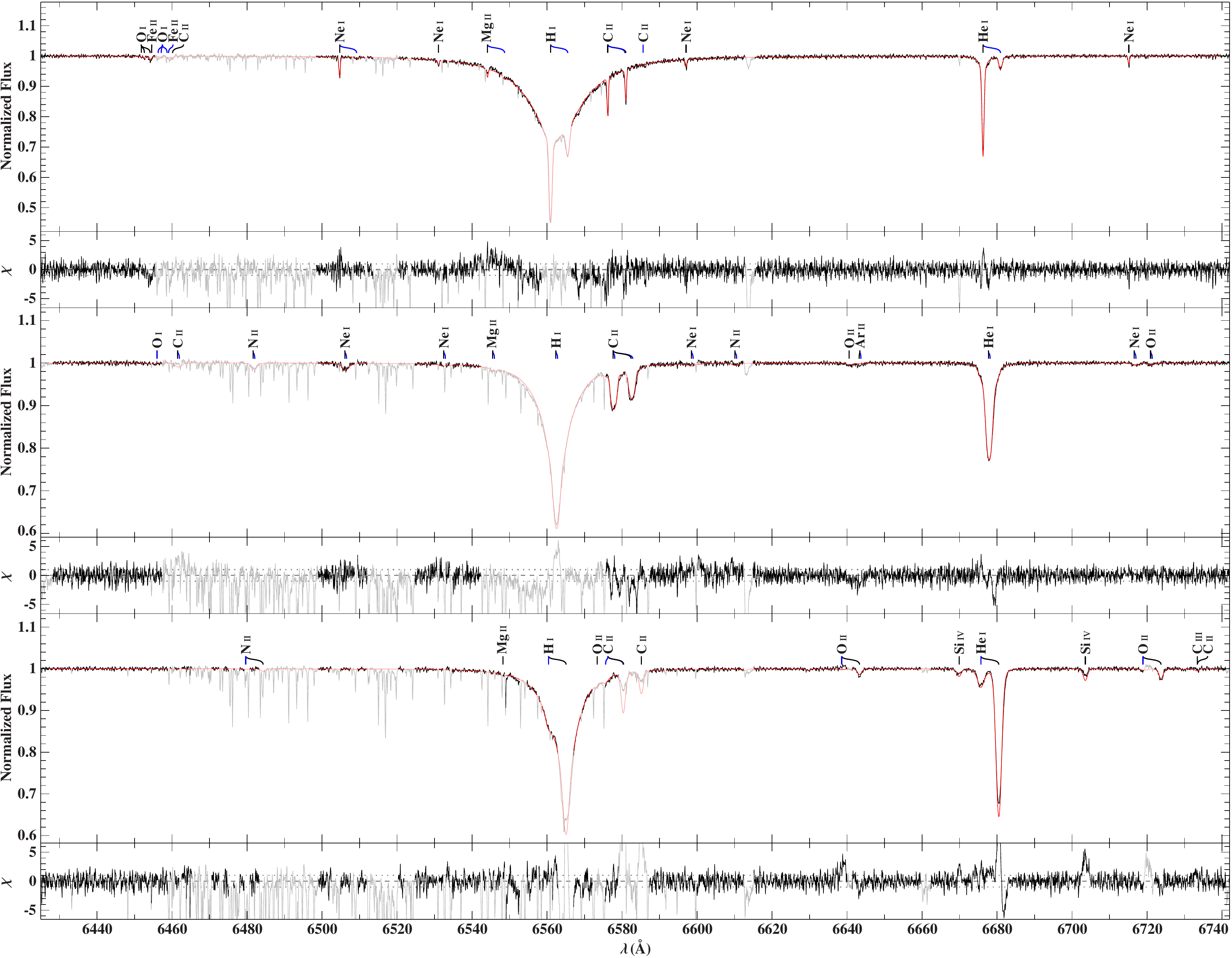}
\caption{Same as Fig.~\ref{fig:binary_spectra_1} in the spectral range $\lambda\lambda$\,6425--6742\,\AA. Strong contamination with telluric lines.}
\label{fig:binary_spectra_9}
\end{figure*}
\renewcommand{\thefigure}{\arabic{figure}}
\setcounter{subfig}{1}
}
%
\subsection{Stellar parameters and mass fractions}\label{subsection:stellar_params}
\begin{table*}
\begin{center}
\tiny
\setlength{\tabcolsep}{0.145cm}
\renewcommand{\arraystretch}{1.3}
\caption{\label{table:stellar_parameters} Stellar parameters and mass fractions of the program stars.}
\begin{tabular}{lcrrcrrcrrcrrcrrcrrcrrcrrcrr}
\hline\hline
\# & & \multicolumn{2}{c}{$M$} & & \multicolumn{2}{c}{$\tau$} & & \multicolumn{2}{c}{$\log(L/L_{\sun})$} & & \multicolumn{2}{c}{$R_{\star}$} & & \multicolumn{2}{c}{$d$} & & \multicolumn{2}{c}{$\Pi^{-1}$} & & \multicolumn{2}{c}{$X$} & & \multicolumn{2}{c}{$Y$} & & \multicolumn{2}{c}{$Z$}\\ 
\cline{3-4} \cline{6-7} \cline{9-10} \cline{12-13} \cline{15-19} \cline{21-22} \cline{24-25} \cline{27-28}
& & \multicolumn{2}{c}{$(M_{\sun})$} & & \multicolumn{2}{c}{(Myr)} & & & & & \multicolumn{2}{c}{$(R_{\sun})$} & & \multicolumn{5}{c}{(pc)} & & & & & & & & &\\
\hline
1    & & $9.3$ & $^{+0.5}_{-0.4}$ & & $12$ & $^{+4}_{-7}$ &  & $3.75$ & $^{+0.08}_{-0.08}$ & & $4.4$ & $^{+0.6}_{-0.6}$ & & $380$ & $^{+60}_{-60}$ & & $270$ & $^{+80}_{-60}$ & & $0.677$ & $^{+0.024}_{-0.022}$ & & $0.311$ & $^{+0.022}_{-0.025}$ & & $0.012$ & $^{+0.002}_{-0.001}$ \\
2    & & $6.7$ & $^{+0.3}_{-0.3}$ & & $29$ & $^{+6}_{-8}$ &  & $3.30$ & $^{+0.08}_{-0.08}$ & & $4.0$ & $^{+0.6}_{-0.5}$ & & $400$ & $^{+70}_{-70}$ & & $400$ & $^{+490}_{-150}$ & & $0.685$ & $^{+0.042}_{-0.045}$ & & $0.301$ & $^{+0.043}_{-0.041}$ & & $0.014$ & $^{+0.002}_{-0.002}$ \\
3    & & $12.9$ & $^{+2.0}_{-0.7}$ & & $1$ & $^{+2}_{-1}$ &  & $4.08$ & $^{+0.14}_{-0.08}$ & & $4.3$ & $^{+0.9}_{-0.3}$ & & $450$ & $^{+90}_{-60}$ & & \ldots & \ldots & & $0.709$ & $^{+0.016}_{-0.019}$ & & $0.279$ & $^{+0.020}_{-0.016}$ & & $0.012$ & $^{+0.001}_{-0.001}$ \\
4p a & & $20.7$ & $^{+2.9}_{-1.8}$ & & $7$ & $^{+2}_{-1}$ &  & $4.96$ & $^{+0.17}_{-0.13}$ & & $11.7$ & $^{+2.3}_{-1.8}$ & & $1040$ & $^{+470}_{-260}$ & & $1000$ & $^{+2300}_{-{\color{white}0}500}$ & & $0.739$ & $^{+0.005}_{-0.003}$ & & $0.252$ & $^{+0.003}_{-0.005}$ & & $0.009$ & $^{+0.001}_{-0.001}$ \\
4s a & & $15.2$ & $^{+1.5}_{-1.2}$ & & $0$ & $^{+1}_{-0}$ & & $4.29$ & $^{+0.11}_{-0.11}$ & & $4.6$ & $^{+0.2}_{-0.3}$ & & \ldots & \ldots & & \ldots & \ldots & & \ldots & \ldots & & \ldots & \ldots & & \ldots & \ldots \\
4p b & & $20.2$ & $^{+1.8}_{-1.5}$ & & $7$ & $^{+1}_{-1}$ &  & $4.92$ & $^{+0.12}_{-0.12}$ & & $10.9$ & $^{+1.7}_{-1.6}$ & & $960$ & $^{+240}_{-200}$ & & $1000$ & $^{+2300}_{-{\color{white}0}500}$ & & $0.770$ & $^{+0.010}_{-0.011}$ & & $0.220$ & $^{+0.011}_{-0.010}$ & & $0.010$ & $^{+0.001}_{-0.002}$ \\
4s b & & $11.5$ & $^{+2.0}_{-0.5}$ & & $0$ & $^{+10}_{-{\color{white}0}0}$ & & $3.93$ & $^{+0.42}_{-0.07}$ & & $3.9$ & $^{+2.9}_{-0.1}$ & & \ldots & \ldots & & \ldots & \ldots & & \ldots & \ldots & & \ldots & \ldots & & \ldots & \ldots \\
4p c & & $21.1$ & $^{+3.1}_{-1.9}$ & & $7$ & $^{+2}_{-2}$ &  & $4.98$ & $^{+0.17}_{-0.13}$ & & $11.8$ & $^{+2.4}_{-1.8}$ & & $1040$ & $^{+420}_{-230}$ & & $1000$ & $^{+2300}_{-{\color{white}0}500}$ & & $0.769$ & $^{+0.009}_{-0.011}$ & & $0.222$ & $^{+0.010}_{-0.009}$ & & $0.010$ & $^{+0.001}_{-0.001}$ \\
4s c & & $13.0$ & $^{+0.6}_{-0.7}$ & & $0$ & $^{+1}_{-0}$ & & $4.08$ & $^{+0.07}_{-0.07}$ & & $4.2$ & $^{+0.1}_{-0.2}$  & & \ldots & \ldots & & \ldots & \ldots & & \ldots & \ldots & & \ldots & \ldots & & \ldots & \ldots \\
5p   & & $5.2$ & $^{+0.2}_{-0.2}$ & & $48$ & $^{+10}_{-16}$ &  & $2.90$ & $^{+0.09}_{-0.09}$ & & $3.4$  & $^{+0.5}_{-0.5}$ & & $470$ & $^{+70}_{-60}$ & & $550$ & $^{+1700}_{-{\color{white}0}240}$ & & $0.660$ & $^{+0.031}_{-0.028}$ & & $0.327$ & $^{+0.028}_{-0.032}$ & & $0.014$ & $^{+0.001}_{-0.001}$ \\
5s   & & $3.5$ & $^{+0.4}_{-0.2}$ & & $49$ & $^{+47}_{-48}$ & & $2.18$ & $^{+0.11}_{-0.13}$ & & $2.3$ & $^{+0.3}_{-0.4}$ & & \ldots & \ldots & & \ldots & \ldots & & \ldots & \ldots & & \ldots & \ldots & & \ldots & \ldots \\
6p   & & $10.4$ & $^{+1.3}_{-1.2}$ & & $20$ & $^{+10}_{-{\color{white}0}4}$ &  & $4.18$ & $^{+0.15}_{-0.15}$ & & $9.7$  & $^{+1.6}_{-1.6}$ & & $510$ & $^{+110}_{-100}$ & & $530$ & $^{+210}_{-120}$ & & $0.700$ & $^{+0.004}_{-0.005}$ & & $0.290$ & $^{+0.004}_{-0.006}$ & & $0.011$ & $^{+0.001}_{-0.001}$ \\
6s   & & $11.1$ & $^{+0.5}_{-2.1}$ & & $18$ & $^{+14}_{-{\color{white}0}2}$ & & $4.29$ & $^{+0.12}_{-0.22}$ & & $13.4$ & $^{+2.4}_{-3.2}$ & & \ldots & \ldots & & \ldots & \ldots & & \ldots & \ldots & & \ldots & \ldots & & \ldots & \ldots \\
\hline
\sun\tablefootmark{a} & & & & & & & & & & & & & & & & & & & & $0.716$ & & & $0.270$ & & & $0.014$ &\\
\hline
\end{tabular}
\tablefoot{Uncertainties cover only the effects induced by variations of $2\%$ in $T_{\mathrm{eff}}$ and $0.1\,\mathrm{dex}$ in $\log(g)$ (see Sect.~\ref{subsection:stellar_params} for details) and are formally taken to be 99\%-confidence limits. Numbering according to Table~\ref{table:program_stars}. Owing to the assumption of a homogeneous chemical composition, abundances of the secondary components ``s'' are tied to the ones of the primaries ``p'' during the analysis. Parallaxes $\Pi$ are from {\sc Hipparcos} \citep{vanleeuwen}, while their original formal errors, which are assumed to be $1\sigma$, are converted to 99\%-confidence intervals here. \tablefoottext{a}{Protosolar nebula values from \citet{sun}.}} 
\end{center}
\end{table*}
The stellar parameters mass $M$, age $\tau$, and luminosity $L$ are derived from $T_{\mathrm{eff}}$, $\log(g)$, and $\varv\,\sin(i)$ by fitting single-star evolutionary tracks that account for stellar rotation ($M \leq 15\,M_{\sun}$: \citealt{georgy_et_al}; otherwise: \citealt{ekstroem_et_al}). The unknown inclination term $\sin(i)$ is replaced by its spherically averaged value $\pi/4$ when matching $\varv\,\sin(i)$ to the equatorial velocity $\varv$ predicted by the evolution tracks. Using the gravitational constant $G$, the stellar radius $R_{\star}$ follows then from the definition of the surface gravity $g = G M R_{\star}^{-2}$. This information combined with the object's photometry can be used to derive the spectroscopic distance $d$ to the star \citep{balmer}. Mass fractions of hydrogen ($X$), helium ($Y$), and metals ($Z$) are directly computed from the deduced atmospheric abundances.

Error propagation for stellar parameters and mass fractions is analogous to the estimation of systematic uncertainties in Sect.~\ref{subsection:uncertainties}. That is, for each pixel surrounded by the solid black line in Fig.~\ref{fig:teff_logg_confmap_chisqr}, they are derived as outlined in this subsection. Minimum and maximum values of the resulting distributions are again interpreted as to define 99\%-confidence intervals.

\begin{figure}
\centering
\includegraphics[width=0.48\textwidth]{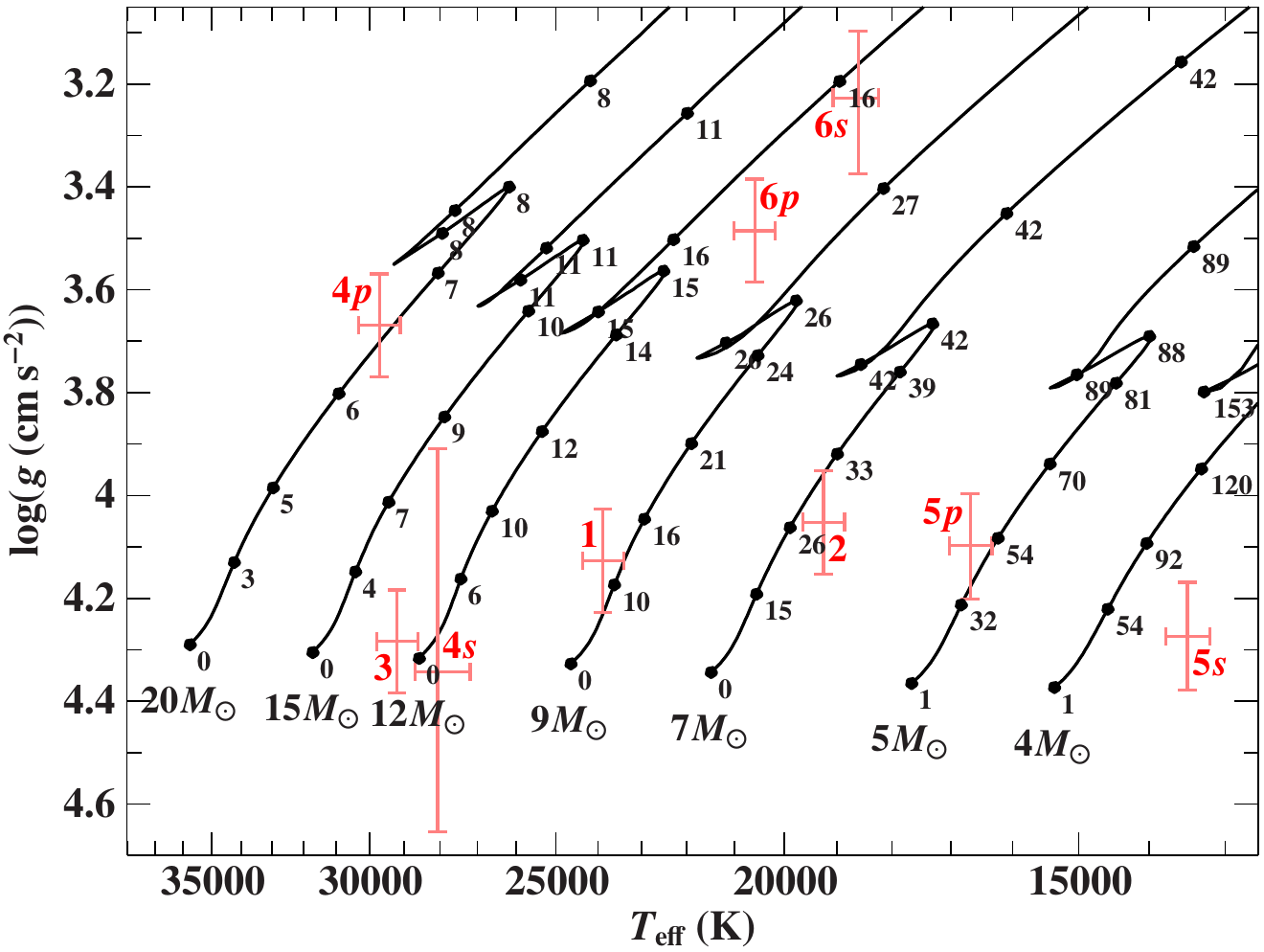}
\caption{Position of the program stars in a $(T_{\mathrm{eff}}, \log(g))$ diagram. Overlaid are evolution tracks for non-rotating stars ($\Omega / \Omega_{\mathrm{crit}} = 0$) of metallicity $Z=0.014$ and different initial masses ($M \leq 15\,M_{\sun}$: \citealt{georgy_et_al}; otherwise: \citealt{ekstroem_et_al}). Black filled circles and numbers mark the age in Myr. Red numbers correspond to those of Table~\ref{table:program_stars}. Error bars indicate 99\%-confidence limits.}
\label{fig:evolution_tracks}
\end{figure}

The program stars' positions in the $(T_{\mathrm{eff}}, \log(g))$ diagram are compared to evolutionary tracks in Fig.~\ref{fig:evolution_tracks} and the resulting stellar parameters and mass fractions are given in Table~\ref{table:stellar_parameters}. Note that the usage of single-star tracks to deduce stellar parameters of binary stars is, of course, justified only if the two binary components have not yet interacted and is otherwise an approximation.
%
\section{Discussion}\label{sec:discussion}
\subsection{Single B- and late O-type stars in Orion}\label{subsection:singlestars}
Focusing on a wide range of effective temperatures, we have selected three slow rotators (HD\,35299, HD\,35912, and HD\,37042) from the sample of \citet{orion_composition} to check our method against previous studies.

As shown in Table~\ref{table:atmospheric_parameters}, our atmospheric parameters have excellent agreement with those derived by \citeauthor{orion_composition}. Similarly, the results for the abundances of helium, carbon, and nitrogen are perfectly consistent with each other within the error bars, even though helium was kept fixed at the solar value in the study of \citeauthor{orion_composition}. The same applies to oxygen and silicon abundances by \citet{orion_fies}. On the other hand, there are systematic discrepancies apparent for neon, magnesium, and iron that can be attributed either to differences in the synthetic models or in the analysis strategy. For instance, several \ion{Mg}{ii} lines, such as $\lambda 4481$\,\AA\ have shown to be very sensitive to the replacement of pre-calculated opacity distribution functions, as used by \citeauthor{orion_composition}, with the more flexible concept of opacity sampling that is coherently used here throughout all computational steps, which explains the deviations in magnesium. The disagreements in neon and iron presumably arise from the underlying analysis techniques and in particular from how the microturbulence parameter is constrained. Nevertheless, it is extremely satisfying to see that the results of the two approaches match so well despite being based on contrary conceptional designs.
\subsection{Spectroscopic binaries}
As a first application to SB2 systems, we have analyzed the composite spectra of three binary systems. While the lines of the two components are sharp and very well separated in our spectrum of HD\,119109, the opposite is true for HD\,213420 (see  Figs.~\ref{fig:binary_spectra_1}--\ref{fig:binary_spectra_9}). In the case of HD\,75821, we have further derived parameters from spectra taken at three distinct orbital phases to investigate its influence on the results. 
\paragraph*{HD\,119109 (\#5):} to our knowledge, there is no hint for binarity in the literature for this system so far. Nevertheless, our spectrum shows that this is doubtlessly a SB2 system owing to the many lines that appear twice in the spectrum (see Figs.~\ref{fig:binary_spectra_1}--\ref{fig:binary_spectra_9}). Given the high quality of our observation and the opportune orbital phase, parameters of both components can be reliably deduced. The system turns out to be composed of two relatively unevolved (see Fig.~\ref{fig:evolution_tracks}), coeval ($\tau_{\mathrm{p}}=48^{+10}_{-16}\,\mathrm{Myr}$, $\tau_{\mathrm{s}}=49^{+47}_{-48}\,\mathrm{Myr}$) late-type B-stars of masses $M_{\mathrm{p}} = 5.2 \pm 0.2\,M_{\sun}$ and $M_{\mathrm{s}} = 3.5^{+0.4}_{-0.2} \,M_{\sun}$ when using single-star evolutionary tracks. The corresponding squared ratio of radii, $(R_{\star,\mathrm{s}} / R_{\star,\mathrm{p}})^2 = 0.45^{+0.29}_{-0.18}$, is consistent with the surface ratio $A_{\mathrm{eff,s}}/A_{\mathrm{eff,p}} = 0.642^{+0.015\mathrm{(stat.)}+0.027\mathrm{(sys.)}}_{-0.013\mathrm{(stat.)}-0.028\mathrm{(sys.)}}$, as is the spectroscopic distance, $d = 470^{+70}_{-60}\,\mathrm{pc}$, with the parallax, $\Pi^{-1} = 550^{+1700}_{-{\color{white}0}240}\,\mathrm{pc}$. The chemical composition resembles that of the single stars studied in Sect.~\ref{subsection:singlestars}. 

Using published radial velocity measurements, \citet{runaway_catalog} have proposed that this object is a runaway star with high probability based on its peculiar space motion. This conclusion should be considered as uncertain as long as the actual system velocity of this binary is unknown.
\paragraph*{HD\,213420 (\#6):} this well-known binary system with a period of about $880$\,days and a radial velocity semi-amplitude of $9\,\mathrm{km\,s^{-1}}$ \citep{SB9} is a clear SB2 system, given the broad absorption features superimposed to He\,{\sc i} $\lambda 4438$\,\AA, $\lambda 6678$\,\AA, C\,{\sc ii} $\lambda 4267$\,\AA, Mg\,{\sc ii} $\lambda 4481$\,\AA, Si\,{\sc iii} $\lambda 4553$\,\AA, $\lambda 4568$\,\AA, and S\,{\sc ii} $\lambda 5454$\,\AA\ (see Figs.~\ref{fig:binary_spectra_1}--\ref{fig:binary_spectra_9}). Although the signatures of the secondary component are weak and thus only detectable in the case of a high S/N, they are apparently sufficient to determine reasonable atmospheric parameters for the companion because the resulting stellar parameters paint a consistent physical picture: In addition to the finding that the ages of both components (with masses $M_{\mathrm{p}} = 10.4^{+1.3}_{-1.2}\,M_{\sun}$, $M_{\mathrm{s}} = 11.1^{+0.5}_{-2.1} \,M_{\sun}$) are in perfect agreement ($\tau_{\mathrm{s}}=20^{+10}_{-{\color{white}0}4}\,\mathrm{Myr}$, $\tau_{\mathrm{p}}=18^{+14}_{-{\color{white}0}2}\,\mathrm{Myr}$), the spectroscopic parameter $A_{\mathrm{eff,s}}/A_{\mathrm{eff,p}} = 0.936^{+0.014+0.069}_{-0.015-0.060}$ lies within the uncertainty interval of the squared ratio of the evolutionary-track radii, $(R_{\star,\mathrm{s}} / R_{\star,\mathrm{p}})^2 = 1.9^{+1.9}_{-1.1}$. The spectroscopic distance of the system, $d = 510^{+110}_{-100}\,\mathrm{pc}$, finally fits to its parallax, $\Pi^{-1} = 530^{+210}_{-120}\,\mathrm{pc}$. Apart from a slight tendency to a lower metallicity (see Table~\ref{table:stellar_parameters}), the chemical composition agrees with the reference stars of Sect.~\ref{subsection:singlestars}.
\paragraph*{HD\,75821 (\#4):} this eclipsing binary has a period of about $26.3$\,days and a radial velocity semi-amplitude of $92\,\mathrm{km\,s^{-1}}$ \citep{KXVel}. 

The spectrum best suited for the spectral analysis is the second one (b) in Table~\ref{table:atmospheric_parameters}, since the spectral line separation is largest in this case, which reveals several pure and unblended features of the companion (see Figs.~\ref{fig:binary_spectra_1}--\ref{fig:binary_spectra_9}). Reliable atmospheric and stellar parameters for both components are, hence, determinable whereby the latter assume that single-star evolutionary tracks are appropriate. Starting from this premise, the system consists of two coeval components ($\tau_{\mathrm{p}}=7^{+1}_{-1}\,\mathrm{Myr}$, $\tau_{\mathrm{s}}\leq 10\,\mathrm{Myr}$): a massive primary ($M_{\mathrm{p}} = 20.2^{+1.8}_{-1.5}\,M_{\sun}$), which is slightly evolved, and a less massive secondary ($M_{\mathrm{s}} = 11.5^{+2.0}_{-0.5}\,M_{\sun}$), which is almost unevolved (see Fig.~\ref{fig:evolution_tracks}). The spectroscopic distance $d = 960^{+240}_{-200}\,\mathrm{pc}$ lies well within the 99\%-uncertainty range of the parallax, $\Pi^{-1} = 1000^{+2300}_{-{\color{white}0}500}\,\mathrm{pc}$. Finally, the spectroscopically deduced effective surface ratio $A_{\mathrm{eff,s}}/A_{\mathrm{eff,p}} = 0.218^{+0.003+0.014}_{-0.002-0.011}$ agrees well with the squared ratio of the evolutionary-track radii, $(R_{\star,\mathrm{s}} / R_{\star,\mathrm{p}})^2 = 0.13^{+0.41}_{-0.04}$, and is further consistent with the photometric light curve \citep{KXVel2}. The elemental abundances of the system are in line with the single stars except for the relatively low helium, nitrogen, and oxygen content (see Table~\ref{table:atmospheric_parameters}). 

The heavily blended and, hence, almost vanishing imprints of the companion on the first (a) and third (c) spectrum are insufficient to properly constrain the secondary component's atmospheric parameters. Instead, unphysical values and large systematic uncertainties, which are induced by variations of $T_{\mathrm{eff}}$ and $\log(g)$ of the primary, are derived for the secondary's $T_{\mathrm{eff}}$ and $\log(g)$. These error margins are, on the one hand, a direct consequence of strong correlations among certain parameters and, on the other hand, related to the fact that contributions of the secondary component barely affect the spectrum at the corresponding orbital phases. In a simplified picture, increasing the primary's $T_{\mathrm{eff}}$ and decreasing its $\log(g)$ at the same time causes the \ion{He}{ii} lines to become considerably deeper than actually observed, while the \ion{He}{i} lines still fit nicely. To compensate for this, $A_{\mathrm{eff,s}}/A_{\mathrm{eff,p}}$ and, hence, the influence of the secondary component, has to be significantly increased to fill the \ion{He}{ii} lines with the continuum which thus weakens them again. However, this makes some spectral lines of the secondary component substantially too strong, which, in turn, is corrected for by smearing them out via a larger $\varv\,\sin(i)$ or $\zeta$ that finally leads to a more uncertain determination of $\varv_{\mathrm{rad}}$, given the high degree of line blending at these particular orbital phases. 

However, the primary's properties and the surface ratio $A_{\mathrm{eff,s}}/A_{\mathrm{eff,p}}$ are nicely recovered in all three orbital phases, which gives us confidence that the presented method is generally able to determine them from one single spectrum.
\section{Summary}\label{sec:summary}
In this paper, a novel objective method to analyze single or composite spectra of early-type stars is presented. It is based on fitting synthetic spectra to observation by using the standard concept of $\chi^2$ minimization, which requires the wavelength-dependent noise of the spectrum to be known. Therefore, a simple but precise way of estimating the local noise has been developed (see Sect.~\ref{subsection:noise_estimation}). To facilitate fast and efficient analyses, we make use of pre-calculated grids of synthetic spectra, instead of computing them on demand during the fitting procedure. To sample the entire multi-dimensional parameter space at once, we exploit the unique spectral properties of early-type stars, such as the low density of lines, which reduces the number of models required by several orders of magnitude. In this way, a simultaneous fit of all parameters is possible which has the great advantage that cumbersome iterations by hand or the risk of missing the global best solution are avoided. Moreover, parameters are not only constrained from a subset of available lines but from all useful features in the spectrum. The extension to composite spectra of double-lined binary systems proves extremely valuable in the future, given the high frequency of SB2 systems among early-type stars \citep[see][]{binaries_1,binaries_2}. In contrast to spectral disentangling techniques like those of \citet{spectral_disentangling_1} or \citet{spectral_disentangling_2}, the method presented here allows for --~at least~-- parameters of the primary and the components' effective surface ratio to be inferred from single-epoch spectra alone.

Statistical and systematic uncertainties of our method are discussed (see Sect.~\ref{subsection:uncertainties}). The former are based here on a clearly defined mathematical measure, namely the $\chi^2$ statistics, while the latter on experience. We show that systematic effects generally dominate in the high-quality regime of our observations. The analysis of a larger sample of stars thus enables us to identify possible shortcomings in our models and to derive results with significantly reduced statistical scatter. 

As a case study, we have determined parameters of three well-known stars in the Orion region that turn out to be in excellent agreement with previous studies. Additionally, three binary systems have been analyzed with all of them yielding very conclusive results. Consequently, we are now in a position to homogeneously analyze large samples of early-type stars in relatively short times. The results of a comprehensive investigation of $63$ nearby mid B-type to late O-type stars will be published in a forthcoming paper.
\begin{acknowledgements}
A.\,I.\ acknowledges support from a research scholarship by the Elite Network of Bavaria and subsequent support by the German Research Foundation (DFG) through grant He1356/45-2. M.-F.\,N.\ acknowledges an FFL stipend from the University of Erlangen-Nuremberg. We are very grateful to Thomas Dauser, Ingo Kreykenbohm, and Fritz-Walter Schwarm for maintaining and improving the Remeis computer cluster which was used for this study, to John E.\ Davis for the development of the {\sc slxfig} module used to prepare the figures in this paper, to J\"orn Wilms for proposing the $\chi^2$ strategy in combination with ISIS, and to Markus Firnstein for providing joint profiles of macroturbulence and stellar rotation obtained from numerical disk integration. We thank Horst Drechsel and Pavel Mayer for drawing our attention to HD\,75821 and for discussions related to it. Mahalo nui loa to Rolf-Peter~Kudritzki and the University of Hawai'i for their great hospitality during A.\,I.'s research stay at the Institute for Astronomy and to the German Academic Exchange Service (DAAD) as well as to the Physics Advanced study program for financial support of this stay on which parts of this paper are based on. We are grateful to Artemio Herrero for a competent and constructive referee report.
\end{acknowledgements}
\bibliographystyle{aa}

\end{document}